\documentclass[12pt,a4paper]{article}
\usepackage{makeidx}
\usepackage{latexsym,amsmath,amssymb,amsfonts,amsthm}

\newtheorem{definition}{Definition}[section]
\newtheorem{corollary}{Corollary}[section]
\newtheorem{theorem}{Theorem}[section]
\newtheorem{remark}{Remark}[section]
\def\nabla{\bigtriangledown}

\begin{document}

\title{ Nonholonomic Deformations of Disk\\
Solutions and Algebroid Symmetries in\\
Einstein and Extra Dimension Gravity }
\author{ Sergiu I. Vacaru\thanks{%
e--mail: vacaru@imaff.cfmac.csic.es } \\
%EndAName
--- \\
{\small Instituto de Matematicas y Fisica Fundamental,}\\
{\small Consejo Superior de Investigaciones Cientificas,}\\
{\small Calle Serrano 123, Madrid 28006, Spain } }
\date{April 20, 2005}
\maketitle

\begin{abstract}
In this article we consider nonholonomic deformations of disk
solutions in general relativity to generic off--diagonal metrics
defining knew classes of exact solutions in 4D and 5D gravity.
These solutions possess Lie algebroid symmetries and local
anisotropy and define certain generalizations of manifolds with
Killing and/ or Lie algebra symmetries. For Lie algebroids, there
are structures functions depending on variables on a base
submanifold and it is possible to work with singular structures
defined by the 'anchor' map. This results in a number of new
physical implications comparing with the usual manifolds
possessing Lie algebra symmetries defined by structure constants.
The spacetimes investigated here have two physically distinct
properties: First, they can give rise to disk type configurations
with angular/ time/ extra dimension gravitational polarizations
and running constants. Second, they define static, stationary or
moving disks in nontrivial solitonic backgrounds, with possible
warped factors, additional spinor and/or noncommutative
symmetries. Such metrics may have nontrivial limits to 4D gravity
with vanishing, or nonzero torsion. The work develops the results
of Ref. \cite{vjhep2} and  emphasizes the solutions with Lie
algebroid symmetries following similar constructions for solutions
with noncommutative symmetries \cite{vnces}.

\newpage


\vskip0.1cm

2000 AMS Subject Classification:\

17B99,\ 53A40,\ 53A99,\ 83C15,\ 83C20,\ 83E99
\end{abstract}

%\newpage

\tableofcontents

\section{Introduction}

Disk solutions have been studied intensively in the past for certain
important reasons: In astrophysics, disk configurations were discussed as
models of certain type galaxies and/or for accretion disks, see e.g. \cite%
{binney}. The Newtonian dust disks are known to be unstable and the same
holds in the relativistic case. This was used for a certain numerical work
when disk type solutions could be taken as exact initial data for numerical
collapse calculations \cite{bawa}. In the relativistic case, the static
disks solutions where obtained as consisting of two counter-rotating streams
of matter with vanishing total angular momentum \cite{morgan}. Infinitely
extended dust disks with finite mass were studied in the static case and
stationary cases \cite{blk}. In analytic form, the first exact solution for
a finite stationary dust disk was constructed and analyzed in Refs. \cite{nm}%
. The solution was given for the rigidly rotating dust disk when the
corresponding boundary value problem for the Einstein equations was solved
with the help of a co--rotating coordinate system. Then, the approach was
extended to electro--vacuum solutions in Ref. \cite{pslet}. A comprehensive
review of black hole -- disk systems and astrophysical disks is presented in
Ref. \cite{rdisk}.

The main difficulties for constructing disk solutions in arbitrary
backgrounds (in general, anisotropically polarized and with extra dimension
dependence and different type of symmetries) is related to technical
analytical and numerical problems of integrating the gravitational and
matter field equations. Our goal in the present paper is to develop a
geometric method of generating such solutions depending on two, three or
four variables and to investigate the properties of a such new class of
spacetimes. We are motivated to do so because any realistic approach to
modern quantum gravity, low--energy string limit and extra dimension physics
and phenomenology are related to nonlinear gravitational configurations,
nonholonomic frames and generalized symmetries. An alternative motivation is
also the study of disk configurations as possible tests of extra dimensions,
torsion fields and topologically nontrivial objects and local anisotropies
induced by off--diagonal interactions in the Einstein or string/brane
gravity and possible implications in modern cosmology and astrophysics.

In this paper, we adopt our previous strategy for constructing
exact solutions by using nonholonomic deformations of certain well
known exact solutions formally defined by diagonalizable (by
coordinate transforms) metric structures
\cite{vjhep2,vsbd,vs,vp,vt,vnces}. We consider both the geometry
of nonholonomic frames (vielbeins) with associated nonlinear
connection (N--connection) structure and the manifolds with
metrics and connections possessing generalized Lie algebroid
symmetries. Although the gravitational algebroid disk symmetries
are discussed only in brief in this paper, they emphasize and
conclude the idea of generalized symmetries for 'off--diagonal
spacetimes; see details in Refs. \cite{galg}, on Lie algebroid
mathematics and recent applications to strings and field theory
and geometric mechanics, and Refs. \cite{valg} on Clifford
algebroids and exact solutions with Lie algebroid symmetry.

We emphasize that the anholonomic frame method applied in this
work is a general one allowing to construct new types of exact
solutions of the Einstein (vacuum and nonvacuum ones) and matter
field equations reduced by generic off--diagonal ansatz to systems
of nonlinear partial differential equations. Surprisingly, this
rigorous geometric technique allows to integrate the equations in
very general form. Such exact solutions depend on classes of
integration functions on one, two and three variables which is
very different from the usual ansatz with spherical (cylindrical
and/or Lie algebra symmetry) reducing the field equations to
algebraic systems or to certain types of nonlinear differential
equations when their general solutions depend on integration
constants.

The gravitational solutions depending on integration constants,
para\-met\-rized by  'diagonalizable' (by coordinate transforms)
metrics and subjected to the limit conditions to result in
asymptotic Minkowski, or constant curvature spacetimes, are
largely applied in modern astrophysics and cosmology (for
instance, the are related to black hole physics and inflational
scenaria). The most important classes of such solutions are
summarized in Ref. \cite{exac} and have been generalized for a
number of modern string/brane gravitational theories.

Nevertheless, some already elaborated geometric, analytic and
numerical methods give rise to more 'sophisticate' classes of
solutions for vacuum and non--vacuum configurations. They possess
new symmetries and depend on classes of integration functions and
may describe a new 'non-perturbative' and 'nonlinear'
gravitational physics. It is an important and difficult task to
investigate the physical implications of such solutions. For
instance, in Ref. \cite{vnces}, we presented a detailed study of
such solutions possessing noncommutative symmetries even they are
constructed in the framework of the 'commutative' Einstein gravity
and its 'commutative' generalizations. Than, it was understood
that the generalized off--diagonal solutions can be also described
and classified by Lie algebroid symmetries \cite{valg}. In this
case, one deals with structure functions instead of Lie algebra
constants, one may consider singular maps and unify certain
concepts of bundle spaces and nonholonomic manifolds, i.e.
manifolds provided with nonintegrable distributions defined by
nonholonomic frames and their deformations. The general aim of
this paper, is to investigate in detail some classes of such
solutions generated by nonholonomic deformations of disk
solutions.

It is suggested that the reader will study in advance the Appendix
containing the main geometric results on the "anholonomic frame
method" before he will consider the content of sections 2--6. We
consider necessary and useful to give a number of technical
details on nonholonomic differential calculus and proofs related
to this method still less known for researches working in gravity
and string theories and their applications in astrophysics and
cosmology.

We organize our presentation as follows:

The section 2 outlines the main results on disk solutions in
general relativity and their conformal transforms and nonholonomic
deformations to exact off--diagonal solutions in 4D gravity. There
are analyzed two vacuum disk configurations with angular
anisotropy: disk solutions with induced torsion and anisotropic
disks in general relativity. Other class of solutions is
constructed in order to investigate possible disk anisotropies
induced by a cosmological constant.

The section 3 is devoted to nonholonomic disk configurations in
bosonic string gravity. There are considered two types of torsion
fields, the nontrivial string torsion ($H$--field) and a
nonholonomically induced torsion, when the dilaton vacuum is
constant and the 5D $H$--field ansatz is self--consistently
related to a nonholonomic gravitational backround in such a manner
that their contributions can be approximated by an effective
cosmological constant acting as a source of local anisotropy. We
analyze four classes of disk like solutions with $H$--field
contributions and nontrivial gravitational 3D solitonic
backgrounds: 1) string--solitonic disks with angular anisotropy;
2) (time) moving string--solitonic disc configurations; 3) disks
with anisotropic polarizations on extra dimension; 4) extra
dimension solitonically propagating disks.

The section 4 contains a study of the Dirac equation on
nonholonomic gravitational disk configurations. There were
constructed two classes of such solutions of the Einstein--Dirac
equations defining nonlinear supperpositions for gravitational
disk -- 3D packages of Dirac--waves: 1) metrics with angular
anisotropy and 2) metrics with time moving nonholonomic spinor
waves.

The section 5 presents the main ideas and formulas for
gravitational Lie algebroids provided with N--connection structure
and illustrate  how metrics with Lie algebroid symmetry can be
defined following the anholonomic frame method. It is shown that
the constructions can be adapted to nonholonomic configurations.

The section 6 contains a detailed discussion, motivation of the
applied geometric methods and conclusion of results.

\section{Nonholonomic Deformations of Disk Solutions}

In this section we consider nonholonomic deformations of the disk solutions
in general relativity to four and five dimensional (in brief, 4D and 5D)
off--diagonal metric ansatz\footnote{%
Such metrics can not be diagonalized by coordinate transforms but
can be written in effective diagonal forms with respect to some
nonholonomic local bases, equivalently, frames, or vielbeins.}
defining also exact solutions in gravity theories, in general,
with nontrivial torsion. We analyze two classes of such spacetimes
and state the conditions when the generic off--diagonal metrics
define exact solutions in vacuum Einstein gravity and/or in
presence of sources of local anisotropy defined by a nontrivial
cosmological constant.

\subsection{Disk solutions in general relativity and their conformal and
an\-ho\-lo\-nomic transforms}

In order to construct 4D disk type solutions one usually uses a metric
ansatz describing the exterior (i.e.\ the vacuum region) of an axisymmetric,
stationary rotating body, written in the Weyl--Lewis--Papapetrou form \cite%
{exac},
\begin{equation}
ds_{[dk]}^{2}=-e^{2U}(dt+ad\phi )^{2}+e^{2(k-U)}(d\widehat{\rho }^{2}+d%
\widehat{\zeta }^{2})+e^{-2U}\widehat{\rho }^{2}d\phi ^{2},  \label{vac1}
\end{equation}%
where $\partial _{t}$ and $\partial _{\phi }$ are two commuting
asymptotically timelike respectively spacelike Killing vectors and the
functions $U(\widehat{\rho },\widehat{\zeta }),a(\widehat{\rho },\widehat{%
\zeta })$ and $k(\widehat{\rho },\widehat{\zeta })$ are taken in a form when
the metric defines a solution of the Einstein equations. For a such ansatz,
the vacuum field equations are equivalent to the Ernst equation for the
potential $q(z,\bar{z})$,
\begin{equation}
q_{z\bar{z}}+\frac{1}{2(z+\bar{z})}(q_{\bar{z}}+q_{z})-\frac{2}{q+\bar{q}}%
q_{z}q_{\bar{z}}=0,  \label{vac1a}
\end{equation}%
where the (complex) Ernst potential is defined by $q=e^{2U}+\mathrm{i}b,$ $%
\bar{q}$ is the complex conjugated coordinate and the real function $b$ is
related to coefficients in the metric via $b_{z}=-(\mathrm{i}/\widehat{\rho }%
)e^{4U}a_{z}.$ The complex variable is defined by $z=\widehat{\rho }+\mathrm{%
i}\widehat{\zeta }$. \footnote{We note that some our denotations
defer from, for instance, those in \cite{nm}.}

Haven chosen a solution of the Ernst equation, the metric function $U$ is
obtained directly from the definition of the Ernst potential whereas $a$ and
$k$ can be constructed from $q$ via quadratures. One uses units where
Newton's gravitational constant $G$ as well as the velocity of light $c$ are
equal to 1. The metric functions $e^{2U}$, $e^{2k}$, and $a/\rho _{0}$
depend uniquely on the normalized coordinates $\widehat{\rho }/\rho _{0}$, $%
\widehat{\zeta }/\rho _{0}$, and the parameter $\mu =2\Omega ^{2}\rho
_{0}^{2}e^{-2V_{0}},$ where $\Omega ,$ $\rho _{0}$ and $V_{0}$ are
respectively the angular velocity, the coordinate radius, and the `surface
potential' $V_{o}(\mu )\equiv U(\widehat{\rho }=0,\widehat{\zeta }=0,\mu $),
respectively. The disk is characterized by the values of coordinates $%
\widehat{\zeta }=0$ and $0\leq \widehat{\rho }\leq \rho _{0}.$

The equation (\ref{vac1a}) is completely integrable. In principle, this also
allows to get explicit solutions for the boundary value problems. For \
infinitesimally thin disks, i. e. for two-dimensionally extended matter
sources, one generates global solutions defined by ordinary differential
equations in the matter region which state boundary data for the vacuum
equations (see details in Refs.\ \cite{prd}). This way one constructs
solutions to the Ernst equation with free functions which have to be
determined by the boundary data. A such solution is unique in the case of
Cauchy data and, in general, is related to singular solutions of the
elliptic field equations, we refer the reader to Refs. \cite{nm} for
explicit constructions and discussions.

In this paper, we consider that the functions $U(\widehat{\rho },\widehat{%
\zeta }),a(\widehat{\rho },\widehat{\zeta })$ and $k(\widehat{\rho },%
\widehat{\zeta })$ for the metric coefficients in (\ref{vac1}) are known
from a 4D disk solution defined in explicit form for some boundary data. Our
purpose is to deform this metric to certain 4D and 5D generic off--diagonal
configurations defining other classes of exact solutions. For such
constructions, it is convenient to consider two auxiliary metrics (in
general, they do not define any solutions of the Einstein equations) which
are conformally equivalent to the mentioned disk solution:

\begin{eqnarray}
ds_{[cd1]}^{2} &=&e^{2(U-k)}ds_{[dk]}^{2}  \label{auxm1} \\
&=&\rho _{0}^{2}d\rho ^{2}+\rho _{0}^{2}d\zeta ^{2}+\rho _{0}^{2}\rho
^{2}e^{-2k}d\varphi ^{2}-e^{2(2U-k)}(d\tau +a\ d\zeta )^{2}  \notag
\end{eqnarray}%
and
\begin{eqnarray}
ds_{[cd2]}^{2} &=&\rho _{0}^{-2}\rho ^{-2}e^{2k}ds_{[cd1]}^{2}  \label{auxm2}
\\
&=&\rho ^{-2}e^{2k}d\rho ^{2}+\rho ^{-2}e^{2k}d\zeta ^{2}+d\varphi ^{2}-\rho
_{0}^{-2}\rho ^{-2}e^{4U}(d\tau +a\ d\zeta )^{2}  \notag
\end{eqnarray}%
where there are introduced the 'normalized' coordinates $\rho =\widehat{\rho
}/\rho _{0}$ and\ $\zeta =\widehat{\zeta }/\zeta _{0}$ and, for simplicity,
we consider that $\rho \neq 0$ for the metric (\ref{auxm2}) (we have to
change the system of coordinates in order analyze the second metric in the
vicinity of $\rho =0).$ We also note that we introduced a new time like
coordinate
\begin{equation*}
t\rightarrow \tau =t+\int \omega ^{-1}(\zeta ,\varphi )\ d\xi (\zeta
,\varphi )
\end{equation*}%
for which
\begin{equation*}
d\tau +a\ d\zeta =dt+a\ d\varphi ,
\end{equation*}%
i. e. $d(\tau -t)=\omega ^{-1}d\xi .$ This is possible, for instance, if we
chose the integrating function $\omega =a^{-1}e^{\zeta -\varphi }$ and put $%
\xi =-e^{\zeta -\varphi }.$

The auxiliary metrics (\ref{auxm1}) and (\ref{auxm2}) can be distinguished
from an off--diagonal metric form (with a trivial extension to the 5th
coordinate),
\begin{eqnarray*}
ds_{[cd]}^{2} &=&\epsilon (dx^{1})^{2}+\ \underline{g}_{\widehat{k}}(x^{%
\widehat{i}})\left( dx^{\widehat{k}}\right) ^{2}+\underline{h}_{a}(x^{%
\widehat{i}})\left( \underline{e}^{a}\right) ^{2}, \\
\underline{e}^{a} &=&dy^{a}+\underline{N}_{i}^{a}(x^{\widehat{i}})dx^{i},
\end{eqnarray*}%
with $\widehat{i},\widehat{k},...=2,3;$ $a,b,...=4,5$ and $\epsilon =\pm 1$
(the sign depends on chosen signature for the dimension labelled by the
coordinate $x^{1}$). The data for the coordinates and nontrivial values of
the metric and local frame (vielbein) coefficients\ in (\ref{auxm1}) are
parametrized
\begin{eqnarray}
x^{1} &=&\chi ,x^{2}=\rho ,x^{3}=\zeta ,y^{4}=v=\varphi ,y^{5}=\tau ,  \notag
\\
\underline{g}_{1} &=&\epsilon ,\underline{g}_{2}=\rho _{0}^{2},\underline{g}%
_{3}=\rho _{0}^{2},\underline{h}_{4}=\rho _{0}^{2}\rho ^{2}e^{-2k},%
\underline{h}_{5}=-e^{2(2U-k)},\underline{N}_{3}^{5}=a,  \label{auxd1a}
\end{eqnarray}%
or
\begin{eqnarray}
x^{1} &=&\chi ,x^{2}=\rho ,x^{3}=\zeta ,y^{4}=v=\tau ,y^{5}=\varphi ,  \notag
\\
\underline{g}_{1} &=&\epsilon ,\underline{g}_{2}=\rho _{0}^{2},\underline{g}%
_{3}=\rho _{0}^{2},\underline{h}_{4}=-e^{2(2U-k)},\underline{h}_{5}=\rho
_{0}^{2}\rho ^{2}e^{-2k},\underline{N}_{3}^{4}=a,  \label{auxd1b}
\end{eqnarray}%
In a similar form, we can define the trivial 5D extensions of (\ref{auxm2})
by considering that
\begin{eqnarray}
x^{1} &=&\varphi ,x^{2}=\rho ,x^{3}=\zeta ,y^{4}=v=\chi ,y^{5}=\tau ,  \notag
\\
\underline{g}_{1} &=&1,\underline{g}_{2}=\rho ^{-2}e^{2k},\underline{g}%
_{3}=\rho ^{-2}e^{2k},\underline{h}_{4}=\epsilon ,\underline{h}_{5}=-\rho
_{0}^{-2}\rho ^{-2}e^{4U},\underline{N}_{3}^{5}=a,  \label{auxd2a}
\end{eqnarray}%
or
\begin{eqnarray}
x^{1} &=&\varphi ,x^{2}=\rho ,x^{3}=\zeta ,y^{4}=v=\tau ,y^{5}=\chi ,  \notag
\\
\underline{g}_{1} &=&1,\underline{g}_{2}=\rho ^{-2}e^{2k},\underline{g}%
_{3}=\rho ^{-2}e^{2k},\underline{h}_{4}=-\rho _{0}^{-2}\rho ^{-2}e^{4U},%
\underline{h}_{5}=\epsilon ,\underline{N}_{3}^{4}=a.  \label{auxd2b}
\end{eqnarray}%
In the above presented data, the coordinate $\chi $ is considered to be the
extra dimensional one.

By introducing $\underline{g}_{i}=(\underline{g}_{1},\underline{g}_{\widehat{%
i}}),$ when indices of type $i,j,k,...$ run values $1,2,3,$ we define a 5D
metric ansatz
\begin{eqnarray}
ds_{[cd]}^{2} &=&\underline{g}_{k}(x^{\widehat{i}})\left( dx^{k}\right) ^{2}+%
\underline{h}_{a}(x^{\widehat{i}})\left( \underline{e}^{a}\right) ^{2},
\label{ans4dd} \\
\underline{e}^{a} &=&dy^{a}+\underline{N}_{i}^{a}(x^{\widehat{i}})dx^{i},
\notag
\end{eqnarray}%
which a particular case of the metric (\ref{ans4d}) considered in Appendix.
We nonholonomically transform (deform) the coefficients of (\ref{ans4dd}),%
\begin{eqnarray}
\underline{g}_{k}(x^{\widehat{i}}) &\rightarrow &g_{k}(x^{\widehat{i}})=\eta
_{k}(x^{\widehat{i}})\underline{g}_{k}(x^{\widehat{i}}),\   \label{polf} \\
\underline{h}_{a}(x^{\widehat{i}}) &\rightarrow &h_{a}(x^{i},v)=\eta
_{a}(x^{i},v)\underline{h}_{a}(x^{\widehat{i}}),  \notag \\
\underline{N}_{i}^{a}(x^{\widehat{i}}) &\rightarrow &N_{i}^{a}(x^{k},v),
\notag
\end{eqnarray}%
with $\eta _{1}=1,$ to a general5D off--diagonal metric ansatz%
\begin{eqnarray}
ds_{[5D]}^{2} &=&g_{k}(x^{\widehat{i}})\left( dx^{k}\right)
^{2}+h_{a}(x^{i},v)\left( e^{a}\right) ^{2},  \label{ans5d} \\
e^{a} &=&dy^{a}+N_{i}^{a}(x^{k},v)dx^{i},  \notag
\end{eqnarray}%
with the coefficients not depending on variable $y^{5}$ but with
respective dependencies on $x^{i}$ (the isotropic, or holonomic
variables) and on $v$ (the anisotropic, or anholonomic variable).
We conventionally introduced the ''polarization'' functions $\eta
_{\alpha }=(\eta _{k},\eta _{a})$ and the
N--connection coefficients $N_{i}^{a}$ (see formulas (\ref{block2}), (\ref%
{ddif}) and (\ref{dder})) which may have corresponding limits $\eta _{\alpha
}\rightarrow 1$ and $N_{i}^{a}\rightarrow \underline{N}_{i}^{a}$ related to
a conformal transform and extra dimension extension of the disk solution (%
\ref{vac1}) to an auxiliar metric (\ref{auxm1}), with parametrizations (\ref%
{auxd1a}) or (\ref{auxd1b}), or to (\ref{auxm2}), with parametrizations (\ref%
{auxd2a}) or (\ref{auxd2b}).

The values $\eta _{\alpha }$ and $N_{i}^{a}$ will be found as certain
general classes of functions solving the gravitational field equations for
some vacuum or nonvacuum models in 4D or 5D gravity. This way we shall
define nonholonomic deformations of the disk solutions in general relativity
to new classes of exact solutions in various types of gravity theories (in
general, with nontrivial torsion and matter field sources and extra
dimensions).

\subsection{A metric ansatz with angular anisotropy}

The idea to deform nonholonomically the Neugebauer--Meinel disk solution %
\cite{nm} to a certain generic off--diagonal vacuum or non--vacuum
gravitational anisotropic (on variable $y^{4}=v=\varphi )$ configuration was
proposed in Ref. \cite{vjhep2}. That approach was valid for small
deformations on $v$ of the Ernst potential and corresponding hyperelliptic
functions. In this subsection, we analyze two classes of such locally
anisotropic solutions in 4D which hold for any type of nonhlonomic
deformations (not only for small ones), see the related geometric formalism
outlined in Appendix \ref{ss4d}.

We nonholonomically deform the ansatz (\ref{auxm1}), defined by data (\ref%
{auxd1a}) reduced to 4D, i. e. without coordinate $x^{1},$ to a 4D generic
off--diagonal metric%
\begin{eqnarray}
\delta s_{[4D\varphi ]}^{2} &=&\rho _{0}^{2}\varpi (\rho ,\zeta )d\rho
^{2}+\rho _{0}^{2}\varpi (\rho ,\zeta )d\zeta ^{2}  \notag \\
&&+\rho _{0}^{2}\rho ^{2}e^{-2\widetilde{k}(\rho ,\zeta ,\varphi )}\delta
\varphi ^{2}-e^{2[2\widetilde{U}(\rho ,\zeta ,\varphi )-\widetilde{k}(\rho
,\zeta ,\varphi )]}\delta \tau ^{2},  \notag \\
\delta \varphi &=&d\varphi +w_{2}(\rho ,\zeta ,\varphi )\ d\rho +w_{3}(\rho
,\zeta ,\varphi )\ d\zeta ,  \label{dsol4d1} \\
\delta \tau &=&d\tau +n_{2}(\rho ,\zeta ,\varphi )\ d\rho +n_{3}(\rho ,\zeta
,\varphi )\ d\zeta ,  \notag
\end{eqnarray}%
where it is emphasized the dependence on the nonholonomic (anisotropic)
coordinate $\varphi.$ The polarizations functions and N--coefficients (\ref%
{polf}) are paramet\-riz\-ed in this form:%
\begin{eqnarray*}
\eta _{2}(\rho ,\zeta ) &=&\varpi (\rho ,\zeta ),\ \eta _{3}(\rho ,\zeta
)=\varpi (\rho ,\zeta ),\eta _{4}(\rho ,\zeta ,\varphi )=e^{-2[\widetilde{k}%
(\rho ,\zeta ,\varphi )-k(\rho ,\zeta )]}, \\
\eta _{5}(\rho ,\zeta ,\varphi ) &=&e^{2[2\widetilde{U}(\rho ,\zeta ,\varphi
)-\widetilde{k}(\rho ,\zeta ,\varphi )]-2[2U(\rho ,\zeta )-k(\rho ,\zeta )]},
\\
N_{2,3}^{4}(\rho ,\zeta ,\varphi ) &=&w_{2,3}(\rho ,\zeta ,\varphi ),\
N_{2,3}^{5}(\rho ,\zeta ,\varphi )=n_{2,3}(\rho ,\zeta ,\varphi ).
\end{eqnarray*}%
For some classes of solutions, we may consider the limits
\begin{equation}
\eta \rightarrow 1,\eta _{4}\rightarrow 1,\eta _{5}\rightarrow
1,w_{2,3}\rightarrow 0,n_{2}\rightarrow 0,N_{3}^{5}=n_{3}(\rho ,\zeta
,\varphi )\rightarrow \underline{N}_{3}^{5}=a(\rho ,\zeta )  \label{limit1}
\end{equation}%
for $\widetilde{k}\rightarrow k$ and $\widetilde{U}\rightarrow U$ which
allows us to use the same boundary conditions as for the non--deformed
metrics and a very similar physical interpretation like in the usual
'locally isotropic' disk solutions. But it is not possible to satisfy all
such limits and to preserve the condition that all metrics related by such
nonholonomic deformations are just solutions of the gravitational field
equations. The rigorous procedure is to find certain values $\eta _{\alpha }$
and $N_{i}^{a}$ for which certain vacuum, or non--vacuum, solutions are
defined and then to analyze which limits of type (\ref{limit1}) are
possible. In general, two exact solutions may be not related by smooth
infinitesimal limits to a fixed type one but they could be related by
certain nontrivial conformal and nonholonomic frame transforms. Such metrics
may also possess different generalized symmetries, for instance, they may
possess Lie  algebroid symmetries, see a detailed discussion in Section \ref%
{sgdal}.

Our aim is to find such functions
\begin{equation*}
\widetilde{U}(\rho ,\zeta ,\varphi ),\widetilde{k}(\rho ,\zeta ,\varphi
),w_{2,3}(\rho ,\zeta ,\varphi ),n_{2,3}(\rho ,\zeta ,\varphi )
\end{equation*}%
when gravitational polarizations (\ref{polf}) define a class of exact
solutions of the 4D Einstein equations for the canonical d--connection (\ref%
{candcon}) with zero or nonzero cosmological constant $\Upsilon
_{2,4}=\Upsilon _{0}=const$ (\ref{emcond4}), see the method presented in
Appendix \ref{ss4d}. We shall also investigate the conditions for the metric
coefficients when the ansatz (\ref{dsol4d1}) defines vacuum solutions for
the Levi--Civita connection, i. e. we shall construct generic off--diagonal
solutions with anisotropic dependence on angular variable $\varphi $ in
general relativity.

\subsubsection{4D vacuum angular anisotropic disk configurations}

For vacuum configurations, the function $\varpi (\rho ,\zeta )$ from (\ref%
{dsol4d1}) can be defined for any $2D$ coordinate transforms $\tilde{x}%
^{2,3}(\rho ,\zeta )$ when
\begin{equation*}
\rho _{0}^{2}\varpi (\rho ,\zeta )d\rho ^{2}+\rho _{0}^{2}\varpi (\rho
,\zeta )d\zeta ^{2}=\varpi (\rho ,\zeta )\left[ \left( d\tilde{x}^{2}\right)
^{2}+\left( d\tilde{x}^{3}\right) ^{2}\right]
\end{equation*}%
and $\psi =\ln |\varpi |$ is any solution of
\begin{equation*}
\psi ^{\bullet \bullet }+\psi ^{^{\prime \prime }}=0
\end{equation*}%
for $\psi ^{\bullet }=\partial \psi /\partial x^{2}$ and $\psi ^{^{\prime
}}=\partial \psi /\partial x^{3},$ see formulas (\ref{con10}) and (\ref%
{auxeq01}) in Appendix. In a particular vacuum case, we can consider $\varpi
=1$ and $\tilde{x}^{2}=\rho ,\tilde{x}^{3}=\zeta .$

\paragraph{Anisotropic disks with induced torsion:}

The functions $\eta _{4}(\rho ,\zeta ,\varphi )$ and $\eta _{5}(\rho ,\zeta
,\varphi ),$ i. e. $h_{4}=\eta _{4}\underline{h}_{4}$ and $h_{5}=\eta _{5}%
\underline{h}_{5},$ for the values $\underline{h}_{4}$ and $\underline{h}%
_{5} $ (\ref{auxd1a}) defined by the disk solution (\ref{vac1}) and the
conformally transformed metric (\ref{vac1a}), are related by formulas (\ref%
{p2}), or (\ref{p1}), solving the equation (\ref{ep2a}), see Theorem \ref%
{texs} in Appendix. For $h_{5}^{\ast }=\partial h_{5}/\partial \varphi \neq
0,$ we can satisfy the condition $\sqrt{|h_{4}|}=h_{[0]}(\sqrt{|h_{5}}%
)^{\ast },$ by taking
\begin{equation}
h_{4}=\rho _{0}^{2}\rho ^{2}e^{-2\widetilde{k}(\rho ,\zeta ,\varphi )}%
\mbox{
and }h_{5}=-\widetilde{k}^{2}e^{2(2U-\widetilde{k})}.  \label{a21}
\end{equation}%
where
\begin{equation}
h_{[0]}(\rho ,\zeta )=\rho _{0}\rho e^{U(\rho ,\zeta )-2k(\rho ,\zeta )}
\label{a20}
\end{equation}%
for some functions $\widetilde{U}(\rho ,\zeta ,\varphi )$ and $\widetilde{k}%
(\rho ,\zeta ,\varphi )$ subjected to the relation $\widetilde{k}=e^{2(%
\widetilde{U}-U)}.$

For a vacuum configuration, we can chose a value of the integration function
$h_{[0]}(\rho ,\zeta )$ when $w_{2,3}=0,$ see the discussion after formula (%
\ref{w}), but this would define more cumbersome formulas for $h_{4}$ and $%
h_{5}.$ It is convenient to consider non--vanishing N--connection
coefficients $w_{2,3}$ defined by differentiating on coordinates a
functional $s$ on $\widetilde{k},U$ and $\rho _{0},\rho ,$
\begin{equation*}
s(\widetilde{k},U,\rho _{0},\rho )=\ln |\sqrt{|h_{4}h_{5}|}/h_{5}^{\ast
}|=\ln |\rho _{0}\rho e^{\widetilde{k}-2U}/\left( \widetilde{k}-1\right)
\widetilde{k}^{\ast }|
\end{equation*}%
computed by using data (\ref{a21}). In result, we get
\begin{equation}
w_{2}=\left( s^{\ast }\right) ^{-1}\frac{\partial s}{\partial \rho }%
\mbox{
and }w_{3}=\left( s^{\ast }\right) ^{-1}\frac{\partial s}{\partial \zeta },
\label{a22}
\end{equation}%
which solve the conditions (\ref{w}). We have to impose the limits $%
w_{2,3}\rightarrow 0$ stating certain limits for $s$ and $\widetilde{k}$ for
large values of $\rho $ and/or $\zeta ,$ in order to consider that the
locally anisotropic solutions will result in the usual disk metric (\ref%
{vac1}) far away from the locally anisotropic disk configuration.

The rest of N--connection coefficients $n_{2,3}$ are computed by introducing
(\ref{a21}) into the first formula (\ref{n})%
\begin{equation}
n_{2,3}=n_{2,3[1]}\left( \rho ,\zeta \right) +n_{2,3[2]}\left( \rho ,\zeta
\right) \int \widetilde{k}^{-3}e^{\widetilde{k}}d\varphi  \label{a23}
\end{equation}%
where we introduced all possible dependencies on $\rho $ and $\zeta $ from $%
h_{4,5}$ in $n_{2,3[2]}.$ The functions $n_{2,3[1,2]}\left( \rho ,\zeta
\right) $ have to be defined from certain boundary/limit conditions for a
class of solutions. We can state, in a more particular case, that $%
n_{2[1]},n_{2[2]}=0$ and write
\begin{equation}
n_{3}=a\left( \rho ,\zeta \right) \left[ 1+n_{3[2]}\left( \rho ,\zeta
\right) \int \widetilde{k}^{-3}e^{\widetilde{k}}d\varphi \right] \rightarrow
a\left( \rho ,\zeta \right)  \label{lim2}
\end{equation}%
for large values of $\rho $ and/or $\zeta $ if we are analyzing some small
''anisotropic'' corrections being proportional to some small values $%
n_{3[2]}.$

Introducing the coefficients (\ref{a21})--(\ref{a23}) \ into (\ref{dsol4d1}%
), we get the off--diagonal metric%
\begin{eqnarray}
\delta s_{[4D\varphi ]}^{2} &=&\rho _{0}^{2}d\rho ^{2}+\rho _{0}^{2}d\zeta
^{2}  \label{sol1} \\
&&+\rho _{0}^{2}\rho ^{2}e^{-2\widetilde{k}(\rho ,\zeta ,\varphi )}\delta
\varphi ^{2}-\widetilde{k}^{2}(\rho ,\zeta ,\varphi )e^{2[2U(\rho ,\zeta )-%
\widetilde{k}(\rho ,\zeta ,\varphi )]}\delta \tau ^{2},  \notag \\
\delta \tau &=&d\tau +a\left( \rho ,\zeta \right) \left[ 1+n_{3[2]}\left(
\rho ,\zeta \right) \int \widetilde{k}^{-3}e^{\widetilde{k}}d\varphi \right]
\ d\zeta ,  \notag \\
\delta \varphi &=&d\varphi +\frac{1}{s^{\ast }}\left( \frac{\partial s}{%
\partial \rho }\ d\rho +\frac{\partial s}{\partial \zeta }\ d\zeta \right) ,%
\mbox{ for }  \notag \\
s &=&\ln \left| \rho _{0}\rho e^{\widetilde{k}-2U}/\left( \widetilde{k}%
-1\right) \widetilde{k}^{\ast }\right| ,  \notag
\end{eqnarray}%
where, for simplicity, there are considered necessary smooth class
coefficients. This ansatz defines a class of vacuum 4D solutions for the
Einstein equations with the Ricci tensor computed for the canonical
d--connection (\ref{candcon}). The metric coefficients depend on an
arbitrary function $\widetilde{k}(\rho ,\zeta ,\varphi ),$ on integration
functions $n_{3[1]}\left( \rho ,\zeta \right) $ and $n_{3[2]}\left( \rho
,\zeta \right) $ and on function $U(\rho ,\zeta )$ defined by the former
Neugebauer--Meinel disk solution (\ref{vac1}). We note that the metric (\ref%
{sol1}) also depends on two another locally isotropic disk functions $%
a\left( \rho ,\zeta \right) $ and $k(\rho ,\zeta )$ as it is stated by the
boundary conditions for the integration functions (\ref{a20}) and (\ref{lim2}%
). Such spacetimes are with nonholonmically induced torsion (its
coefficients, defined by the off--diagonal metric terms, may be computed by
introducing the N--coefficients (\ref{a22}) and (\ref{a23}) into the
formulas (\ref{dtorsb})). In the next sections, we shall use certain ansatz
generalizing (\ref{sol1}) in order to investigate certain locally
anisotropic disk solutions in string gravity with nontrivial torsion. We
note that such metrics may be used for various calculus concerning physical
effects with nonholonomic frames and tests of general relativity.

\paragraph{Anisotropic disks in general relativity:}

We can impose certain constraints on the coefficients of the ansatz (\ref%
{dsol4d1}) in order to generate a class of exact locally anisotropic disk
solutions just for the vacuum Einstein gravity. In this case we must satisfy
the conditions (\ref{aux32}) and (\ref{aux31}), see Appendix, stating that
the nontrivial coefficients of the Ricci tensor for the Levi--Civita
connection are equal to the corresponding coefficients of the Ricci tensor
for the canonical d--connection. We restrict the ansatz for such
coefficients when $h_{5}^{\ast }=0$ and $w_{k}^{\ast \ast }=0$ but $%
n_{k}^{\ast }\neq 0.$ This is possible if we put $w_{k}=0,$ see the
discussion of the formula (\ref{w}) in Appendix, and consider any functions $%
\widetilde{U}$ and $\widetilde{k}$ subjected to the condition
\begin{equation}
2\widetilde{U}(\rho ,\zeta ,\varphi )-\widetilde{k}(\rho ,\zeta ,\varphi )=%
\widetilde{k}_{0}(\rho ,\zeta )  \label{ax0a}
\end{equation}%
which result in $h_{5}^{\ast }=0.$ In this case, any values of type
\begin{equation}
h_{4}(\rho ,\zeta ,\varphi )=\rho _{0}^{2}\rho ^{2}e^{-2\widetilde{k}(\rho
,\zeta ,\varphi )}\mbox{ and
}h_{5}(\rho ,\zeta )=-e^{2\widetilde{k}_{0}(\rho ,\zeta )}  \label{ax1a}
\end{equation}%
solve the equation (\ref{ep2a}). The nontrivial N--coefficients $n_{k}$ have
to be computed by the second formula in (\ref{n}),%
\begin{equation*}
n_{2,3}(\rho ,\zeta ,\varphi )=n_{2,3[1]}\left( \rho ,\zeta \right)
+n_{2,3[2]}\left( \rho ,\zeta \right) \int e_{4}^{-2\widetilde{k}(\rho
,\zeta ,\varphi )}d\varphi ,
\end{equation*}%
where we have introduced into $n_{k[2]}\left( \rho ,\zeta \right) $ the
factor $\rho _{0}^{2}\rho ^{2}.$ We have to constrain the functions $%
n_{k[1,2]}\left( \rho ,\zeta \right) $ and $\widetilde{k}(\rho ,\zeta
,\varphi )$ in a such way as to satisfy (\ref{aux31}) which, for $w_{k}=0,$
holds for any $n_{2}$ and $n_{3}$ subjected to the condition
\begin{equation}
n_{2}^{\prime }-n_{3}^{\bullet }=0.  \label{ax2a}
\end{equation}%
If we take
\begin{equation}
n_{3}=a\left( \rho ,\zeta \right) \left[ 1+n_{3[2]}\left( \rho ,\zeta
\right) \int e_{4}^{-2\widetilde{k}(\rho ,\zeta ,\varphi )}d\varphi \right]
\rightarrow a\left( \rho ,\zeta \right)  \label{lim3}
\end{equation}%
for small values $n_{3[2]},$ we have to consider some small coefficients $%
n_{2[1,2]}\left( \rho ,\zeta \right) $ defined in a such way that $%
n_{2}^{\prime }=n_{3}^{\bullet }.$ It is also possible to state that for
large values of $\rho $ and/or $\zeta $ one has
\begin{equation*}
2\widetilde{U}(\rho ,\zeta ,\varphi )-\widetilde{k}(\rho ,\zeta ,\varphi
)\rightarrow 2U(\rho ,\zeta )-k(\rho ,\zeta )=\widetilde{k}_{0}(\rho ,\zeta
).
\end{equation*}

Introducing the values (\ref{ax0a})--(\ref{lim3}) in (\ref{dsol4d1}), we get
a vacuum 4D Einstein metric%
\begin{eqnarray}
\delta s_{[4D\varphi E]}^{2} &=&\rho _{0}^{2}d\rho ^{2}+\rho _{0}^{2}d\zeta
^{2}  \label{sol1a} \\
&&+\rho _{0}^{2}\rho ^{2}e^{-2\widetilde{k}(\rho ,\zeta ,\varphi )}d\varphi
^{2}-e^{2\widetilde{k}_{0}(\rho ,\zeta )}\ \delta \tau ^{2},  \notag \\
\delta \tau &=&d\tau +\left[ n_{2[1]}\left( \rho ,\zeta \right)
+n_{2[2]}\left( \rho ,\zeta \right) \int e_{4}^{-2\widetilde{k}(\rho ,\zeta
,\varphi )}d\varphi \right] d\rho  \notag \\
&&+a\left( \rho ,\zeta \right) \left[ 1+n_{3[2]}\left( \rho ,\zeta \right)
\int e_{4}^{-2\widetilde{k}(\rho ,\zeta ,\varphi )}d\varphi \right] \ d\zeta
,  \notag
\end{eqnarray}%
for any set of integration functions satisfying the limit conditions to the
metric (\ref{auxm1}). We note that even we may consider some small values $%
n_{2[1]}\left( \rho ,\zeta \right) $ and/ or $n_{2[2]}\left( \rho ,\zeta
\right) $ such functions can not be both zero because only (\ref{sol1a}) \
is a solution of the vacuum gravitational equations but not the conformally
transformed metric (\ref{auxm1}). With respect to the nonholonomic frames (%
\ref{ddif}) the new locally anisotropic disk metric has the coefficients
very similar to the usual disk solution but with a certain dependence on
functions polarized on $\varphi $. For a more restricted class of solutions,
we can search for any periodic polarizations of type
\begin{equation*}
\widetilde{k}(\rho ,\zeta ,\varphi )\simeq k(\rho ,\zeta )\cos \omega
_{0}\varphi
\end{equation*}%
when $\omega _{0}$ has to be defined experimentally or computed from certain
models of gravitational interactions with nonlinear self-polarization,
quantum fluctuations or extra dimension corrections. An alternative way is
to treat such solutions as usual disk solutions but self--consistently
mapped into nontrivial off--diagonal background defined by a solitonic
gravitational wave like we shall consider in details in the next sections.

\subsubsection{Disk anisotropy induced by cosmological constants}

Let us consider how the disk solution (\ref{vac1}), by applying a chain of
conformal (\ref{vac1a}) and static nonholonmic trasforms, can be generalized
to define exact off--diagonal solutions of the Einstein equations with
cosmological constant. We consider nontrivial sources $\Upsilon
_{2}=\Upsilon _{4}=\lambda _{0}=const$ (\ref{emcond4}) for the 4D ansatz
(without dependence on $x^{1}$) of the Einstein equations (\ref{ep1a})--(\ref%
{ep4a}), see Appendix. The existence and a form of solution is stated by the
Corollary \ref{corgsol1},
\begin{eqnarray}
\delta s_{[4D\varphi \lambda ]}^{2} &=&\rho _{0}^{2}\varpi (\rho ,\zeta
)d\rho ^{2}+\rho _{0}^{2}\varpi (\rho ,\zeta )d\zeta ^{2}+  \label{sol2} \\
&&\left[ f^{\ast }\left( \rho ,\zeta ,\varphi \right) \right] ^{2}|\varsigma
_{4}\left( \rho ,\zeta ,\varphi \right) |\left( \delta \varphi \right) ^{2}-%
\left[ f\left( \rho ,\zeta ,\varphi \right) -f_{0}(\rho ,\zeta )\right]
^{2}\left( \delta \tau \right) ^{2},  \notag \\
\delta \varphi &=&d\varphi +w_{2}\left( \rho ,\zeta ,\varphi \right) d\rho
+w_{3}\left( \rho ,\zeta ,\varphi \right) d\zeta ,\   \notag \\
\delta \tau &=&d\tau +n_{2}\left( \rho ,\zeta ,\varphi \right) d\rho
+n_{3}\left( \rho ,\zeta ,\varphi \right) d\zeta ,  \notag
\end{eqnarray}%
which is the 4D version of the metric (\ref{gensol1}) for $x^{2}=\rho
,x^{3}=\zeta ,y^{4}=v=\varphi $ and $y^{5}=\tau $ and $\epsilon
_{2,3}=1,\epsilon _{4}=1$ and $\epsilon _{5}=-1.$

The function $\varpi (\rho ,\zeta )$ from (\ref{sol2}) can be defined as a
solution of
\begin{equation}
\left( \ln |\varpi |\right) ^{\bullet \bullet }+\left( \ln |\varpi |\right)
^{\prime \prime }=2\lambda _{0},  \label{ao0}
\end{equation}%
see (\ref{auxeq01}). One computes
\begin{equation*}
\varsigma _{4}\left( \rho ,\zeta ,\varphi \right) =\varsigma _{4[0]}\left(
\rho ,\zeta \right) -\frac{\lambda _{0}}{8}h_{0}^{2}(\rho ,\zeta )\int
f^{\ast }\left( \rho ,\zeta ,\varphi \right) \left[ f\left( \rho ,\zeta
,\varphi \right) -f_{0}(\rho ,\zeta )\right] d\varphi ,
\end{equation*}%
and, in result, the N--connection coefficients $N_{2,3}^{4}=w_{2,3}(\rho
,\zeta ,\varphi )$ and $N_{2,3}^{5}=n_{2,3}(\rho ,\zeta ,\varphi )$ can be
expressed
\begin{equation}
w_{2}=-(\frac{\partial \varsigma _{4}}{\partial \varphi })^{-1}(\frac{%
\partial \varsigma _{4}}{\partial \rho })\mbox{ and }w_{3}=-(\frac{\partial
\varsigma _{4}}{\partial \varphi })^{-1}(\frac{\partial \varsigma _{4}}{%
\partial \zeta }),  \label{ao1}
\end{equation}%
and
\begin{equation}
n_{2,3}\left( \rho ,\zeta ,\varphi \right) =n_{2,3[1]}\left( \rho ,\zeta
\right) +n_{2,3[2]}\left( \rho ,\zeta \right) \int \frac{\left[ f^{\ast
}\left( \rho ,\zeta ,\varphi \right) \right] ^{2}\varsigma _{4}\left( \rho
,\zeta ,\varphi \right) }{\left[ f\left( \rho ,\zeta ,\varphi \right)
-f_{0}(\rho ,\zeta )\right] ^{3}}d\varphi .  \label{ao2}
\end{equation}%
The class of solutions defined by a such ansatz (\ref{sol2}) depends on
integration functions $h_{0}(\rho ,\zeta ),f_{0}(\rho ,\zeta ),$ $\varsigma
_{4[0]}\left( \rho ,\zeta \right) $ and $n_{2,3[1,2]}\left( \rho ,\zeta
\right) ,$ on a solution $\varpi (\rho ,\zeta )$ \ of (\ref{ao0}) and on a
general function $f\left( \rho ,\zeta ,\varphi \right) $ with $f^{\ast }\neq
0$ describing a class of backgrounds to which the disk solution can be
mapped self--consistently. The relation to the usual disk solution can be
established by considering that
\begin{eqnarray*}
\underline{h}_{4} &=&\rho _{0}^{2}\rho ^{2}e^{-2k}=|\varsigma _{4[0]}\left(
\rho ,\zeta \right) |\mbox{ and }\underline{h}_{5}=-e^{2(2U-k)}=-\left[
f_{0}(\rho ,\zeta )\right] ^{2}, \\
\underline{N}_{3}^{5} &=&a=n_{3[1]}\left( \rho ,\zeta \right)
\end{eqnarray*}%
like in (\ref{auxd1a}). Such '$\varphi $--anisotropic' solutions are with
nontrivial polarizations $\eta _{2,3}=\varpi $ induced by the cosmological
constant which modify the former values $\underline{g}_{2}=\rho _{0}^{2},%
\underline{g}_{3}=\rho _{0}^{2}.$

One can be imposed certain constraints when the Ricci tensor for the
Levi--Civita connection has the same coefficients as the Ricci tensor for
the canonical d--connection (all such coefficients being computed with
respect to N--adapted vierbeins, see (\ref{dder4}) and (\ref{ddif4})). \ For
instance, for $h_{5}=\left[ f\left( \rho ,\zeta ,\varphi \right) -f_{0}(\rho
,\zeta )\right] ^{2},$ one holds the condition $h_{5}^{\ast }\neq 0.$ In
this case, we must solve the equations (\ref{aux32}) by any $n_{\widehat{k}}$
and $w_{\widehat{k}}$ when $n_{\widehat{k}}^{\ast }=0$ and
\begin{equation}
w_{\widehat{k}}^{\ast \ast }+\frac{w_{\widehat{k}}^{\ast }}{2}\frac{%
h_{5}^{\ast }}{h_{5}}=0.  \label{ax5}
\end{equation}%
We can take $n_{3}=$ $n_{3[1]}\left( \rho ,\zeta \right) =a\left( \rho
,\zeta \right) $ for $n_{2,3[2]}\left( \rho ,\zeta \right) =0$ and any $%
n_{2}=$ $n_{2[1]}\left( \rho ,\zeta \right) $ satisfying the condition $%
n_{2}^{\prime }-n_{3}^{\bullet }=0.$ Such conditions can be solved, for
instance, for
\begin{equation}
f=f_{0}\left( \rho ,\zeta \right) \Phi (\varphi )  \label{par01}
\end{equation}%
when we fix some of the integration functions to be related by%
\begin{equation*}
|f_{0}h_{0}|^{2}\varsigma _{4[0]}\left( \rho ,\zeta \right) =-4
\end{equation*}
(we may consider that this can be obtained in result of some 2D coordinate
transforms). In this case, we get
\begin{equation*}
h_{5}=-f_{0}^{2}(1-\Phi )^{2}\mbox{ and }\varsigma _{4}\left( \rho ,\zeta
,\varphi \right) =\varsigma _{4[0]}\left( \rho ,\zeta \right) \varsigma
(\Phi )
\end{equation*}%
for a functional $\varsigma $ on $\Phi ,$
\begin{equation*}
\varsigma \{\Phi \}=1-\lambda _{0}(1-\Phi )^{2}.
\end{equation*}%
Introducing such parametrizations in (\ref{ao1}), we find
\begin{equation}
w_{2}=-\widetilde{\Phi }\frac{\partial }{\partial \rho }\ln |\varsigma
_{4[0]}|\mbox{ and }w_{3}=-\widetilde{\Phi }\frac{\partial }{\partial \zeta }%
\ln |\varsigma _{4[0]}|,  \label{ax7}
\end{equation}%
for $\widetilde{\Phi }\doteqdot 1/\left( \ln |\varsigma \{\Phi \}|\right)
^{\ast }.$ This transform the equations (\ref{ax5}) into the equation
\begin{equation}
\widetilde{\Phi }^{\ast \ast }-\frac{1}{2}\widetilde{\Phi }^{\ast }(1-\Phi
)=0  \label{ax6}
\end{equation}%
for a function $\Phi (\varphi ).$ We can approximate, for small values of $%
\lambda _{0},$ that
\begin{equation*}
\Phi (\varphi )\approx p_{0}+p_{3}\varphi ^{3}+....
\end{equation*}%
for some constants $p_{0}$ and $p_{3}$ related in a form to solve (\ref{ax6}%
). Because the functions $w_{2,3}$ defined by (\ref{ax7}), for $n_{\widehat{k%
}}^{\ast }=0,$ satisfy the conditions (\ref{aux31}) from Appendix, we get
the result that the locally anisotropic metric (\ref{sol2}) can be
constrained to define a class of off--diagonal solutions, with anisotropy on
$\varphi ,$ in Einstein gravity with cosmological constant $\lambda _{0}.$
The anisotropic terms in such metrics are induced by nontrivial values of $%
\lambda _{0}.$ In a similar way, one can use other types of generation
functions $f(\rho ,\zeta ,\varphi )$ than the parametrization (\ref{par01})
with explicit or non--explicit solutions of the constraints for the
integration functions.

It should be emphasized that if the limits (\ref{limit1}) are satisfied, one
transforms some classes of solutions (vacuum, or nonvacuum, ones, defined by
nonholonomic configurations) into a metric which is conformally equivalent
to the disk solution, see the metric (\ref{auxm1}), but the last one is not
just a vacuum solution. We can obtain another type of vacuum solution (with
Killing symmetry) after a conformal map. Such nonholonomic maps relate
certain class of metrics with generalized Lie algebroid symmetry (see
Section \ref{sgdal}) to metrics with Killing symmetry. The limit case may
connected, by certain conformal and nonlinear coordinate transforms, to a
well known exact (disk) solution which allows to state the boundary
conditions and compare corresponding physical properties and symmetries both
for the deformed and undeformed metrics.

Finally, in this section, we conclude that is possible to ''map'' a known
disk solution, via supperpositons of conformal and nonholonomic frame
transforms, into a new class of exact solutions, conventionally called disks
with locally anisotropic polarizations or disks in nonholonmic backgrounds.

\section{Disks in String Gravity with Solitonic Bac\-kgrounds}

Since the physical effects related to gravitational disk and matter field
configurations may be used as tests of different type of theories of
gravity, one may look for implication of generalized disk type solutions in
string gravity with nontrivial solitonic backrounds. In this section, we
construct and analyze some classes of such solutions.

\subsection{String gravity with nonholonomic $H$--fields}

Let us show how the data (\ref{auxd1a})--(\ref{auxd2b}) for the ansatz (\ref%
{ans4dd}), can be nonholonomically deformed, see relations (\ref{polf}), to
5D generic off--diagonal metrics (\ref{ans5d}) defining locally anisotropic
disk like configurations in string gravity. For simplicity, we consider a
bosonic string model with trivial constant dilaton fields and a so--called $%
H $--field torsion, $\mathbf{H}_{\nu \lambda \rho }$;\ see, for instance,
Refs. \cite{sgr} for the basic results in string gravity. \

We start with an ansatz for the $H$--field, see (\ref{aux51a}) in Appendix,
\begin{equation}
\mathbf{H}_{\nu \lambda \rho }=\widehat{\mathbf{Z}}_{\ \nu \lambda \rho }+%
\widehat{\mathbf{H}}_{\nu \lambda \rho }=\lambda _{H}\sqrt{|\mathbf{g}%
_{\alpha \beta }|}\varepsilon _{\nu \lambda \rho },  \label{ans61}
\end{equation}%
where $\varepsilon _{\nu \lambda \rho }$ is completely antisymmetric and $%
\lambda _{H}=const,$ which satisfies the field equations for $\mathbf{H}%
_{\nu \lambda \rho },$ see (\ref{aux51b}). \ The ansatz (\ref{ans61}) is not
just for the $H$--field like in the former approaches, when the
contributions of a such field with coefficients were defined with respect to
a coordinate basis and ''summarized'' into an effective cosmological
constant. In our case, we consider a deformation of $\widehat{\mathbf{H}}%
_{\nu \lambda \rho }$ by a locally anisotropic background with
nonholonomically induced distorsion $\widehat{\mathbf{Z}}_{\ \nu \lambda
\rho }$ (\ref{aux53})\ defined by the d--torsions (\ref{dtorsb}) for the
canonical d--connection (\ref{candcon}),\footnote{%
see Definition \ref{ddcon} in Appendix, on definition of d--connections and
related discussion on d--tensors} to an effective $\mathbf{H}_{\nu \lambda
\rho },$ which can be approximated by a cosmological constant but only with
respect to N--adapted frames (\ref{ddif}) and (\ref{dder}). Re--defining the
constructions with respect to coordinate frames, we get not just a
cosmological constant but a more general source defined both by $\widehat{%
\mathbf{Z}}_{\ \nu \lambda \rho }$ and $\widehat{\mathbf{H}}_{\nu \lambda
\rho }.$

With respect to the N--adapted frames, the source in (\ref{aux51c}) is
effectively diagonalized,
\begin{equation*}
\mathbf{\Upsilon }_{\alpha }^{[\mathbf{H}]\beta }=\{\frac{\lambda _{H}^{2}}{2%
},\frac{\lambda _{H}^{2}}{4},\frac{\lambda _{H}^{2}}{4},\frac{\lambda
_{H}^{2}}{4},\frac{\lambda _{H}^{2}}{4}\},
\end{equation*}%
and the equations (\ref{ep1a}), (\ref{ep2a}) transform respectively into
\begin{eqnarray}
R_{2}^{2} &=&R_{3}^{3}=\frac{1}{2g_{2}g_{3}}[\frac{g_{2}^{\bullet
}g_{3}^{\bullet }}{2g_{2}}+\frac{(g_{3}^{\bullet })^{2}}{2g_{3}}%
-g_{3}^{\bullet \bullet }+\frac{g_{2}^{^{\prime }}g_{3}^{^{\prime }}}{2g_{3}}%
+\frac{(g_{2}^{^{\prime }})^{2}}{2g_{2}}-g_{2}^{^{\prime \prime }}]=-\frac{%
\lambda _{H}^{2}}{4}  \label{eq64a} \\
S_{4}^{4} &=&S_{5}^{5}=-\frac{1}{2h_{4}h_{5}}\left[ h_{5}^{\ast \ast
}-h_{5}^{\ast }\left( \ln \sqrt{|h_{4}h_{5}|}\right) ^{\ast }\right] =-\frac{%
\lambda _{H}^{2}}{4},  \label{eq64b}
\end{eqnarray}%
when the off--diagonal terms $w_{i}$ and $n_{i}$ are defined correspondingly
by equations (\ref{gensol1w}) and (\ref{gensol1n}). The solution of (\ref%
{eq64a})\ can be found as for (\ref{ao0}), see also (\ref{auxeq01}) in
Appendix, when $\psi =\ln |g_{2}|=\ln |g_{3}|$ is a solution of
\begin{equation}
\epsilon _{2}\psi ^{\bullet \bullet }+\epsilon _{3}\psi ^{\prime \prime }=-%
\frac{\lambda _{H}^{2}}{2},  \label{aux73}
\end{equation}%
where, for simplicity we choose the h--variables $x^{2}=\tilde{x}^{2}$ and $%
x^{3}=\tilde{x}^{3},$ with $\psi ^{\bullet }=\partial \psi /\partial x^{2}$
and $\psi ^{^{\prime }}=\partial \psi /\partial x^{3}$ and $\epsilon
_{2,3}=\pm 1.$

The Corollary \ref{corgsol1} defines a general class of 5D solutions
\begin{eqnarray}
\delta s^{2} &=&\epsilon _{1}(dx^{1})^{2}+e^{\psi (x^{2},x^{3})}\left[
\epsilon _{2}(dx^{2})^{2}+\epsilon _{3}(dx^{3})^{2}\right] +\epsilon
_{4}h_{0}^{2}(x^{i})\times  \label{str5m} \\
&&\left[ f^{\ast }\left( x^{i},v\right) \right] ^{2}|\varsigma _{4}\left(
x^{i},v\right) |\left( \delta v\right) ^{2}+\epsilon _{5}\left[ f\left(
x^{i},v\right) -f_{0}(x^{i})\right] ^{2}\left( \delta y^{5}\right) ^{2},
\notag \\
\delta v &=&dv+w_{k}\left( x^{i},v\right) dx^{k},\ \delta
y^{5}=dy^{5}+n_{k}\left( x^{i},v\right) dx^{k},  \notag
\end{eqnarray}%
with $\psi (x^{2},x^{3})$ defined from (\ref{aux73}) and the N--connection
coefficients $N_{i}^{4}=w_{i}(x^{k},v)$ and $N_{i}^{5}=n_{i}(x^{k},v)$
computed in the form
\begin{equation}
w_{i}=-\frac{\partial _{i}\varsigma _{4}\left( x^{k},v\right) }{\varsigma
_{4}^{\ast }\left( x^{k},v\right) }  \label{str5mw}
\end{equation}%
and
\begin{equation}
n_{k}=n_{k[1]}\left( x^{i}\right) +n_{k[2]}\left( x^{i}\right) \int \frac{%
\left[ f^{\ast }\left( x^{i},v\right) \right] ^{2}}{\left[ f\left(
x^{i},v\right) -f_{0}(x^{i})\right] ^{3}}\varsigma _{4}\left( x^{i},v\right)
dv,  \label{str5mn}
\end{equation}%
for
\begin{equation}
\varsigma _{4}\left( x^{i},v\right) =\varsigma _{4[0]}\left( x^{i}\right) -%
\frac{\epsilon _{4}\lambda _{H}^{2}}{16}h_{0}^{2}(x^{i})\int f^{\ast }\left(
x^{i},v\right) \left[ f\left( x^{i},v\right) -f_{0}(x^{i})\right] dv,
\label{gf}
\end{equation}%
defining an exact solution of the system of string gravity equations (\ref%
{eq64a}), (\ref{eq64b}) and (\ref{ep3a}),(\ref{ep4a}), with nonholonomic
variables and integration functions $n_{k[1,2]}\left( x^{i}\right)
,\varsigma _{4[0]}\left( x^{i}\right) ,h_{0}^{2}(x^{i})$ and $f_{0}(x^{i}).$
This solution is similar to the 4D metric (\ref{sol2}), but written for 5D
ansatz with arbitrary signatures.

We can choose the function $f\left( x^{2},x^{3},v\right) $ from (\ref{str5m}%
) as it would be a solution of the Kadomtsev--Petviashvili (KdP) equation %
\cite{kdp},
\begin{equation}
f^{\bullet \bullet }+\epsilon \left( f^{\prime }+6ff^{\ast }+f^{\ast \ast
\ast }\right) ^{\ast }=0,\ \epsilon =\pm 1,  \label{kdpe}
\end{equation}%
or, for another locally anisotropic background, to satisfy the $(2+1)$%
--dimensi\-o\-nal sine--Gordon (SG) equation,
\begin{equation}
-f^{\bullet \bullet }+f^{\prime \prime }+f^{\ast \ast }=\sin f,  \label{sg3e}
\end{equation}%
see Refs. \cite{soliton} on gravitational solitons and theory of solitons.
In this case, we define a class of 5D nonholonomic spacetimes with string
corrections by $H$--field self--consistently interacting with 3D
gravitational solitons. Such solutions generalized those considered in Refs. %
\cite{v1} for 4D and 5D gravity.

\subsection{5D disk configurations with string $H$--field and solitonic
backgro\-unds}

The class of solutions (\ref{str5m}) has various parametrizations defining
nonholonomic deformations of the disk solution (\ref{vac1}) to certain type
of generic 5D off--diagonal solutions of bosonic string gravity. We analyze
four examples of such generic off--diagonal metrics.

\subsubsection{String--solitonic disks with angular $\protect\varphi $%
--anisotropy}

We parametrize the ansatz (\ref{str5m}) in a such form that it is a metric
of type (\ref{ans5d}) generated from the data (\ref{auxd1a}) by nonholonomic
deformations of type (\ref{polf}), when the angular coordinate $\varphi $ is
considered to be that one on which the solutions depend anisotropically. For
parametrizations of 5D coordinates%
\begin{equation*}
x^{1}=\chi ,x^{2}=\rho ,x^{3}=\zeta ,y^{4}=v=\varphi ,y^{5}=\tau ,
\end{equation*}%
one obtains the metric
\begin{eqnarray}
\delta s^{2} &=&\epsilon _{1}(d\chi )^{2}+e^{\psi (\rho ,\zeta )}\left[
(d\rho )^{2}+(d\zeta )^{2}\right] +h_{0}^{2}(\rho ,\zeta )\times
\label{sol3a} \\
&&\left[ f^{\ast }\left( \rho ,\zeta ,\varphi \right) \right] ^{2}|\varsigma
_{4}\left( \rho ,\zeta ,\varphi \right) |\left( \delta \varphi \right) ^{2}-%
\left[ f\left( \rho ,\zeta ,\varphi \right) -f_{0}(\rho ,\zeta )\right]
^{2}\left( \delta \tau \right) ^{2},  \notag \\
\delta \varphi  &=&d\varphi +w_{1}\left( \chi ,\rho ,\zeta ,\varphi \right)
d\chi +w_{2}\left( \chi ,\rho ,\zeta ,\varphi \right) d\rho +w_{3}\left(
\chi ,\rho ,\zeta ,\varphi \right) d\zeta ,\   \notag \\
\delta \tau  &=&d\tau +n_{1}\left( \chi ,\rho ,\zeta ,\varphi \right) d\chi
+n_{2}\left( \chi ,\rho ,\zeta ,\varphi \right) d\rho +n_{3}\left( \chi
,\rho ,\zeta ,\varphi \right) d\zeta ,  \notag
\end{eqnarray}%
where $\psi (\rho ,\zeta )$ is to generated by a solution of (\ref{aux73}),\
\ which for $\epsilon _{2,3}=1$ is a 2D Poisson equation. The N--connection
coefficients $w_{i}$ (\ref{str5mw}) and $n_{i}$ (\ref{str5mn}) are defined
by
\begin{equation*}
\varsigma _{4}\left( \rho ,\zeta ,\varphi \right) =\varsigma _{4[0]}\left(
\rho ,\zeta \right) -\frac{\lambda _{H}^{2}}{16}h_{0}^{2}(\rho ,\zeta )\int
f^{\ast }\left( \rho ,\zeta ,\varphi \right) \left[ f\left( \rho ,\zeta
,\varphi \right) -f_{0}(\rho ,\zeta )\right] d\varphi .
\end{equation*}%
The boundary conditions can related to the 4D disk solution (\ref{vac1}) via
data
\begin{equation*}
\underline{g}_{1}=\epsilon _{1},\underline{g}_{2}=\rho _{0}^{2},\underline{g}%
_{3}=\rho _{0}^{2},\underline{h}_{4}=\rho _{0}^{2}\rho ^{2}e^{-2k},%
\underline{h}_{5}=-e^{2(2U-k)},\underline{N}_{3}^{5}=a,
\end{equation*}%
when the gravitational polarizations $\eta _{\alpha }$ are computed by using
the metric coefficients of (\ref{sol3a}),
\begin{eqnarray}
\mathbf{g}_{\alpha \beta } &=&diag\left\{ g_{i}=\eta _{i}\underline{g}%
_{1},h_{a}=\eta _{a}\underline{h}_{a}\right\} =diag\{1,e^{\psi (\rho ,\zeta
)},e^{\psi (\rho ,\zeta )},  \label{ex02} \\
&&h_{0}^{2}(\rho ,\zeta )\left[ f^{\ast }\left( \rho ,\zeta ,\varphi \right) %
\right] ^{2}|\varsigma _{4}\left( \rho ,\zeta ,\varphi \right) |,\left[
f\left( \rho ,\zeta ,\varphi \right) -f_{0}(\rho ,\zeta )\right] ^{2}\}
\notag
\end{eqnarray}%
stated with respect to N--adapted dual funfbein $\mathbf{e}^{\alpha }=(d\chi
,d\rho ,d\zeta ,\delta \varphi ,\delta \tau ).$

The constructed stationary '$\varphi $--anisotropic' solutions may be
constrained to be a 4D configuration trivially imbedded in 5D if we put $%
w_{1}=n_{1}=0$ and do not consider dependencies on extra dimension
coordinate $\chi .$ In this case, the class of metrics with $\eta _{\alpha
}\approx 1$ and $N_{1}^{a}\approx \underline{N}_{3}^{5}$ are very similar to
the conformal transform of the disk solution, i. e. to (\ref{vac1a}), but
contain certain polarizations induced nonholonomically by the $H$--field in
string gravity, approximated together with the N--anholonomic torsion to
result in an effective cosmological constant $\lambda _{H}.$ These solutions
can be considered to define some locally anisotropically polarized disks
imbedded self--consistently in 5D curved background defined by the function $%
f\left( \rho ,\zeta ,\varphi \right) .$ A such background can be of 3D
solitonic nature if we chose $f$ to be any solution of (\ref{kdpe}),
redefined for the signatures $\epsilon _{2,3,4}=1.$ If the string
contributions are normalized in a form to have $\lambda _{0}=\lambda
_{H}^{2}/8,$ we may reproduce all geometric and physical properties of the
solution (\ref{sol2}).

We note that the symmetry properties of (\ref{sol3a}) should be derived as
certain nonholonomic deformations of the 4D Killing disk symmetries. For
various classes of integration functions and stated polarizations $\eta
_{a}, $ the new type of spacetimes can be characterized by noncommutative
symmetries and/or Lie algebroid symmetries, see section \ref{sgdal}.

\subsubsection{Moving string--solitonic disk configurations}

It is possible to deform nonholonomically the metric (\ref{auxm1}) in a form
generating a class of disk like solutions in string gravity with
polarizations depending anisotropically on the time like coordinate $\tau .$
One uses the coordinate parametrizations $x^{1}=\chi ,x^{2}=\rho
,x^{3}=\zeta ,y^{4}=v=\tau $ and $y^{5}=\varphi $ and the data (\ref{auxd1b}%
), when the undeformed (nontrivial) metric and N--connection coefficients
are $\underline{g}_{1}=\epsilon _{1},\underline{g}_{2}=\rho _{0}^{2},$ $%
\underline{g}_{3}=\rho _{0}^{2},$ $\underline{h}_{4}=-e^{2(2U-k)},$ $%
\underline{h}_{5}=\rho _{0}^{2}\rho ^{2}e^{-2k},$ $\underline{N}_{3}^{4}=a.$
The new class of solutions of the system of string gravity equations (\ref%
{eq64a}), (\ref{eq64b}) and (\ref{ep3a}),(\ref{ep4a}) is parametrized in the
form
\begin{eqnarray}
\delta s^{2} &=&\epsilon _{1}(d\chi )^{2}+e^{\psi (\rho ,\zeta )}\left[
(d\rho )^{2}+(d\zeta )^{2}\right] +h_{0}^{2}(\rho ,\zeta )\times
\label{sol3b} \\
&&\left[ f^{\ast }\left( \rho ,\zeta ,\tau \right) \right] ^{2}|\varsigma
_{4}\left( \rho ,\zeta ,\tau \right) |\left( \delta \tau \right) ^{2}-\left[
f\left( \rho ,\zeta ,\tau \right) -f_{0}(\rho ,\zeta )\right] ^{2}\left(
\delta \varphi \right) ^{2},  \notag \\
\delta \tau &=&d\varphi +w_{1}\left( \chi ,\rho ,\zeta ,\tau \right) d\chi
+w_{2}\left( \chi ,\rho ,\zeta ,\tau \right) d\rho +w_{3}\left( \chi ,\rho
,\zeta ,\tau \right) d\zeta ,\   \notag \\
\delta \varphi &=&d\tau +n_{1}\left( \chi ,\rho ,\zeta ,\tau \right) d\chi
+n_{2}\left( \chi ,\rho ,\zeta ,\tau \right) d\rho +n_{3}\left( \chi ,\rho
,\zeta ,\tau \right) d\zeta ,  \notag
\end{eqnarray}%
where $\psi (\rho ,\zeta )$ is also generated by a solution of (\ref{aux73}%
),\ \ which for $\epsilon _{2,3}=1$ is a 2D Poisson equation and the
N--connection coefficients $w_{i}$ (\ref{str5mw}) and $n_{i}$ (\ref{str5mn})
are computed by using
\begin{equation*}
\varsigma _{4}\left( \rho ,\zeta ,\tau \right) =\varsigma _{4[0]}\left( \rho
,\zeta \right) -\frac{\lambda _{H}^{2}}{16}h_{0}^{2}(\rho ,\zeta )\int
f^{\ast }\left( \rho ,\zeta ,\tau \right) \left[ f\left( \rho ,\zeta ,\tau
\right) -f_{0}(\rho ,\zeta )\right] d\tau .
\end{equation*}%
We can restrict the class of solutions for the 4D metrics trivially imbedded
in 5D when $w_{1}=n_{1}=0$ and the dependence on $\chi $ is not considered
for the metric coefficients. The gravitational polarizations $\eta _{\alpha
} $ follows from the metric coefficients in (\ref{sol3b}),
\begin{eqnarray*}
\mathbf{g}_{\alpha \beta } &=&diag\left\{ g_{i}=\eta _{i}\underline{g}%
_{1},h_{a}=\eta _{a}\underline{h}_{a}\right\} =diag\{1,e^{\psi (\rho ,\zeta
)},e^{\psi (\rho ,\zeta )}, \\
&&h_{0}^{2}(\rho ,\zeta )\left[ f^{\ast }\left( \rho ,\zeta ,\tau \right) %
\right] ^{2}|\varsigma _{4}\left( \rho ,\zeta ,\tau \right) |,\left[ f\left(
\rho ,\zeta ,\tau \right) -f_{0}(\rho ,\zeta )\right] ^{2}\}
\end{eqnarray*}%
In a more explicit form, we can choose $f\left( \rho ,\zeta ,\tau \right) $
to be a solution of the 3D signe--Gordon equation (\ref{sg3e}). This way,
there are generating metrics describing 4D locally anisotropically polarized
disks moved away in time by certain solitonic waves in 5D spacetime. \ In a
similar form, one had been constructed solutions defining solitonically
moved black holes \cite{vs}.

\subsubsection{Disks with anistoropic polarizations on extra dimension}

By nonholonomic deformations of (\ref{auxm2}) we can generate disk like
solutions in string gravity with $\chi $--anisotropy. We consider\ the data (%
\ref{auxd2a}) when nontrivial (nondeformed) metric coefficients are $%
\underline{g}_{1}=1,$ $\underline{g}_{2}=\rho ^{-2}e^{2k},$ $\underline{g}%
_{3}=\rho ^{-2}e^{2k},$ $\underline{h}_{4}=\epsilon ,$ $\underline{h}%
_{5}=-\rho _{0}^{-2}\rho ^{-2}e^{4U}$ and $\underline{N}_{3}^{5}=a$ for the
local coordinate parametrization $x^{1}=\varphi ,x^{2}=\rho ,x^{3}=\zeta
,y^{4}=v=\chi $ and $y^{5}=\tau .$ For such data, the solutions follow from (%
\ref{str5m}),%
\begin{eqnarray}
\delta s^{2} &=&d\varphi ^{2}+e^{\psi (\rho ,\zeta )}d\rho ^{2}+e^{\psi
(\rho ,\zeta )}d\zeta ^{2}+\epsilon _{4}h_{0}^{2}(\rho ,\zeta )\times
\label{sol3c} \\
&&\left[ f^{\ast }\left( \rho ,\zeta ,\chi \right) \right] ^{2}|\varsigma
_{4}\left( \rho ,\zeta ,\chi \right) |\delta \chi ^{2}-\left[ f\left( \rho
,\zeta ,\chi \right) -f_{0}(\rho ,\zeta )\right] ^{2}\delta \tau ^{2},
\notag \\
\delta \chi &=&d\chi +w_{1}\left( \varphi ,\rho ,\zeta ,\chi \right)
d\varphi +w_{2}\left( \varphi ,\rho ,\zeta ,\chi \right) d\rho +w_{3}\left(
\varphi ,\rho ,\zeta ,\chi \right) d\zeta ,  \notag \\
\delta \tau &=&d\tau +n_{1}\left( \varphi ,\rho ,\zeta ,\chi \right)
d\varphi +n_{2}\left( \varphi ,\rho ,\zeta ,\chi \right) d\rho +n_{3}\left(
\varphi ,\rho ,\zeta ,\chi \right) d\zeta .  \notag
\end{eqnarray}%
The function $\psi (\rho ,\zeta )$ is a solution of the 2D Poisson equation (%
\ref{aux73}). The N--connection coefficients $w_{i}$ (\ref{str5mw}) and $%
n_{i}$ (\ref{str5mn}) are defined by
\begin{equation*}
\varsigma _{4}\left( \rho ,\zeta ,\chi \right) =\varsigma _{4[0]}\left( \rho
,\zeta \right) -\frac{\lambda _{H}^{2}}{16}h_{0}^{2}(\rho ,\zeta )\int
f^{\ast }\left( \rho ,\zeta ,\chi \right) \left[ f\left( \rho ,\zeta ,\chi
\right) -f_{0}(\rho ,\zeta )\right] d\chi .
\end{equation*}%
The gravitational polarizations $\eta _{\alpha }$ are are defined by using
the metric coefficients of (\ref{sol3c}) and 'nondeformed' data,
\begin{eqnarray*}
\mathbf{g}_{\alpha \beta } &=&diag\left\{ g_{i}=\eta _{i}\underline{g}%
_{i},h_{a}=\eta _{a}\underline{h}_{a}\right\} =diag\{1,e^{\psi (\rho ,\zeta
)},e^{\psi (\rho ,\zeta )}, \\
&&h_{0}^{2}(\rho ,\zeta )\left[ f^{\ast }\left( \rho ,\zeta ,\chi \right) %
\right] ^{2}|\varsigma _{4}\left( \rho ,\zeta ,\chi \right) |,\left[ f\left(
\rho ,\zeta ,\chi \right) -f_{0}(\rho ,\zeta )\right] ^{2}\}.
\end{eqnarray*}

The metric (\ref{sol3c}) defines a stationary disk like 5D solution in
string gravity with generic dependence on extra dimension coordinate. We can
consider a subclass of solutions when $w_{1}=n_{1}=0$ and the integration
functions are chosen in a such form that one does not have dependence on
angular coordinate $\varphi .$\ Such spacetimes can not be reduced to
certain trivial imbedding of 4D configurations into 5D ones but can model a
4D disk like structure for small nonholonomic deformations. The extra
dimension background may be defined for various types of generating
functions $f\left( \rho ,\zeta ,\chi \right) $ which can be of solitonic or
other nature, with explicit dependence on the extra dimension coordinate.

\subsubsection{Disks solitonically propagating in extra dimensions}

There is another class of nonholonomic deformations of (\ref{auxm2}) being
very different from the constructions with $\chi $--anisotropy from the
previous subsection. One considers the (\ref{auxd2b}) with the local
coordinates $x^{1}=\varphi ,$ $x^{2}=\rho ,$ $x^{3}=\zeta ,$ $y^{4}=v=\tau ,$
$y^{5}=\chi $ when the (nondeformed) nonzero metric coefficients are $%
\underline{g}_{2}=\rho ^{-2}e^{2k},\underline{g}_{3}=\rho ^{-2}e^{2k},%
\underline{h}_{4}=-\rho _{0}^{-2}\rho ^{-2}e^{4U},\underline{h}_{5}=\epsilon
$ and $\underline{N}_{3}^{4}=a.$

The solution with anisotropic dependence on $\tau $ is given by the ansatz (%
\ref{str5m}) stated for the mentioned parametrizations of coordinates,%
\begin{eqnarray}
\delta s^{2} &=&d\varphi ^{2}+e^{\psi (\rho ,\zeta )}d\rho ^{2}+e^{\psi
(\rho ,\zeta )}d\zeta ^{2}+\epsilon _{4}h_{0}^{2}(\rho ,\zeta )\times
\label{sol3d} \\
&&\left[ f^{\ast }\left( \rho ,\zeta ,\tau \right) \right] ^{2}|\varsigma
_{4}\left( \rho ,\zeta ,\tau \right) |\delta \tau ^{2}-\left[ f\left( \rho
,\zeta ,\tau \right) -f_{0}(\rho ,\zeta )\right] ^{2}\delta \chi ^{2},
\notag \\
\delta \tau &=&d\tau +w_{1}\left( \varphi ,\rho ,\zeta ,\tau \right)
d\varphi +w_{2}\left( \varphi ,\rho ,\zeta ,\tau \right) d\rho +w_{3}\left(
\varphi ,\rho ,\zeta ,\tau \right) d\zeta ,  \notag \\
\delta \chi &=&d\chi +n_{1}\left( \varphi ,\rho ,\zeta ,\tau \right)
d\varphi +n_{2}\left( \varphi ,\rho ,\zeta ,\tau \right) d\rho +n_{3}\left(
\varphi ,\rho ,\zeta ,\tau \right) d\zeta .  \notag
\end{eqnarray}%
This metric is characterized by gravitational polarizations $\eta _{\alpha }$
relating the 'nondeformed' data with the deformed ones,
\begin{eqnarray*}
\mathbf{g}_{\alpha \beta } &=&diag\left\{ g_{i}=\eta _{i}\underline{g}%
_{i},h_{a}=\eta _{a}\underline{h}_{a}\right\} =diag\{1,e^{\psi (\rho ,\zeta
)},e^{\psi (\rho ,\zeta )}, \\
&&h_{0}^{2}(\rho ,\zeta )\left[ f^{\ast }\left( \rho ,\zeta ,\tau \right) %
\right] ^{2}|\varsigma _{4}\left( \rho ,\zeta ,\tau \right) |,\left[ f\left(
\rho ,\zeta ,\tau \right) -f_{0}(\rho ,\zeta )\right] ^{2}\}.
\end{eqnarray*}%
For this class of metrics, the function $\psi (\rho ,\zeta )$ is also a
solution of the 2D Poisson equation (\ref{aux73}). The N--connection
coefficients $w_{i}$ and $n_{i}$ are computed by similar formulas (\ref%
{str5mw}) and, respectively, (\ref{str5mn}) but for
\begin{equation*}
\varsigma _{4}\left( \rho ,\zeta ,\tau \right) =\varsigma _{4[0]}\left( \rho
,\zeta \right) -\frac{\lambda _{H}^{2}}{16}h_{0}^{2}(\rho ,\zeta )\int
f^{\ast }\left( \rho ,\zeta ,\tau \right) \left[ f\left( \rho ,\zeta ,\tau
\right) -f_{0}(\rho ,\zeta )\right] d\tau .
\end{equation*}

The solution (\ref{sol3d}) does not depend \ explicitly on the extra
dimension coordinate. Nevertheless, it is a generic five dimensional one and
does not have straightforward limits to configurations containing 4D metrics
as trivial imbedding. The extra dimension background may be defined for
various types of generating functions $f\left( \rho ,\zeta ,\tau \right) $
which can be of solitonic or other nature. We can choose $f\left( \rho
,\zeta ,\tau \right) $ to be a solution of the 3D signe--Gordon equation (%
\ref{sg3e}), or inversely, to be a solution of Kadomtsev--Petviashvili (KdP)
equation (\ref{kdpe}). This way we construct generic off--diagonal 5D
solutions in string gravity with 3D soliton backgrounds.

\section{Disks in Extra Dimensions and Spinors}

The geometry of spinors on N--anholonomic spaces was elaborated in Refs. %
\cite{nhsp}. The main problem was to give a rigorous definition of spinors
on spaces provided with nonintregrable distributions defined by
N--connections.\footnote{%
In the first works from \cite{nhsp}, the problem of rigorous definition of
nonholonmic spinor structures was solved for Finsler and Lagrange spaces but
the question of constructing a theory of nonholonomic spinor and Dirac
operators is also important for (pseudo) Riemannian and Riemann--Cartan
spaces \cite{v0408121}.} In this section, we consider exact solutions of the
5D Einstein--Dirac equations defining certain spinor wave packages on
locally anisotropic spaces derived by nonholnomic transforms of disk
configurations and giving explicit examples of N--anholonomic spacetimes. We
analyze two classes of such solutions.

\subsection{Dirac equation on nonholnomic gravitational disk configurations}

We consider 5D nonholonomic deformations of (\ref{auxm1}) starting from the
data (\ref{auxd1a}), or (\ref{auxd1b}), when the extra dimension coordinate
is time--like. For a d--metric (\ref{ans5d}) with coefficients
\begin{equation*}
g_{\alpha \beta }(u)=(g_{ij}(u),h_{ab}(u))=(1,g_{\widehat{i}}(u),h_{a}(u)),
\end{equation*}%
where $\widehat{i}=1,2;i=0,1,2;a=3,4,$ defined with respect to the
N--adapted basis (\ref{dder}), we can easily define the funfbein (pentad)
fields
\begin{eqnarray}
e_{\underline{\mu }} &=&e_{\underline{\mu }}^{\mu }\mathbf{e}_{\mu }=\{e_{%
\underline{i}}=e_{\underline{i}}^{i}\delta _{i},e_{\underline{a}}=e_{%
\underline{a}}^{a}\partial _{a}\},  \label{pentad1} \\
\ e^{\underline{\mu }} &=&e_{\mu }^{\underline{\mu }}\mathbf{e}^{\mu }=\{e^{%
\underline{i}}=e_{i}^{\underline{i}}d^{i},e^{\underline{a}}=e_{a}^{%
\underline{a}}\delta ^{a}\}  \notag
\end{eqnarray}%
satisfying the conditions
\begin{eqnarray*}
g_{ij} &=&e_{i}^{\underline{i}}e_{j}^{\underline{j}}g_{\underline{i}%
\underline{j}}\mbox{ and }h_{ab}=e_{a}^{\underline{a}}e_{b}^{\underline{b}%
}h_{\underline{a}\underline{b}}, \\
g_{\underline{i}\underline{j}} &=&diag[-1,11]\mbox{ and }h_{\underline{a}%
\underline{b}}=diag[1,-1]\mbox{ or }=diag[-1,1].
\end{eqnarray*}
We write
\begin{equation*}
e_{i}^{\underline{i}}=\sqrt{\left| g_{i}\right| }\delta _{i}^{\underline{i}}%
\mbox{
and }e_{a}^{\underline{a}}=\sqrt{\left| h_{a}\right| }\delta _{a}^{%
\underline{a}},
\end{equation*}%
where $\delta _{i}^{\underline{i}}$ and $\delta _{a}^{\underline{a}}$ are
Kronecker's symbols.

The Dirac spinor fields on such nonholonomically deformed spaces are
parametrized
\begin{equation*}
\Psi \left( u\right) =[\Psi ^{\overline{\alpha }}\left( u\right) ]=[\psi ^{%
\widehat{I}}\left( u\right) ,\chi _{\widehat{I}}\left( u\right) ],
\end{equation*}%
where $\widehat{I}=0,1,$ are defined with respect to the 4D Euclidean
tangent subspace belonging the tangent space to\ the 5D N--anholonomic
manifold $\mathbf{V}.$ The $4\times 4$ dimensional gamma matrices $\gamma ^{%
\underline{\alpha }^{\prime }}=[\gamma ^{\underline{2}^{\prime }},\gamma ^{%
\underline{3}^{\prime }},\gamma ^{\underline{4}^{\prime }},\gamma ^{%
\underline{5}^{\prime }}]$ are defined \ in the usual way, in order to
satisfy the relation
\begin{equation}
\left\{ \gamma ^{\underline{\alpha }^{\prime }},\,\gamma ^{\underline{\beta }%
^{\prime }}\right\} =2g^{\underline{\alpha }\underline{^{\prime }\beta }%
^{\prime }},  \label{gammarel}
\end{equation}%
where $\left\{ \gamma ^{\underline{\alpha }^{\prime }}\,\gamma ^{\underline{%
\beta }^{\prime }}\right\} $ is a symmetric commutator, $g^{\underline{%
\alpha }\underline{^{\prime }\beta }^{\prime }}=(1,1,-1,1),$ or $g^{%
\underline{\alpha }\underline{^{\prime }\beta }^{\prime }}=(1,1,1,-1),$
which generates a Clifford algebra distinguished on two holonomic and two
anholonomic directions. It is possible to extend the relations (\ref%
{gammarel}) for unprimed indices $\alpha ,\beta ...$ by a conventional
completing the set of primed gamma matrices with a matrix $\gamma ^{%
\underline{1}},$ i. .e. write $\gamma ^{\underline{\alpha }}=[\gamma ^{%
\underline{1}},\gamma ^{\underline{2}},\gamma ^{\underline{3}},\gamma ^{%
\underline{4}},\gamma ^{\underline{5}}]$ when
\begin{equation*}
\left\{ \gamma ^{\underline{\alpha }},\,\gamma ^{\underline{\beta }}\right\}
=2g^{\underline{\alpha }\underline{\beta }}.
\end{equation*}

The coefficients of N--anholonomic gamma matrices can be computed with
respect to anholonomic bases (\ref{dder}) by using respectively the funfbein
coefficients
\begin{equation*}
\widehat{\gamma }^{\beta }(u)=e_{\underline{\beta }}^{\beta }(u)\gamma ^{%
\underline{\beta }}.
\end{equation*}%
A covariant derivation of the Dirac spinor field, $\overrightarrow{%
\bigtriangledown }_{\alpha }\Psi ,$ can be defined by using pentad
(equivalently, funfbein) decompositions of the d--metric (\ref{ans5d}),

\begin{equation}
\overrightarrow{\bigtriangledown }_{\alpha }\Psi =\left[ \mathbf{e}_{\alpha
}+\frac{1}{4}\mathbf{S}_{\underline{\alpha }\underline{\beta }\underline{%
\gamma }}\left( u\right) ~e_{\alpha }^{\underline{\alpha }}\left( u\right)
\gamma ^{\underline{\beta }}\gamma ^{\underline{\gamma }}\right] \Psi ,
\label{covspindder}
\end{equation}%
where there are introduced N--elongated partial derivatives. The
coefficients
\begin{equation*}
(\mathbf{S}_{\underline{\alpha }\underline{\beta }\underline{\gamma }}\left(
u\right) =\left( \mathbf{D}_{\gamma }e_{\underline{\alpha }}^{\alpha
}\right) e_{\underline{\beta }\alpha }e_{\underline{\gamma }}^{\gamma }
\end{equation*}%
are transformed into rotation Ricci d--coefficients $\mathbf{S}_{\underline{%
\alpha }\underline{\beta }\underline{\gamma }}$ which together with the
d--covariant derivative $\mathbf{D}_{\gamma }$ are defined by anholonomic
pentads and anholonomic deformations of the Christoffel symbols. In the
canonical case, we should take the operator of canonical d--connection $%
\widehat{\mathbf{D}}_{\gamma }$ with coefficients (\ref{candcon}).

For diagonal d--metrics, the funfbein coefficients can be taken in their
turn in diagonal form and the corresponding gamma matrix $\widehat{\gamma }%
^{\alpha }\left( u\right) $ for anisotropic curved spaces are proportional
to the usual gamma matrix in flat spaces $\gamma ^{\underline{\gamma }}.$
The Dirac equations on nonholonomiv manifolds are written in the simplest
form with respect to anholonomic frames,
\begin{equation}
(i\widehat{\gamma }^{\alpha }\left( u\right) \overrightarrow{%
\bigtriangledown }_{\alpha }-M)\Psi =0,  \label{diraceq}
\end{equation}%
where $M$ is the mass constant of the Dirac field. The Dirac equation is the
Euler--Lagrange equation for the Lagrangian
\begin{eqnarray}
\mathcal{L}^{(1/2)}\left( u\right) &= & \sqrt{\left| g\right| }\{[\Psi
^{+}\left( u\right) \widehat{\gamma }^{\alpha }\left( u\right)
\overrightarrow{\bigtriangledown }_{\alpha }\Psi \left( u\right)
\label{direq} \\
&{}&-(\overrightarrow{\bigtriangledown }_{\alpha }\Psi ^{+}\left( u\right) )%
\widehat{\gamma }^{\alpha }\left( u\right) \Psi \left( u\right) ]-M\Psi
^{+}\left( u\right) \Psi \left( u\right) \},  \notag
\end{eqnarray}%
where by $\Psi ^{+}\left( u\right) $ we denote the complex conjugation and
transposition of the column$~\Psi \left( u\right) .$ Varying $\mathcal{L}%
^{(1/2)}$ on d--metric (\ref{direq}) we obtain the symmetric
energy--moment\-um d--tensor
\begin{eqnarray}
\Upsilon _{\alpha \beta }\left( u\right) &=&\frac{i}{4}[\Psi ^{+}\left(
u\right) \widehat{\gamma }_{\alpha }\left( u\right) \overrightarrow{%
\bigtriangledown }_{\beta }\Psi \left( u\right) +\Psi ^{+}\left( u\right)
\widehat{\gamma }_{\beta }\left( u\right) \overrightarrow{\bigtriangledown }%
_{\alpha }\Psi \left( u\right)  \notag \\
&{}&-(\overrightarrow{\bigtriangledown }_{\alpha }\Psi ^{+}\left( u\right) )%
\widehat{\gamma }_{\beta }\left( u\right) \Psi \left( u\right) -(%
\overrightarrow{\bigtriangledown }_{\beta }\Psi ^{+}\left( u\right) )%
\widehat{\gamma }_{\alpha }\left( u\right) \Psi \left( u\right) ].
\label{diracemd}
\end{eqnarray}

We can verify that for a such class of metrics one holds the conditions $%
\widehat{\mathbf{D}}_{\gamma }e_{\underline{\alpha }}^{\alpha }=0$ resulting
in zero Ricci rotation coefficients,
\begin{equation*}
\mathbf{S}_{\underline{\alpha }\underline{\beta }\underline{\gamma }}\left(
u\right) =0\mbox{ and }\overrightarrow{\bigtriangledown }_{\alpha }\Psi =%
\mathbf{e}_{\alpha }\Psi ,
\end{equation*}%
and the N--anholonomic Dirac equations (\ref{diraceq}) transform into
\begin{equation}
(i\widehat{\gamma }^{\alpha }\left( u\right) \mathbf{e}_{\alpha }-M)\Psi =0.
\label{diraceq1}
\end{equation}%
Further simplifications are possible for Dirac fields depending only on
coordinates $(x^{1}=\chi ,x^{2}=\rho ,x^{3}=\zeta )$, i. e. $\Psi =\Psi
(x^{k})$ when the equation (\ref{diraceq1}) transforms into
\begin{equation*}
(i\gamma ^{\underline{1}}\partial _{\chi }+i\gamma ^{\underline{2}}\frac{1}{%
\sqrt{\left| g_{2}\right| }}\partial _{2}+i\gamma ^{\underline{3}}\frac{1}{%
\sqrt{\left| g_{3}\right| }}\partial _{3}-M)\Psi =0.
\end{equation*}%
The equation (\ref{diraceq1}) simplifies substantially in $\widetilde{x}$%
--coordinates
\begin{equation*}
\left( \chi ,\widetilde{x}^{2}=\widetilde{x}^{2}(\rho ,\zeta ),\widetilde{x}%
^{3}=\widetilde{x}^{3}(\rho ,\zeta )\right) ,
\end{equation*}%
defined in a form when are be satisfied the conditions
\begin{equation}
\frac{\partial }{\partial \widetilde{x}^{2}}=\frac{1}{\sqrt{\left|
g_{2}\right| }}\partial _{2}\mbox{ and }\frac{\partial }{\partial \widetilde{%
x}^{3}}=\frac{1}{\sqrt{\left| g_{3}\right| }}\partial _{3}  \label{zetacoord}
\end{equation}%
We get
\begin{equation}
(i\gamma _{\underline{1}}\frac{\partial }{\partial \chi }+i\gamma _{%
\underline{2}}\frac{\partial }{\partial \widetilde{x}^{2}}+i\gamma _{%
\underline{3}}\frac{\partial }{\partial \widetilde{x}^{3}}-M)\Psi (\chi ,%
\widetilde{x}^{2},\widetilde{x}^{3})=0.  \label{diraceq2}
\end{equation}%
The equation (\ref{diraceq2}) describes the wave function of a Dirac
particle of mass $M$ propagating in a three dimensional Minkowski flat plane
which is imbedded as a N--adapted distribution into a 5D spacetime when the
extra dimension coordinate may be considered to by time--like.

The solution $\Psi =\Psi (\chi ,\widetilde{x}^{2},\widetilde{x}^{3})$ of (%
\ref{diraceq2}) is searched in the form
\begin{equation*}
\Psi =\left\{
\begin{array}{rcl}
\Psi ^{(+)}(\widetilde{x}) & = & \exp {[-i(k_{1}\chi +k_{2}\widetilde{x}%
^{2}+k_{3}\widetilde{x}^{3})]}\varphi ^{1}(k) \\
&  & \mbox{for positive energy;} \\
\Psi ^{(-)}(\widetilde{x}) & = & \exp {[i(k_{1}\chi +k_{2}\widetilde{x}%
^{2}+k_{3}\widetilde{x}^{3})]}\chi ^{1}(k) \\
&  & \mbox{for negative energy,}%
\end{array}%
\right.
\end{equation*}%
with the condition that $k_{1}$ is identified with the positive energy and $%
\varphi ^{1}(k)$ and $\chi ^{1}(k)$ are constant bispinors. To satisfy the
Klein--Gordon equation we must have
\begin{equation*}
k^{2}=-\left( {k_{1}}\right) ^{2}{+}\left( {k_{2}}\right) ^{2}{+}\left( {%
k_{3}}\right) ^{2}=M^{2}.
\end{equation*}%
The Dirac equation implies
\begin{equation*}
(\sigma ^{i}k_{i}-M)\varphi ^{1}(k)\mbox{ and }(\sigma ^{i}k_{i}+M)\chi
^{1}(k),
\end{equation*}%
where $\sigma ^{i}(i=1,2,3)$ are Pauli matrices corresponding to a
realization of gamma matrices as to a form of splitting to usual Pauli
equations for the bispinors $\varphi ^{1}(k)$ and $\chi ^{1}(k).$

In the rest frame for the horizontal plane parametrized by coordinates $%
\widetilde{x}=\{\chi ,\widetilde{x}^{2},\widetilde{x}^{3}\}$ there are four
independent solutions of the Dirac equation,
\begin{equation*}
\varphi _{(1)}^{1}(M,0)=\left(
\begin{array}{c}
1 \\
0 \\
0 \\
0%
\end{array}%
\right) ,\ \varphi _{(2)}^{1}(M,0)=\left(
\begin{array}{c}
0 \\
1 \\
0 \\
0%
\end{array}%
\right) ,\
\end{equation*}%
\begin{equation*}
\chi _{(1)}^{1}(M,0)=\left(
\begin{array}{c}
0 \\
0 \\
1 \\
0%
\end{array}%
\right) ,\ \chi _{(2)}^{1}(M,0)=\left(
\begin{array}{c}
0 \\
0 \\
0 \\
1%
\end{array}%
\right) .
\end{equation*}

We consider wave packets of type (for simplicity, we can use only
superpositions of positive energy solutions)
\begin{equation}
{}\Psi ^{(+)}(\zeta )=\int \frac{d^{3}p}{2\pi ^{3}}\frac{M}{\sqrt{%
M^{2}+(k^{2})^{2}}}\sum_{[\alpha ]=1,2,3}b(p,[\alpha ])\varphi ^{\lbrack
\alpha ]}(k)\exp {[-ik_{i}\widetilde{x}^{i}]}  \label{packet}
\end{equation}%
when the coefficients $b(p,[\alpha ])$ define a current (the group velocity)
\begin{equation}
J^{2}\equiv \sum_{\lbrack \alpha ]=1,2,3}\int \frac{d^{3}p}{2\pi ^{3}}\frac{M%
}{\sqrt{M^{2}+(k^{2})^{2}}}|b(p,[\alpha ])|^{2}\frac{p^{2}}{\sqrt{%
M^{2}+(k^{2})^{2}}}{}  \notag
\end{equation}%
with $|p^{2}|\sim M$ and the energy--momentum d--tensor (\ref{diracemd}) has
nontrivial coefficients
\begin{equation}
\Upsilon _{1}^{1}=2\Upsilon ({\widetilde{x}}^{2},{\widetilde{x}}%
^{3})=k_{1}\Psi ^{+}\gamma _{a}\Psi ,\Upsilon _{2}^{2}=-k_{2}\Psi ^{+}\gamma
_{2}\Psi ,\Upsilon _{3}^{3}=-k_{3}\Psi ^{+}\gamma _{3}\Psi \   \label{compat}
\end{equation}%
where the holonomic coordinates can be reexpressed ${\widetilde{x}}^{i}={%
\widetilde{x}}^{i}(x^{i}).$ We must take two or more waves in the packet and
choose such coefficients $b(p,[\alpha ]),$ satisfying corresponding
algebraic equations, in order to get from (\ref{compat}) the equalities
\begin{equation}
\Upsilon _{2}^{2}=\Upsilon _{3}^{3}=\Upsilon ({\widetilde{x}}^{2},{%
\widetilde{x}}^{3})=\Upsilon (x^{2}=\rho ,x^{3}=\zeta ),  \label{compat1}
\end{equation}%
required by the conditions (\ref{diracemd}).

We note that the ansatz for the 5D metric (\ref{ans5d}) and 4D spinor fields
depending on 3D h--coordinates $\widetilde{x}=\{\chi ,\widetilde{x}^{2},%
\widetilde{x}^{3}\}$ reduces the Dirac equation to the usual one projected
on a flat 3D spacetime. This configuration is N--adapted, because all
coefficients are computed with respect to N--adapted frames. The spinor
sourse $\Upsilon (x^{2},x^{3})$ induces an additional nonholonomic
deformation.

\subsection{Anisotropic gravitational disk--spinor fields}

\label{salgs}In this subsection, we construct two new classes of solutions
of the Einstein--Dirac fields generalizing the 5D disk metrics in string
gravity (\ref{str5m}) to contain contributions from the energy momentum
d--tensor
\begin{equation*}
\Upsilon _{\beta }^{\alpha }=\{2\Upsilon (\rho ,\zeta ),\Upsilon (\rho
,\zeta ),\Upsilon (\rho ,\zeta ),0,0\}
\end{equation*}%
for a Dirac wave packet satisfying the conditions (\ref{compat}) and (\ref%
{compat1}). This preserves the same gravitational field equations (\ref{ep1a}%
), i.e. (\ref{eq64a}), (\ref{ep3a}) and (\ref{ep4a}) but modifies (\ref%
{eq64b}) to
\begin{equation}
S_{4}^{4}=S_{5}^{5}=-\frac{1}{2h_{4}h_{5}}\left[ h_{5}^{\ast \ast
}-h_{5}^{\ast }\left( \ln \sqrt{|h_{4}h_{5}|}\right) ^{\ast }\right] =-\frac{%
\lambda _{H}^{2}}{4}-\Upsilon (\rho ,\zeta ).  \label{eq64bs}
\end{equation}%
As a particular case we can put $\lambda _{H}=0.$ Following the Corollary %
\ref{corgsol1}, see Appendix, the solution of the spinorially modified
equations can be written in the form
\begin{eqnarray}
\delta s^{2} &=&-(d\chi )^{2}+e^{\psi (\rho ,\zeta )}\left[ (d\rho
)^{2}+(d\zeta )^{2}\right] +\epsilon _{4}h_{0}^{2}(x^{i})\left[ f^{\ast
}\left( \rho ,\zeta ,v\right) \right] ^{2}\times  \label{str5msp} \\
&&|\varsigma _{4}\left( \rho ,\zeta ,v\right) |\left( \delta v\right)
^{2}+\epsilon _{5}\left[ f\left( \rho ,\zeta ,v\right) -f_{0}(\rho ,\zeta )%
\right] ^{2}\left( \delta y^{5}\right) ^{2},  \notag \\
\delta v &=&dv+w_{1}\left( \chi ,\rho ,\zeta ,v\right) d\chi +w_{2}\left(
\chi ,\rho ,\zeta ,v\right) d\rho +w_{3}\left( \chi ,\rho ,\zeta ,v\right)
d\zeta ,\   \notag \\
\delta y^{5} &=&dy^{5}+n_{1}\left( \chi ,\rho ,\zeta ,v\right) d\chi
+n_{2}\left( \chi ,\rho ,\zeta ,v\right) d\rho +n_{3}\left( \chi ,\rho
,\zeta ,v\right) d\zeta .  \notag
\end{eqnarray}%
From this ansatz, we can distinguish two classes of solutions: 1) If we take
$y^{4}=v=\varphi $ and $y^{5}=\tau ,$ with $\epsilon _{4}=1$ and $\epsilon
_{5}=-1,$ we generate a spinor source generalization of (\ref{sol3a}). 2)
For $y^{4}=v=\tau $ and $y^{5}=\varphi ,$ with $\epsilon _{4}=-1$ and $%
\epsilon _{5}=1,$ we obtain a spinor generalization of (\ref{sol3b}).

We note that the formulas for the N--connection coefficients,\\ $%
N_{i}^{4}=w_{i}(x^{k},v)$ and $N_{i}^{5}=n_{i}(x^{k},v),$ are
derived for a
different generation function (\ref{gf}), than that used in (\ref{str5mw}%
) and (\ref{str5mn}). In our case, we have
\begin{eqnarray}
\varsigma _{4}\left( \rho ,\zeta ,v\right) &=&\varsigma _{4[0]}\left( \rho
,\zeta \right) -\frac{\epsilon _{4}}{4}h_{0}^{2}(\rho ,\zeta )\times
\label{gfa} \\
&&\int \left[ -\frac{\lambda _{H}^{2}}{4}-\Upsilon (\rho ,\zeta )\right]
f^{\ast }\left( \rho ,\zeta ,v\right) \left[ f\left( \rho ,\zeta ,v\right)
-f_{0}(\rho ,\zeta )\right] dv.  \notag
\end{eqnarray}%
Substituting this value, respectively, in (\ref{gensol1w}) and (\ref%
{gensol1n}), we can compute
\begin{equation}
w_{i}=-\frac{\partial _{i}\varsigma _{4}\left( \rho ,\zeta ,v\right) }{%
\varsigma _{4}^{\ast }\left( \rho ,\zeta ,v\right) }  \label{str5mwsp}
\end{equation}%
and
\begin{equation}
n_{k}=n_{k[1]}\left( \chi ,\rho ,\zeta \right) +n_{k[2]}\left( \chi ,\rho
,\zeta \right) \int \frac{\left[ f^{\ast }\left( \rho ,\zeta ,v\right) %
\right] ^{2}}{\left[ f\left( \rho ,\zeta ,v\right) -f_{0}(\rho ,\zeta )%
\right] ^{3}}\varsigma _{4}\left( \rho ,\zeta ,v\right) dv.  \label{str5mnsp}
\end{equation}

The function $\psi =\ln |g_{2}|=\ln |g_{3}|$ from \ (\ref{str5msp}) is just
the solution of (\ref{eq64a})\ with signatures $\epsilon _{2}=\epsilon
_{3}=1,$ $\psi $ is a solution of
\begin{equation*}
\psi ^{\bullet \bullet }+\psi ^{\prime \prime }=-\frac{\lambda _{H}^{2}}{2},
\end{equation*}%
with $\psi ^{\bullet }=\partial \psi /\partial \rho $ and $\psi ^{^{\prime
}}=\partial \psi /\partial \zeta .$

We can fix a disk--solitonic background on which packages of 3D Dirac wave
propagate self--consistently if we take the function $f\left( \rho ,\zeta
,v\right) $ from the coefficients (\ref{str5mnsp}), computed by formulas (%
\ref{str5mwsp}) and (\ref{str5mnsp}), to be a solution of the 3D solitonic
equation (\ref{kdpe}) or (\ref{sg3e}).

\section{Lie Algebroid Symmetries and Disk Configurations}

\label{sgdal}The nonholonomically deformed disk solutions do not possess, in
general, Killing symmetries. In this section we show that it is possible to
distinguish certain configurations with Lie algebroid symmetry. We restrict
our considerations to the class of Riemann--Cartan spacetimes defining Lie
algebroid configuratons \cite{valg}. On details on Lie algebroids geometry
and applications, we refer to \cite{galg}.

\subsection{Gravitational Lie algebroids and N--connections}

Let us consider a vector bundle $\mathcal{E}=(E,\pi ,M),$ with a surjective
map $\pi :E\longrightarrow M$ of the total spaces $E$ to the base manifold $%
M,$ where dimensions are finite ones, $\dim E=n+m$ and $\dim M=n.$ The Lie
algebroid structure $\mathcal{A}\doteqdot (E,\left[ \cdot ,\cdot \right]
,\rho )$ is defined by an anchor map $\rho :\ E\rightarrow TM$ \ ($TM$ is
the tangent bundle to $M$) and a Lie bracket on the $C^{\infty }(M)$--module
of sections of $E,$ denoted $Sec(E),$ such that
\begin{equation*}
\left[ X,fY\right] =f\left[ X,Y\right] +\rho (X)(f)Y
\end{equation*}%
for any $X,Y\in Sec(E)$ and $f\in C^{\infty }(M).$

For applications to gravity and string theories and spaces privided with
nonholonomic frame (vielbein) structure, one works on a general manifold $%
\mathbf{V},~$\ $\ dim\mathbf{V}=n+m,$ which is a (pseudo) Riemannian
spacetime, or a more general one with nontrivial torsion. A Lie algebroid
structure can be modelled locally on a nonholonomic spacetime $\mathbf{V}$
defined by a Whitney type sum%
\begin{equation}
T\mathbf{V=}h\mathbf{V}\oplus v\mathbf{V}  \label{whit}
\end{equation}%
stating a splitting into certain conventional horizontal (h) and vertical
(v) subspaces. This distribution is in general nonitegrable (nonholonomic)
and defines a N--connection structure. A such manifold is called
N--anholonomic (see Ref. \cite{valg,v0408121}). An anchor structure is
defined as a map $\ \widehat{\rho }:\ \mathbf{V}\rightarrow h\mathbf{V}$ and
the Lie bracket structure is considered on the spaces of sections $Sec(v%
\mathbf{V}).$ For the purposes of this paper, we can consider
Riemann--Cartan nonholonomic manifolds admitting a locally fibered structure
induced by the splitting (\ref{whit}) when the Lie algebroid constructions
are usual ones but with formal substitutions $E\rightarrow $ $\ \mathbf{V}$
and $M\rightarrow h\mathbf{V.}$

One defines the Lie algebroid structure in local form by its structure
functions $\rho _{a}^{i}(x)$ and $C_{ab}^{f}(x)$ defining the relations
\begin{eqnarray}
\ \rho (v_{a}) &=&\rho _{a}^{i}(x)\ e_{i}=\rho _{a}^{\underline{i}}(x)\
\partial _{\underline{i}},  \label{anch} \\
\lbrack v_{a},v_{b}] &=&C_{ab}^{c}(x)\ v_{c}  \label{liea}
\end{eqnarray}%
and subjected to the structure equations
\begin{equation}
\rho _{a}^{j}\frac{\partial \rho _{b}^{i}}{\partial x^{j}}-\rho _{b}^{j}%
\frac{\partial \rho _{a}^{i}}{\partial x^{j}}=\rho _{c}^{j}C_{ab}^{c}~%
\mbox{\ and\ }\sum\limits_{cyclic(a,b,c)}\left( \rho _{a}^{j}\frac{\partial
C_{bc}^{d}}{\partial x^{j}}+C_{af}^{d}C_{bc}^{f}\right) =0.  \label{lasa}
\end{equation}%
We shall omit underlying of coordinate indices if it will not result in
ambiguities. Such equations are standard ones for the Lie algebroids but
defined on N--anholonomic manifolds. In brief, we call such spaces to be Lie
N--algebroids. We can consider that spacetimes provided with Lie algebroid
structure generalize the class of manifolds possessub Lie algebra symmetry
to a case when the structure constants transform into structure functions
depending on holonomic coordinates, see (\ref{lasa}). On such spaces, we can
work with singular structures defined by the so--called 'anchor' map.

The Lie algebroid and N--connection structures can be established in a
mutually compatible form by introducing the ''N--adapted'' anchor%
\begin{equation}
{}\widehat{\mathbf{\rho }}_{a}^{j}(x,u)\doteqdot \mathbf{e}_{\ \underline{j}%
}^{j}(x,u)\mathbf{e}_{a}^{\ \underline{a}}(x,u)\ \rho _{\underline{a}}^{%
\underline{j}}(x)  \label{bfanch}
\end{equation}%
and ''N--adapted'' (boldfaced) structure functions
\begin{equation}
\mathbf{C}_{ag}^{f}(x,u)=\mathbf{e}_{\ \underline{f}}^{f}(x,u)\mathbf{e}%
_{a}^{\ \underline{a}}(x,u)\mathbf{e}_{g}^{\ \underline{g}}(x,u)\ C_{%
\underline{a}\underline{g}}^{\underline{f}}(x),  \label{bfstrf}
\end{equation}%
respectively, into formulas (\ref{anch}), (\ref{liea}) and (\ref{lasa}). One
follows that the Lie algebroids on N--anholonomic manifolds are defined by
the corresponding sets of functions ${}\widehat{\mathbf{\rho }}_{a}^{j}(x,u)$
and $\mathbf{C}_{ag}^{f}(x,u)$ with additional dependencies on v--variables $%
u^{b}$ for the N--adapted structure functions. For such generalized Lie
N--algebroids, the structure relations became
\begin{eqnarray}
{}\widehat{\mathbf{\rho }}(v_{b}) &=&{}\widehat{\mathbf{\rho }}%
_{b}^{i}(x,u)\ e_{i},  \label{anch1d} \\
\lbrack v_{d},v_{b}] &=&\mathbf{C}_{db}^{f}(x,u)\ v_{f}  \label{lie1d}
\end{eqnarray}%
and the structure equations of the Lie N--algebroid are written
\begin{eqnarray}
{}\widehat{\mathbf{\rho }}_{a}^{j}e_{j}({}\widehat{\mathbf{\rho }}%
_{b}^{i})-{}\widehat{\mathbf{\rho }}_{b}^{j}e_{j}({}\widehat{\mathbf{\rho }}%
_{a}^{i}) &=&{}\widehat{\mathbf{\rho }}_{e}^{j}\mathbf{C}_{ab}^{e},
\label{lased} \\
\sum\limits_{cyclic(a,b,e)}\left( {}\widehat{\mathbf{\rho }}_{a}^{j}e_{j}(%
\mathbf{C}_{be}^{f})+\mathbf{C}_{ag}^{f}\mathbf{C}_{be}^{g}-\mathbf{C}%
_{b^{\prime }e^{\prime }}^{f^{\prime }}{}\widehat{\mathbf{\rho }}_{a}^{j}%
\mathbf{Q}_{f^{\prime }bej}^{fb^{\prime }e^{\prime }}\right) &=&0,  \notag
\end{eqnarray}%
for $\mathbf{Q}_{f^{\prime }bej}^{fb^{\prime }e^{\prime }}=\mathbf{e}_{\
\underline{b}}^{b^{\prime }}\mathbf{e}_{\ \underline{e}}^{e^{\prime }}%
\mathbf{e}_{f^{\prime }}^{\ \underline{f}}~e_{j}(\mathbf{e}_{b}^{\
\underline{b}}\mathbf{e}_{e}^{\ \underline{e}}\mathbf{e}_{\ \underline{f}%
}^{f})$ with the values $\mathbf{e}_{\ \underline{b}}^{b^{\prime }}$ and $%
\mathbf{e}_{f^{\prime }}^{\ \underline{f}}$ defined by the \ N--connection.
The Lie N--algebroid structure is characterized by the coefficients ${}%
\widehat{\mathbf{\rho }}_{b}^{i}(x,u)$ and $\mathbf{C}_{db}^{f}(x,u)$ stated
with respect to the N--adapted frames (\ref{dder}) \ and (\ref{ddif}).

\subsection{Disks with Lie Algebroid symmetry}

Let us analyze the conditions when a class of metrics of type (\ref{ans5d})
models a Lie algebroid provided with N--connection structure.\ We re--write
the metric in dual form,
\begin{eqnarray*}
\mathbf{g} &=&\mathbf{g}^{\alpha \beta }\left( \mathbf{u}\right) \mathbf{e}%
_{\alpha }\otimes \mathbf{e}_{\beta }=g^{ij}\left( \mathbf{u}\right)
e_{i}\otimes e_{j}+h^{ab}\left( \mathbf{u}\right) \ v_{b}\otimes \ v_{b}, \\
&=&\left( g_{i}\right) ^{-1}e_{i}\otimes e_{i}+\left( h_{a}\right) ^{-1}\
v_{a}\otimes \ v_{a}
\end{eqnarray*}%
where $v_{a}\doteqdot $ $e_{a\ }^{\ \underline{a}}(x,u)\ \partial /\partial
u^{\underline{a}}$ satisfy the Lie N--algebroid conditions $%
v_{a}v_{b}-v_{b}v_{a}=\mathbf{C}_{ab}^{d}(x,u)v_{d}$ of type (\ref{lie1d})
with $e_{i}$ being of type (\ref{dder}). The functions $g_{i}$ and $h_{a}$
are any data defining a nonholonomic deformation of a disk solutions
considered in the previous sections.

It is possible to introduce an anchor map (\ref{anch1d}) in the form
\begin{equation}
\ \widehat{\mathbf{\rho }}_{a^{\prime }}^{i}\left( \mathbf{u}\right) =%
\mathbf{e}_{\ \underline{i}}^{i}(\mathbf{u})\mathbf{e}_{a^{\prime }}^{\
\underline{a}}(\mathbf{u})\ \rho _{\underline{a}}^{\underline{i}}(x),
\label{anchbf}
\end{equation}%
defined in the form (\ref{bfanch}) by some matrices of type (\ref{vt1}) and (%
\ref{vt2}), and write the canonical relation
\begin{equation}
g^{ij}\left( \mathbf{u}\right) =h^{a^{\prime }b^{\prime }}\left( \mathbf{u}%
\right) \ \widehat{\mathbf{\rho }}_{a^{\prime }}^{i}\left( \mathbf{u}\right)
\ \widehat{\mathbf{\rho }}_{b^{\prime }}^{j}\left( \mathbf{u}\right) .
\label{cananch1}
\end{equation}%
As a result, the h--component of the d--metric is represented%
\begin{eqnarray*}
g^{ij}\left( \mathbf{u}\right) e_{i}\otimes e_{j} &=&h^{a^{\prime }b^{\prime
}}\left( \mathbf{u}\right) e_{\ \underline{i}}^{i}\left( \mathbf{u}\right) \
e_{a^{\prime }}^{\ \underline{a}}(x)\rho _{\underline{a}}^{\underline{i}%
}(x)e_{\ \underline{j}}^{j}\left( \mathbf{u}\right) e_{b^{\prime }}^{\
\underline{b}}(x)\rho _{\underline{b}}^{\underline{j}}(x)e_{i}\otimes e_{j}
\\
&=&h^{a^{\prime }b^{\prime }}\left( \mathbf{u}\right) \rho _{a^{\prime }}^{%
\underline{i}}(x)\rho _{b^{\prime }}^{\underline{j}}(x)\frac{\partial }{%
\partial x^{\underline{i}}}\otimes \frac{\partial }{\partial x^{\underline{j}%
}},
\end{eqnarray*}%
where $\rho _{a^{\prime }}^{\underline{i}}(x)=e_{a^{\prime }}^{\ \underline{a%
}}(x)\rho _{\underline{a}}^{\underline{i}}(x).$ For effective diagonal
metrics, with respect to N--adapted frames, the formulas (\ref{anchbf}) and (%
\ref{cananch1}) symplify substantially. We conclude that the ansatz (\ref%
{ans5d}) admits a Lie algebroid type structure with the structure functions $%
\rho _{\underline{a}}^{\underline{i}}(x)$ and $C_{\underline{a}\underline{b}%
}^{\underline{d}}(x)$ if and only if the contravariant h--component of the
corresponding d--metric, with respect to the local coordinate basis, can be
parametrized in the form
\begin{equation}
g^{\underline{i}\underline{j}}\left( \mathbf{u}\right) =h^{a^{\prime
}b^{\prime }}\left( \mathbf{u}\right) \rho _{a^{\prime }}^{\underline{i}%
}(x)\rho _{b^{\prime }}^{\underline{j}}(x),  \label{cananch2}
\end{equation}%
i.e.
\begin{equation*}
\left( g_{i}\right) ^{-1}\left( \mathbf{u}\right) =\sum_{a}\left(
h_{a}\right) ^{-1}\left( \mathbf{u}\right) \left[ \rho _{a}^{i}(x)\right]
^{2}
\end{equation*}%
The anchor $\rho _{a^{\prime }}^{\underline{i}}(x)$ may be treated as a
vielbein transform depending on $x$--coordinates lifting the horizontal
components of the contravariant metric on the v--subspace. The Lie type
structure functions $C_{ab}^{d}(x,u)$ define certain anholonomy relations
for the basis $v_{a}.$ On a general Lie N--algebroids we can consider any
set of coefficients $\ \widehat{\mathbf{\rho }}_{a^{\prime }}^{i}\left(
\mathbf{u}\right) $ and $\mathbf{C}_{ab}^{d}(x,u)$ not obligatory subjected
to the data (\ref{anchbf}).

On N--anholonomic manifolds, it is more convenient to work with the
N--adapted relations (\ref{cananch1}) than with (\ref{cananch2}). \ For
effectively diagonal d--metrics, such anchor map conditions must be
satisfied both for $\eta _{\alpha }=1$ and nontrivial values of $\eta
_{\alpha },$ i. e.
\begin{eqnarray}
g^{i} &=&h^{4}\left( \ \widehat{\mathbf{\rho }}_{4}^{i}\right)
^{2}+h^{5}\left( \ \widehat{\mathbf{\rho }}_{5}^{i}\right) ^{2},
\label{eqanch} \\
\underline{g}^{i} &=&\underline{g}^{4}\left( \ \widehat{\mathbf{\rho }}%
_{4}^{i}\right) ^{2}+\underline{g}^{5}\left( \ \widehat{\mathbf{\rho }}%
_{5}^{i}\right) ^{2},  \notag
\end{eqnarray}%
for $\eta _{\alpha }\rightarrow 1,$ where $g^{i}=1/g_{i},h^{a}=1/h_{a},\eta
^{\alpha }=1/$ $\eta _{\alpha }$ and $\underline{g}^{\alpha }=1/\underline{g}%
_{\alpha }.$ The real solutions of (\ref{eqanch}) are
\begin{equation}
\left( \ \widehat{\mathbf{\rho }}_{4}^{i}\right) ^{2}=\underline{g}^{i}%
\underline{g}_{4}H_{4}^{i},\ \left( \ \widehat{\mathbf{\rho }}%
_{5}^{i}\right) ^{2}=-\underline{g}^{i}\underline{g}_{5}H_{4}^{i},
\label{anch5}
\end{equation}%
where%
\begin{equation*}
H_{a}^{i}=\eta _{a}\frac{1-\eta _{4}/\eta _{i}}{\eta _{5}-\eta _{4}},
\end{equation*}%
$\eta _{1}=1,$ for any parametrization of solutions with a set of values of $%
\underline{g}^{\alpha }$ and $\eta ^{\alpha }$ for which $\left( \ \widehat{%
\mathbf{\rho }}_{a}^{i}\right) ^{2}\geq 0.$ The nontrivial anchor
coefficients can be related to a general solution of type (\ref{ans5d}) via
gravitational polarizatons $\eta _{\alpha }$ from (\ref{polf}) when $\
g_{k}=\eta _{k}\underline{g}_{k}$ and $h_{a}=\eta _{a}\underline{h}_{a}.$ In
a particular case, we may compute the anchor nontrivial components (\ref%
{anch5}) for an explicit solutions, for instance, taking the data (\ref{ex02}%
).

The final step in Lie algebroid classification of such classes of metrics is
to define $\mathbf{C}_{ab}^{d}(x,u)$ from the algebraic relations defined by
the first equation in (\ref{lased}) with given values for $\ \widehat{%
\mathbf{\rho }}_{a}^{i},$ see (\ref{anch5}), and defined N--elongated
operators $e_{i}.$ In result, the second equation in (\ref{lased}) will be
satisfied as a consequence of the first one. This restrict the classes of
possible v--frames, $v_{b}=e_{b}^{\ \underline{b}}(x,u)\partial /\partial u^{%
\underline{b}},$ where $e_{b}^{\ \underline{b}}(x,u)$ have to satisfy the
algebraic relations (\ref{lie1d}). We conclude, that the Lie N--algebroid
structure imposes certain algebraic constraints on the coefficients of
vielbein transforms.

Finally, we note that the Lie algebroid classification of N--anholonomic
spaces can be re--written in terms of Clifford algebroids elaborated in Ref. %
\cite{valg}. A such approach is more appropriate in spinor gravity
and for Einstein--Dirac fields. In general, the spinor
constructions can be included in the left or right minimal ideals
of Clifford algebras generalized to Clifford spaces (C--spaces,
elaborated in details in Refs. \cite{castro}), in our case
provided with additional N--connections and/or algebroid
structure.

\section{ Discussion and Conclusions}

In this paper we constructed explicit exact solutions describing
nonholonmic deformations of four dimensional (in brief, 4D) disk
solutions to generic off--diagonal metrics in 4D and 5D gravity.
The solutions depend on three and, respectively, four coordinates,
posses local anisotropies and generalized symmetries. We
identified the circumstances under which they may resemble the
usual disk solutions but with certain nonlinear polarizations in
the static and stationary cases and/or propagating in time  or
extra dimension. There were also analyzed the conditions when the
solutions may be constrained to define vacuum and nonvacuum
configurations in Einstein gravity.

Our results and the specific properties of the applied geometric
method, in more detail, are as follows:

Let us firstly consider the main features of the so--called
anholonomic frame method elaborated in our works
\cite{vjhep2,vsbd,vs,vp,vt,vnces} and its differences from well
known approaches to constructing disk, instanton, black hole or
toroidal solutions in gravity theories:

The very complicated structure of the Einstein equations and
matter field equations in a general curved spacetime gives none
hope to find general solutions of such systems of nonlinear
partial differential equations. The bulk of known physically
important  solutions in gravity (see, for instance, Ref.
\cite{exac}) were constructed by some particular types of metric
and (for non--vacuum configurations) matter field ansatz reducing
the field equations to any systems of algebraic or nonlinear
ordinary differential equations. The solutions of such equations
depend on (integration) constants which are
physically defined from some symmetry prescriptions \footnote{%
for instance, one impose the spherical or cylindrical symmetry of
spacetime, or some Lie algebra symmetries like in anisotropic
cosmology models} and some boundary (asymptotic) conditions like
the request to get in the limit the Minkowski flat spacetime and
the Newton gravitational low. Such solutions are usually
parametrized by diagonal metrics, vielbeins and connections which
in corresponding coordinates depends on one time, or one space,
like coordinate, or there are considered some stationary
(rotating) and/or wave type dependencies.

In our works, we used more general ansatz for the gravitational
and matter fields, see metric (\ref{metric5}) in Appendix, where
(for convenience) there are summarized the main results on the
anholonomic frame method. In 4D and 5D, such generic off--diagonal
metric ansatz
\footnote{%
they can not be diagonalized by any coordinate transforms} depend
respectively on three and four variables (coordinates). The
constructions can be also performed by analogy for higher or lower
dimensions. In the vacuum case, a such ansatz reduces the Einstein
equations to a system of nonlinear partial differential equations,
see Theorem \ref{t5dr}, which can be integrated in very general
form, see Theorem \ref{texs}. To prove such results, we applied
new geometric concepts and methods which came from the Finsler and
Lagrange geometry \cite{ma} but further adapted to purposes of
gravity theories on
nonholonmic (pseudo) Riemannian and Rieman--Cartan--Weyl spacetimes \cite%
{vnces,v1,v2}. The new classes of solutions depend not only on integration
constants but also on certain types of integration functions of one, two and
three variables, for 4D, and on functions on four variables, in 5D. This is
a general property of solutions of systems of partial differential equations.

By applying the anholonomic frame method, many rigorous and sophisticate
solutions of Einstein's equations have been constructed, see \cite%
{vjhep2,vsbd,vs,vp,vt,vnces,valg,v2} and there presented references. A part
of such solutions have been considered for physically relevant situations
(like locally anisotropic black hole, wormhole, Taub NUT ... solutions).
Nevertheless, there is still a problem of complete understanding the
physical meaning of solutions depending not only on arbitrary physical
constants but on classes of functions arising by integrating systems of
nonlinear partial differential equations.

The first fundamental property of such solutions is that they
describe  classes of spacetimes characterized by certain types of
nonholonomy relations for vielbeins with associated nonlinear
(N--connection) structure defined by generic off--diagonal metric
terms. In this approach, one deals with induced torsions and the
gravitational and matter field interactions are modelled on
nonholonmomic manifolds, i.e. on manifolds with prescribed
nonholonomic, equivalently, anholonomic, or nonintegrable
distributions. As a matter of principle, the constructions can be
equivalently redefined with respect to coordinate frames with
vanishing of induced torsions but the formulas became very
cumbersome and non adapted to the fundamental geometric objects.
On such nonholonomic spaces, it is convenient to work with more
general classes of linear connections (not only with the
Levi--Civita one) which allows us to apply the method in string
and gauge gravity models when the torsion fields are not trivial.
The (psudo) Riemannian configurations can be emphasized by certain
special classes of constraints on the nonholonomic frame
structure. But even in such cases, the N--connection may be
nontrivial and the constructions depend on some classes of
integration functions.

The second fundamental property of the nonholonomic spacetimes (defined by
generic off--diagonal metrics and nonholonomic frames) is that they are
characterized by more general symmetries than in the case of usual Killing
spacetimes with spherical, or cylindrical, or with Lie algebra symmetries.
In Ref. \cite{vnces}, we analyzed in details some examples of exact
solutions characterized by noncommutative symmetries which (surprisingly)
are present even on real 'off--diagonal' spacetimes and can be emphasized if
the nonholonomic deformations of metrics are associated to certain types of
Siberg--Witten transforms.

Developing such ideas, we proposed to characterize the new classes of
generic off--diagonal solutions also by Lie algebroid symmetries, or
corresponding generalizations to Clifford algebroids \cite{valg} for
Einstein--Dirac systems. Roughly speaking, the concept of Lie algebroid
generalizes the concept of Lie algebra to a case when the structure
constants transform into structure functions depending on a base manifold
coordinates and one can work with singular structures defined by the
so--called 'anchor' map (see the main concepts and results in Refs. \cite%
{galg}). In a slight different interpretation, a spacetime with
Lie algebroid symmetry may be considered as a geometric
construction modelling the fibered and/or bundle spaces on
nonholonomic manifols if anholonomic frames and their deformations
are introduces into consideration. This allows us to work with
classes of solutions characterized by structure functions and
nonholonomic frame structures associated to integration functions
defined by constructing more general type of solutions of the
Einstein equations.

Of course, we may impose further constraints and see what happen
when such ansatz depend on one time, or space/extra dimension,
like variable and the Einstein equations transform into some
systems of algebraic or nonlinear ordinary differential equations.
These are very restricted cases when we loose certain quality and
fundamental properties of the generic gravitational and matter
field equations. The real world and gravity can not be completely
analyzed only by ansatz resulting in ordinary differential
equations, or algebraic systems. In general, the gravitational and
matter field interactions are described, by multi--dimensional
nonlinear and nonperturbative effects which can not be derived
only from solutions of ordinary differential equations. The
anholonomic frame method is a geometric one giving the possibility
to construct exact solutions of gravitational and matter field
equations reduced to systems of nonlinear partial equations on
2--4 variables. Here, we note that it is possible to consider more
particular classes of integration functions when 'far away' from
such nonholonomic gravitational--matter configurations the
spacetime will have a Minkowski or (anti) de Sitter limit. But for
certain finite, or infinite, regions, the nonholonomic spacetimes
will be characterized, for instance, by algebroid and/or
noncommutative symmetries.

A special interest presents the classes of metrics with algebroid
symmetries which can be defined by a chain of conformal and/or
nonholonomic deformations of a well known class of exact solutions
describing certain physical interesting situations. In this work
we investigated in details how a given disk solution (for
instance, any Neugebauer--Meinel one \cite{nm}) can be
nonholonomically extended to another type of solutions, depending
correspondingly on classes of functions on 2--4 variables, in
nonholonomic 4D and 5D gravity. We considered examples from
general gravity, with nontrivial cosmological constant, bosonic
string gravity with so--called $H$--field torsion. There were also
analyzed various type of nontrivial gravitational 3D solitonic
background deformations and nonholonomic Einstein--Dirac solutions
with disk like symmetries. For small deformations, such new
solutions seem to preserve the properties of usual disk solutions
but with additional anisotropic polarizations and new type of
symmetries. In general, the disk 'character' of solutions is
broken by nonholonomic transforms.

Finally, we note that in our further works we shall be interested
in constructing and investigation of certain exact solutions
describing nonlinear supperpositions of some black
ellipsoid/torus/hole, or wormhole, configurations with a polarized
disk configuration, or to say that a disk configuration is moving
self--consistently in a solitonic background in 4D or extra
dimension.  This would be a further development of results in
\cite{vsbd,vs,vp,vt} in order to generate solutions with
nontrivial Lie algebroid and/or noncommutative symmetries and see
their applications in modern cosmology and astrophysics.

\vskip6pt

\textbf{Acknowledgement: } The work is supported by a sabbatical
fellowship of the Ministry of Education and Research of Spain.

\appendix

\section{Vielbeins and N--Connections}

In this section we recall some basic facts on the geometry of nonholonomic
frames (equivalently, vielbeins) with associated nonlinear connection
(N--connection) structure in Riemann--Cartan spaces, see Ref. \cite{v2} for
more general constructions in generalized Finsler--affine geometry.

A spacetime is modelled as a manifold $V^{n+m}$ of dimension $n+m,$ with $%
n\geq 2$ and $m\geq 1.$ The local coordinates are labelled in the form $%
u^{\alpha }=(x^{i},y^{a})$ when Greek indices $\alpha ,\beta ,...$ split
into subclasses like $\alpha =\left( i,a\right) ,$ $\beta =\left( j,b\right)
...$ where the Latin indices (the so--called horizontal, h, ones) $i,j,k,...$
run values $1,2,...n$ and $\ $(the vertical, v, ones) $a,b,c,...$ run values
$n+1,n+2,$ ..., $n+m.$ We denote by $\pi ^{\top }:TV^{n+m}\rightarrow TV^{n}$
the differential of a map $\pi :V^{n+m}\rightarrow V^{n}$ defined by fiber
preserving morphisms of the tangent bundles $TV^{n+m}$ and $TV^{n}.$ The
kernel of $\pi ^{\top }$is just the vertical subspace $vV^{n+m}$ with a
related inclusion mapping $i:vV^{n+m}\rightarrow TV^{n+m}.$

\begin{definition}
A nonlinear connection (N--connection) $\mathbf{N}$ on space $V^{n+m}$ is
defined by the splitting on the left of an exact sequence
\begin{equation*}
0\rightarrow vV^{n+m}\rightarrow TV^{n+m}\rightarrow
TV^{n+m}/vV^{n+m}\rightarrow 0,
\end{equation*}%
i. e. by a morphism of submanifolds $\mathbf{N:\ \ }TV^{n+m}\rightarrow
vV^{n+m}$ such that $\mathbf{N\circ i}$ is the unity in $vV^{n+m}.$
\end{definition}

Equivalently, a N--connection is defined by a Whitney sum of horizontal (h)
subspace $\left( hV^{n+m}\right) $ and vertical (v) subspaces,
\begin{equation}
TV^{n+m}=hV^{n+m}\oplus vV^{n+m}.  \label{whitney}
\end{equation}%
A spacetime provided with N--connection structure is denoted $\mathbf{V}%
^{n+m}.$ This is a nonholonomic manifold because the distribution (\ref%
{whitney}), in general, is not integrable, i.e. nonholonomic (equivalently,
anholonomic). In brief, we call such spaces to be N--anholonomic because
their nonholonomy is defined by a N--connection structure. We shall use
boldfaced indices for the geometric objects adapted to a N--connec\-ti\-on.%
\footnote{%
We also use boldfaced symbols in order to emphasize that some geometric
objects are defined with respect to vielbeins with associated N--connection
structure.}

Locally, a N--connection is defined by its coefficients $N_{i}^{a}\left(
u\right) =N_{i}^{a}(x,y),$ i. e.
\begin{equation*}
\mathbf{N}=N_{i}^{a}(u)d^{i}\otimes \partial _{a},
\end{equation*}%
characterized by the N--connection curvature
\begin{equation*}
\mathbf{\Omega }=\frac{1}{2}\Omega _{ij}^{a}d^{i}\wedge d^{j}\otimes
\partial _{a},
\end{equation*}%
with coefficients%
\begin{equation}
\Omega _{ij}^{a}=\delta _{\lbrack j}N_{i]}^{a}=\frac{\partial N_{i}^{a}}{%
\partial x^{j}}-\frac{\partial N_{j}^{a}}{\partial x^{i}}+N_{i}^{b}\frac{%
\partial N_{j}^{a}}{\partial y^{b}}-N_{j}^{b}\frac{\partial N_{i}^{a}}{%
\partial y^{b}}.  \label{ncurv}
\end{equation}%
The linear connections are defined as a particular case when the
coefficients are linear on $y^{a},$ i. e. $N_{i}^{a}(u)=\Gamma
_{bj}^{a}(x)y^{b}.$

A general metric structure may be written in the form
\begin{equation}
\mathbf{g}=\mathbf{g}_{\alpha \beta }\left( u\right) \mathbf{e}^{\alpha
}\otimes \mathbf{e}^{\beta }=g_{ij}\left( u\right) d^{i}\otimes
d^{j}+h_{ab}\left( u\right) \delta ^{a}\otimes \delta ^{b},  \label{block2}
\end{equation}%
where
\begin{equation}
\ \mathbf{e}^{\beta }=\left( d^{i},\delta ^{a}\right) \equiv \delta
u^{\alpha }=\left( \delta x^{i}=dx^{i},\delta y^{a}=dy^{a}+N_{i}^{a}\left(
u\right) dx^{i}\right)   \label{ddif}
\end{equation}%
is dual to
\begin{equation}
\mathbf{e}_{\alpha }=\left( \delta _{i},\partial _{a}\right) \equiv \frac{%
\delta }{\delta u^{\alpha }}=\left( \frac{\delta }{\delta x^{i}}=\partial
_{i}-N_{i}^{a}\left( u\right) \partial _{a},\frac{\partial }{\partial y^{a}}%
\right) .  \label{dder}
\end{equation}%
For constructions on spaces provided with N--connection structure, we have
to consider 'N--elongated' operators instead of usual partial derivatives,
like $\delta _{j}$ in (\ref{ncurv}) and (\ref{dder}). \

We emphasize that a set of coefficients $N_{i}^{a}$ defines a special class
of frame transforms parametrized by matrices of type

\begin{eqnarray}
\mathbf{e}_{\alpha }^{\ \underline{\alpha }} &=&\left[
\begin{array}{cc}
e_{i}^{\ \underline{i}}(u) & N_{i}^{b}(u)e_{b}^{\ \underline{a}}(u) \\
0 & e_{a}^{\ \underline{a}}(u)%
\end{array}%
\right] ,  \label{vt1} \\
\mathbf{e}_{\ \underline{\beta }}^{\beta } &=&\left[
\begin{array}{cc}
e_{\ \underline{i}}^{i\ }(u) & -N_{k}^{b}(u)e_{\ \underline{i}}^{k\ }(u) \\
0 & e_{\ \underline{a}}^{a\ }(u)%
\end{array}%
\right] .  \label{vt2}
\end{eqnarray}%
In a particular case, one put $e_{i}^{\ \underline{i}}=\delta _{i}^{%
\underline{i}}$ and $e_{a}^{\ \underline{a}}=\delta _{a}^{\underline{a}}$
with $\delta _{i}^{\underline{i}}$ and $\delta _{a}^{\underline{a}}$ being
the Kronecker symbols, defining a global splitting of $\mathbf{V}^{n+m}$
into ''horizontal'' and ''vertical'' subspaces (respectively, h- and
v--subspaces) with the vielbein structure%
\begin{equation*}
\mathbf{e}_{\alpha }=\mathbf{e}_{\alpha }^{\ \underline{\alpha }}\partial _{%
\underline{\alpha }}\mbox{ and }\mathbf{e}_{\ }^{\beta }=\mathbf{e}_{\
\underline{\beta }}^{\beta }du^{\underline{\beta }},
\end{equation*}%
where we underline the indices related the local coordinate basis $\partial
_{\underline{\alpha }}=\partial /\partial u^{\underline{\alpha }}$ and $du^{%
\underline{\beta }}.$ We shall omit underlining of indices if this will not
result in ambiguities.

The metric (\ref{block2}) can be equivalently written in ''off--diagonal''
form
\begin{equation*}
\mathbf{g}=\underline{g}_{\alpha \beta }\left( u\right) du^{\alpha }\otimes
du^{\beta },
\end{equation*}%
for the coefficients
\begin{equation}
\underline{g}_{\alpha \beta }=\left[
\begin{array}{cc}
g_{ij}+N_{i}^{a}N_{j}^{b}h_{ab} & N_{j}^{e}h_{ae} \\
N_{i}^{e}h_{be} & h_{ab}%
\end{array}%
\right]  \label{ansatz}
\end{equation}%
computed with respect to a coordinate co--basis $du^{\alpha }=\left(
dx^{i},dy^{a}\right) .$

The N--coframe (\ref{ddif}) satisfies the anholonomy relations
\begin{equation}
\left[ \mathbf{e}_{\alpha },\mathbf{e}_{\beta }\right] =\mathbf{e}_{\alpha }%
\mathbf{e}_{\beta }-\mathbf{e}_{\beta }\mathbf{e}_{\alpha }=\mathbf{w}_{\
\alpha \beta }^{\gamma }\left( u\right) \mathbf{e}_{\gamma }  \label{anhr}
\end{equation}%
with nontrivial anholonomy coefficients $\mathbf{w}_{\beta \gamma }^{\alpha
}\left( u\right) $ computed
\begin{equation}
\mathbf{w}_{~ji}^{a}=-\mathbf{w}_{~ij}^{a}=\Omega _{ij}^{a},\ \mathbf{w}%
_{~ia}^{b}=-\mathbf{w}_{~ai}^{b}=\partial _{a}N_{i}^{b}.  \label{anhc}
\end{equation}%
It should be noted that, in general, a metric (\ref{ansatz})\ is generic
off--diagonal, i. e. it can not be diagonalized by any coordinate
transforms. In general, the anholonomy coefficients (\ref{anhc}), defined by
the off--diagonal reprezentation (\ref{ansatz}), or equivalently by the
block reprezentation (\ref{block2}), do not vanish.

The geometric constructions can be adapted to the N--connection structure.
For instance, one can be elaborated this approach to the theory of linear
connections:

\begin{definition}
\label{ddcon}A distinguished connection (d--connection) $\mathbf{D}=\{%
\mathbf{\Gamma }_{\beta \gamma }^{\alpha }\}$ on $\mathbf{V}^{n+m}$ is a
linear connection conserving under parallelism the Whitney sum (\ref{whitney}%
).
\end{definition}

One also uses the term d--tensor for the decompositions of tensor with
respect to N--adapted bases and say that (\ref{block2}) is a d--metric
structure. \ The N--adapted components $\mathbf{\Gamma }_{\beta \gamma
}^{\alpha }$ of a d--connection $\mathbf{D=\{D}_{\alpha }\}$ (equivalently,
a covariant derivative) are defined by the equations
\begin{equation}
\mathbf{\Gamma }_{\ \alpha \beta }^{\gamma }\left( u\right) =\left( \mathbf{D%
}_{\alpha }\mathbf{e}_{\beta }\right) \rfloor \mathbf{e}^{\gamma }.
\label{dcon1}
\end{equation}%
The operations of h- and v-covariant derivations, $D_{k}^{[h]}=%
\{L_{jk}^{i},L_{bk\;}^{a}\}$ and $D_{c}^{[v]}=\{C_{jk}^{i},C_{bc}^{a}\}$ are
introduced by corresponding h- and v--parametrizations of (\ref{dcon1}),%
\begin{equation*}
L_{jk}^{i}=\left( \mathbf{D}_{k}\delta _{j}\right) \rfloor d^{i},\quad
L_{bk}^{a}=\left( \mathbf{D}_{k}\partial _{b}\right) \rfloor \delta
^{a},~C_{jc}^{i}=\left( \mathbf{D}_{c}\delta _{j}\right) \rfloor d^{i},\quad
C_{bc}^{a}=\left( \mathbf{D}_{c}\partial _{b}\right) \rfloor \delta ^{a}.
\end{equation*}%
The components $\mathbf{\Gamma }_{\ \alpha \beta }^{\gamma }=\left(
L_{jk}^{i},L_{bk}^{a},C_{jc}^{i},C_{bc}^{a}\right) $ completely define a \
linear connection $\mathbf{D}$ in N--adapted form the global splitting of $%
\mathbf{V}^{n+m}$ into h- and v--subspaces. We can consider a corresponding
d--connection 1--form
\begin{equation*}
\mathbf{\Gamma }_{\ \alpha }^{\gamma }=\mathbf{\Gamma }_{\ \alpha \beta
}^{\gamma }\mathbf{e}^{\beta }
\end{equation*}%
and say that a d--connection $\mathbf{D}_{\alpha }$ is compatible with a
metric $\mathbf{g}$ if%
\begin{equation}
\mathbf{Dg}=0.  \label{mc}
\end{equation}

Using the covariant derivative $\mathbf{D},$ we can de fine the torsion
tensor%
\begin{equation}
\ \mathcal{T}^{\alpha }\doteqdot \mathbf{De}^{\alpha }=d\mathbf{e}^{\alpha }+%
\mathbf{\Gamma }_{\ \beta }^{\gamma }\wedge \mathbf{e}^{\beta }  \label{dt}
\end{equation}%
and the curvature tensor
\begin{equation}
\ \mathcal{R}_{\ \beta }^{\alpha }\doteqdot \mathbf{D\Gamma }_{\ \beta
}^{\alpha }=d\mathbf{\Gamma }_{\ \beta }^{\alpha }-\mathbf{\Gamma }_{\ \beta
}^{\gamma }\wedge \mathbf{\Gamma }_{\ \ \gamma }^{\alpha },  \label{dc}
\end{equation}%
where ''$\wedge "$ denotes the antisymmetric product of forms.

If a spacetime $\mathbf{V}^{n+m}$ is provided with both N--connection $%
\mathbf{N}$\ and d--metric $\mathbf{g}$ structures, there is a unique linear
symmetric and torsionless connection $\bigtriangledown ,$ called the
Levi--Civita connection.\ This connection is metric compatible, i. e. $%
\bigtriangledown _{\gamma }\mathbf{g}_{\alpha \beta }=0$ $\ $for $\mathbf{g}%
_{\alpha \beta }=\left( g_{ij},h_{ab}\right) ,$ see (\ref{block2}), with the
coefficients
\begin{equation*}
\ ^{\bigtriangledown }\mathbf{\Gamma }_{\alpha \beta \gamma }=\mathbf{g}%
\left( \mathbf{e}_{\alpha },\bigtriangledown _{\gamma }\mathbf{e}_{\beta
}\right) =\mathbf{g}_{\alpha \tau }\ ^{\bigtriangledown }\mathbf{\Gamma }_{\
\beta \gamma }^{\tau },\,
\end{equation*}%
computed as
\begin{equation}
\ \ ^{\bigtriangledown }\mathbf{\Gamma }_{\alpha \beta \gamma }=\frac{1}{2}%
\left[ \mathbf{e}_{\beta }\mathbf{g}_{\alpha \gamma }+\mathbf{e}_{\gamma }%
\mathbf{g}_{\beta \alpha }-\mathbf{e}_{\alpha }\mathbf{g}_{\gamma \beta }+%
\mathbf{g}_{\alpha \tau }\mathbf{w}_{\gamma \beta }^{\tau }+\mathbf{g}%
_{\beta \tau }\mathbf{w}_{\alpha \gamma }^{\tau }-\mathbf{g}_{\gamma \tau }%
\mathbf{w}_{\beta \alpha }^{\tau }\right]   \label{lcsym}
\end{equation}%
with respect to N--frames $\mathbf{e}_{\beta }$ (\ref{dder}) and N--coframes
$\mathbf{e}_{\ }^{\alpha }$ (\ref{ddif}), this formula is proved for any
nonholonomic frames, for instance, in Refs. \cite{stw}. We note that the
Levi--Civita connection is not adapted to the N--connection structure: its
h-- and v-- coefficients can not defined in a form preserved under
coordinate and frame transforms.

There is a type of d--connections which are similar to the Levi--Civita
connection and satisfy certain metricity conditions, such metrics being
adapted to the N--connection. One considers the so--called canonical
d--connection $\widehat{\mathbf{D}}$\ \ $\mathbf{=}\left( \widehat{D}^{[h]},%
\widehat{D}^{[v]}\right) ,$ equivalently $\widehat{\mathbf{\Gamma }}_{\
\alpha }^{\gamma }=\widehat{\mathbf{\Gamma }}_{\ \alpha \beta }^{\gamma }%
\mathbf{e}^{\beta },$\ which ''minimally'' extends the Levi--Civita
connection in order to be N--adapted, metric compatible and defined only by
the coefficients of metric (\ref{ansatz}) (equivalently by the block
reprezentation (\ref{block2})), see a general proof in \cite{v2}. It is
defined by the components $\widehat{\mathbf{\Gamma }}_{\ \alpha \beta
}^{\gamma }=(\widehat{L}_{jk}^{i},\widehat{L}_{bk}^{a},$ $\widehat{C}%
_{jc}^{i},\widehat{C}_{bc}^{a}),$ where%
\begin{eqnarray}
\widehat{L}_{jk}^{i} &=&\frac{1}{2}g^{ir}\left( \frac{\delta g_{jk}}{\delta
x^{k}}+\frac{\delta g_{kr}}{\delta x^{j}}-\frac{\delta g_{jk}}{\delta x^{r}}%
\right) ,  \label{candcon} \\
\widehat{L}_{bk}^{a} &=&\frac{\partial N_{k}^{a}}{\partial y^{b}}+\frac{1}{2}%
h^{ac}\left( \frac{\delta h_{bc}}{\delta x^{k}}-\frac{\partial N_{k}^{d}}{%
\partial y^{b}}h_{dc}-\frac{\partial N_{k}^{d}}{\partial y^{c}}h_{db}\right)
,  \notag \\
\widehat{C}_{jc}^{i} &=&\frac{1}{2}g^{ik}\frac{\partial g_{jk}}{\partial
y^{c}},  \notag \\
\widehat{C}_{bc}^{a} &=&\frac{1}{2}h^{ad}\left( \frac{\partial h_{bd}}{%
\partial y^{c}}+\frac{\partial h_{cd}}{\partial y^{b}}-\frac{\partial h_{bc}%
}{\partial y^{d}}\right) .  \notag
\end{eqnarray}%
This connection satisfies the torsionless conditions for the h--
and v--subspa\-ces, when, respectively,
$\widehat{T}_{jk}^{i}=\widehat{T}_{bc}^{a}=0.$ On such subspaces,
it has the usual properties of the Levi--Civita connection but, in
general, it contains certain additional coefficients induced by
the nonholonomic frame structure and N--connection coefficients
(see below \ the formulas (\ref{dtorsb}) but re--defined for $\widehat{L}%
_{jk}^{i},\widehat{C}_{.ja}^{i},$ and $\widehat{C}_{bc}^{a}).$

The coefficients of the Levi--Civita connection $\ ^{\bigtriangledown }%
\mathbf{\Gamma }_{\ \beta \gamma }^{\tau }$ and of the canonical
d--connection \ $\widehat{\mathbf{\Gamma }}_{\ \beta \gamma }^{\tau }$\ are
related by the formulas%
\begin{equation}
^{\bigtriangledown }\mathbf{\Gamma }_{\ \beta \gamma }^{\tau }=\left(
\widehat{L}_{jk}^{i},\widehat{L}_{bk}^{a}-\frac{\partial N_{k}^{a}}{\partial
y^{b}},\widehat{C}_{jc}^{i}+\frac{1}{2}g^{ik}\Omega _{jk}^{a}h_{ca},\widehat{%
C}_{bc}^{a}\right) ,  \label{lcsyma}
\end{equation}%
where $\Omega _{jk}^{a}$\ \ is the N--connection curvature\ (\ref{ncurv}).
The proof of this results consists from a straightforward computation of the
coefficients of $^{\bigtriangledown }\mathbf{\Gamma }_{\ \beta \gamma
}^{\tau }$ (\ref{lcsym}) with respect to the nonholonomic bases (\ref{ddif})
and (\ref{dder}) and a re--definition of coefficients following (\ref%
{candcon}).

The torsion $\mathbf{T}_{\ \beta \gamma }^{\alpha }=(T_{\ jk}^{i},T_{\
ja}^{i},T_{\ ij}^{a},T_{\ bi}^{a},T_{\ bc}^{a})$ of a d--connection $\mathbf{%
\Gamma }_{\alpha \beta }^{\gamma }$ (\ref{dcon1}) is defined by
corresponding d--torsions
\begin{eqnarray}
T_{\ jk}^{i} &=&L_{jk}^{i}-L_{kj}^{i},\ T_{ja}^{i}=C_{.ja}^{i},\ T_{\
ji}^{a}=\frac{\delta N_{i}^{a}}{\delta x^{j}}-\frac{\delta N_{j}^{a}}{\delta
x^{i}}=\Omega _{ji}^{a},  \notag \\
T_{\ bi}^{a} &=&=P_{\ bi}^{a}=\frac{\partial N_{i}^{a}}{\partial y^{b}}%
-L_{.bj}^{a},\ T_{\ bc}^{a}=S_{\ bc}^{a}=C_{bc}^{a}-C_{cb}^{a}\
\label{dtorsb}
\end{eqnarray}%
computed in explicit form by distinguishing the formulas (\ref{dt}) with
respect to the N--adapted vielbeins (\ref{dder}) and (\ref{ddif}). We note
that on (pseudo) Riemanian spacetimes the d--torsions can be induced by the
N--connection coefficients and reflect the anholonomic character of the
N--adapted vielbein structure. Such objects vanish when we transfer our
considerations with respect to holonomic bases for a trivial N--connection
and zero ''vertical'' dimension.

The curvature $\mathbf{R}_{\ \beta \gamma \tau }^{\alpha }=(R_{\
hjk}^{i},R_{\ bjk}^{a},P_{\ jka}^{i},P_{\ bka}^{c},S_{\ jbc}^{i},S_{\
bcd}^{a})$ of a d--con\-nec\-ti\-on $\mathbf{\Gamma }_{\alpha \beta
}^{\gamma }$ (\ref{dcon1}) is defined by corresponding d--curvatures
\begin{eqnarray}
R_{\ hjk}^{i} &=&\frac{\delta L_{.hj}^{i}}{\delta x^{k}}-\frac{\delta
L_{.hk}^{i}}{\delta x^{j}}%
+L_{.hj}^{m}L_{mk}^{i}-L_{.hk}^{m}L_{mj}^{i}-C_{.ha}^{i}\Omega _{.jk}^{a},
\label{dcurv} \\
R_{\ bjk}^{a} &=&\frac{\delta L_{.bj}^{a}}{\delta x^{k}}-\frac{\delta
L_{.bk}^{a}}{\delta x^{j}}%
+L_{.bj}^{c}L_{.ck}^{a}-L_{.bk}^{c}L_{.cj}^{a}-C_{.bc}^{a}\ \Omega
_{.jk}^{c},  \notag
\end{eqnarray}%
\begin{eqnarray*}
P_{\ jka}^{i} &=&\frac{\partial L_{.jk}^{i}}{\partial y^{k}}-\left( \frac{%
\partial C_{.ja}^{i}}{\partial x^{k}}%
+L_{.lk}^{i}C_{.ja}^{l}-L_{.jk}^{l}C_{.la}^{i}-L_{.ak}^{c}C_{.jc}^{i}\right)
+C_{.jb}^{i}P_{.ka}^{b}, \\
P_{\ bka}^{c} &=&\frac{\partial L_{.bk}^{c}}{\partial y^{a}}-\left( \frac{%
\partial C_{.ba}^{c}}{\partial x^{k}}+L_{.dk}^{c%
\,}C_{.ba}^{d}-L_{.bk}^{d}C_{.da}^{c}-L_{.ak}^{d}C_{.bd}^{c}\right)
+C_{.bd}^{c}P_{.ka}^{d},
\end{eqnarray*}%
\begin{eqnarray*}
S_{\ jbc}^{i} &=&\frac{\partial C_{.jb}^{i}}{\partial y^{c}}-\frac{\partial
C_{.jc}^{i}}{\partial y^{b}}+C_{.jb}^{h}C_{.hc}^{i}-C_{.jc}^{h}C_{hb}^{i}, \\
S_{\ bcd}^{a} &=&\frac{\partial C_{.bc}^{a}}{\partial y^{d}}-\frac{\partial
C_{.bd}^{a}}{\partial y^{c}}+C_{.bc}^{e}C_{.ed}^{a}-C_{.bd}^{e}C_{.ec}^{a}.
\end{eqnarray*}%
Such formulas follow from an explicit coefficient calculus of (\ref{dc})
with respect to the N--adapted vielbeins (\ref{dder}) and (\ref{ddif}). They
are equivalent to the formulas given in \cite{stw,exac} but rewritten in a
form adapted to vielbein transforms (\ref{vt1}) and (\ref{vt2}).

The Ricci tensor
\begin{equation*}
\mathbf{R}_{\alpha \beta }\doteqdot \mathcal{R}_{\ \alpha \beta \tau }^{\tau
}
\end{equation*}%
is characterized by four d--tensor components $\mathbf{R}_{\alpha \beta
}=(R_{ij},R_{ia},R_{ai},S_{ab}),$ where%
\begin{eqnarray}
R_{ij} &\doteqdot &R_{\ ijk}^{k},\quad R_{ia}\doteqdot -\ ^{2}P_{ia}=-P_{\
ika}^{k},  \label{dricci} \\
R_{ai} &\doteqdot &\ ^{1}P_{ai}=P_{\ aib}^{b},\quad S_{ab}\doteqdot S_{\
abc}^{c}.  \notag
\end{eqnarray}%
It should be emphasized that because, in general, $^{1}P_{ai}\neq ~^{2}P_{ia}
$ the Ricci d--tensors are non symmetric (this a nonholonmic frame effect).\
A such tensor became symmetric with respect to holonomic vielbeins and for
the Levi--Civita connection.

Contracting with the inverse to a d--metric of type (\ref{block2}) in $%
\mathbf{V}^{n+m},$ we can introduce the scalar curvature of a d--connection $%
\mathbf{D,}$
\begin{equation}
{\overleftarrow{\mathbf{R}}}\doteqdot \mathbf{g}^{\alpha \beta }\mathbf{R}%
_{\alpha \beta }\doteqdot R+S,  \label{dscal}
\end{equation}%
where $R\doteqdot g^{ij}R_{ij}$ and $S\doteqdot h^{ab}S_{ab}$ and compute
the Einstein tensor
\begin{equation}
\mathbf{G}_{\alpha \beta }\doteqdot \mathbf{R}_{\alpha \beta }-\frac{1}{2}%
\mathbf{g}_{\alpha \beta }{\overleftarrow{\mathbf{R}}.}  \label{deinst}
\end{equation}%
In the vacuum case, $\mathbf{G}_{\alpha \beta }=0,$ that the Ricci
d--tensors (\ref{dricci}) vanish.

\section{N--Anholonomic Frames and String Gravity}

The Einstein equations for the canonical d--connection $\widehat{\mathbf{%
\Gamma }}_{\ \alpha \beta }^{\gamma }$ (\ref{candcon}),
\begin{equation}
\widehat{\mathbf{R}}_{\alpha \beta }-\frac{1}{2}\mathbf{g}_{\alpha \beta }%
\overleftarrow{\mathbf{\hat{R}}}=\kappa \mathbf{\Upsilon }_{\alpha \beta },
\label{einsteq}
\end{equation}%
are defined for a general source of matter fields and possible string
corrections, $\mathbf{\Upsilon }_{\alpha \beta }.$ The model contains a
nontrivial torsion $\widehat{\mathbf{T}}_{\ \alpha \beta }^{\gamma }$ with
d--torsions computed by introducing the components of (\ref{candcon}) into
formulas (\ref{dtorsb}).

We can express the 1--form of the canonical d--connection $\widehat{\mathbf{%
\Gamma }}_{\ \alpha }^{\gamma }$ via the deformation of the Levi--Civita
connection $\ ^{\bigtriangledown }\mathbf{\Gamma }_{\ \ \alpha }^{\gamma },$

\begin{equation*}
\widehat{\mathbf{\Gamma }}_{\ \alpha }^{\gamma }=\ ^{\bigtriangledown }%
\mathbf{\Gamma }_{\ \ \alpha }^{\gamma }+\widehat{\mathbf{Z}}_{\ \alpha
}^{\gamma }
\end{equation*}%
where%
\begin{equation}
\widehat{\mathbf{Z}}_{\alpha \beta }=\mathbf{e}_{\beta }\rfloor \widehat{%
\mathbf{T}}_{\alpha }-\mathbf{e}_{\alpha }\rfloor \widehat{\mathbf{T}}%
_{\beta }+\frac{1}{2}\left( \mathbf{e}_{\alpha }\rfloor \mathbf{e}_{\beta
}\rfloor \widehat{\mathbf{T}}_{\gamma }\right) \mathbf{e}^{\gamma }.
\label{aux53}
\end{equation}
This mean that it is possible to split \ all geometric objects into
(pseudo)\ Riemannian and post--Riemannian pieces, for instance,
\begin{eqnarray}
\widehat{\mathcal{R}}_{\ \beta }^{\alpha } &=&\ ^{\ }\mathcal{R}_{\ \beta
}^{\alpha }+\ ^{\bigtriangledown }\mathcal{Z}_{\ \beta }^{\alpha },
\label{dist1} \\
\ ^{\bigtriangledown }\mathcal{Z}_{\ \beta }^{\alpha } &=&\bigtriangledown
\widehat{\mathbf{Z}}_{\ \ \beta }^{\alpha }+\widehat{\mathbf{Z}}_{\ \ \gamma
}^{\alpha }\wedge \widehat{\mathbf{Z}}_{\ \ \beta }^{\gamma }.  \notag
\end{eqnarray}%
We conclude that the Einstein equations (\ref{einsteq}) for the canonical
d--connecti\-on $\widehat{\mathbf{\Gamma }}_{\ \alpha }^{\gamma }$
constructed for a d--metric $\mathbf{g}_{\alpha \beta }=[g_{ij},h_{ab}]$ (%
\ref{block2}) and N--connection $N_{i}^{a}$ are equivalent to the
gravitational field equations for the Einstein--Cartan theory with torsion $%
\widehat{\mathbf{T}}_{\ \alpha }^{\gamma }$ defined by the N--connection.

The Einstein gravity theory is defined by the condition that the deformation
tensor $\ ^{\bigtriangledown }\mathcal{Z}_{\ \beta }^{\alpha }$ (\ref{dist1}%
) is algebraically constrained that its coefficients satisfy the equations%
\begin{equation}
\ ^{\bigtriangledown }\mathcal{Z}_{\ \alpha \beta \tau }^{\tau }=0
\label{cdist}
\end{equation}%
which, as a matter of principle, for vacuum configurations, can be solved
for nonzero values of $\widehat{\mathbf{Z}}_{\ \alpha }^{\gamma }.$  In the
presence of general matter sources, we may have to impose the condition $%
\widehat{\mathbf{Z}}_{\ \alpha }^{\gamma }=0$ because distorsions of the
covariant derivatives may be contained in the field equations and
energy--momentum of matter.

The Einstein equations (\ref{einsteq}) can be decomposed into h-- and
v--compo\-nents following from (\ref{dricci}) and (\ref{dscal}),%
\begin{eqnarray}
\widehat{R}_{ij}-\frac{1}{2}g_{ij}\left( \widehat{R}+\widehat{S}\right)  &=&%
\mathbf{\Upsilon }_{ij},  \label{ep1} \\
\widehat{S}_{ab}-\frac{1}{2}h_{ab}\left( \widehat{R}+\widehat{S}\right)  &=&%
\mathbf{\Upsilon }_{ab},  \label{ep2} \\
\ ^{1}P_{ai} &=&\mathbf{\Upsilon }_{ai},  \label{ep3} \\
\ -\ ^{2}P_{ia} &=&\mathbf{\Upsilon }_{ia}.  \label{ep4}
\end{eqnarray}%
The sources of such equations have to be defined as certain matter field
contributions or corrections, for instance, from string gravity. For
instance, in the sigma model for bosonic string (see, \cite{sgr}), the
background connection is taken not the Levi--Civita one but a certain
deformation by the strength (torsion) tensor
\begin{equation*}
H_{\mu \nu \rho }\doteqdot e_{\mu }B_{\nu \rho }+e_{\rho }B_{\mu \nu
}+e_{\nu }B_{\rho \mu }
\end{equation*}%
of an antisymmetric field $B_{\nu \rho }.$ The connection is of type
\begin{equation*}
\mathcal{D}_{\mu }=\bigtriangledown _{\mu }+\frac{1}{2}H_{\mu \nu }^{\quad
\rho }.
\end{equation*}%
For trivial dilaton configurations, we may write
\begin{eqnarray*}
R_{\mu \nu } &=&-\frac{1}{4}H_{\mu }^{\ \lambda \rho }H_{\nu \lambda \rho },
\\
\bigtriangledown _{\lambda }H_{\ \mu \nu }^{\lambda } &=&0.
\end{eqnarray*}%
The equations may be re--defined with respect to N--adapted frames (\ref%
{ddif}) and (\ref{dder}) for ''boldfaced'' values.

Here we consider, for simplicity, a model with zero dilaton field but with
nontrivial $H$--field related to the d--torsions induced by the
N--connection and canonical d--connection when a class of generic
off--diagonal metrics can be derived from the bosonic string theory if $%
\mathbf{H}_{\nu \lambda \rho }$ and $\mathbf{B}_{\nu \rho }$ are related to
the d--torsions components $\widehat{\mathbf{T}}_{\ \alpha \beta }^{\gamma }.
$ We can take a special ansatz for $B$--field,
\begin{equation*}
\mathbf{B}_{\nu \rho }=\left[ B_{ij},B_{ia},B_{ab}\right] ,
\end{equation*}%
and consider that
\begin{equation}
\mathbf{H}_{\nu \lambda \rho }=\widehat{\mathbf{Z}}_{\ \nu \lambda \rho }+%
\widehat{\mathbf{H}}_{\nu \lambda \rho }  \label{aux51a}
\end{equation}%
where $\widehat{\mathbf{Z}}_{\ \nu \lambda \rho }$ is the distorsion of the
Levi--Civita connection induced by $\widehat{\mathbf{T}}_{\ \alpha \beta
}^{\gamma },$ see (\ref{aux53}), and $\widehat{\mathbf{H}}_{\nu \lambda \rho
}$ is generated by nonholonomic deformations of $\mathbf{H}_{\nu \lambda
\rho }.$ In this case, the induced by N--connection torsion structure is
related to the antisymmetric $H$--field and correspondingly to the $B$%
--field from string theory. The equations
\begin{equation}
\bigtriangledown ^{\nu }\mathbf{H}_{\nu \lambda \rho }=\bigtriangledown
^{\nu }(\widehat{\mathbf{Z}}_{\ \nu \lambda \rho }+\widehat{\mathbf{H}}_{\nu
\lambda \rho })=0  \label{aux51}
\end{equation}%
impose certain dynamical restrictions to the N--connection coefficients $%
N_{i}^{a}$ and d--metric $\mathbf{g}_{\alpha \beta }=[g_{ij},h_{ab}]$ \
contained in $\widehat{\mathbf{T}}_{\ \alpha \beta }^{\gamma }.$ If it is
prescribed the canonical d--connection $\widehat{\mathbf{D}}$ on the
background space, we can state a model with the equations (\ref{aux51})
written in the form
\begin{equation}
\widehat{\mathbf{D}}^{\nu }\mathbf{H}_{\nu \lambda \rho }=\widehat{\mathbf{D}%
}^{\nu }(\widehat{\mathbf{Z}}_{\ \nu \lambda \rho }+\widehat{\mathbf{H}}%
_{\nu \lambda \rho })=0,  \label{aux51b}
\end{equation}%
where $\widehat{\mathbf{H}}_{\nu \lambda \rho }$ are computed for stated
values of $\widehat{\mathbf{T}}_{\ \alpha \beta }^{\gamma }.$ For trivial
N--connections, when $\widehat{\mathbf{Z}}_{\ \nu \lambda \rho }\rightarrow 0
$ and $\widehat{\mathbf{D}}^{\nu }\rightarrow \bigtriangledown ^{\nu },$ the
$\widehat{\mathbf{H}}_{\nu \lambda \rho }$ transform into usual $H$--fields.

The dynamics of gravity of nonholonomic string corrections by the field $%
\mathbf{H}_{\nu \lambda \rho }$ is defined by the system of field equations
\begin{equation}
\widehat{\mathbf{R}}_{\alpha \beta }-\frac{1}{2}\mathbf{g}_{\alpha \beta }%
\overleftarrow{\mathbf{\hat{R}}}=-\frac{1}{4}\mathbf{H}_{\alpha }^{\ \nu
\rho }\mathbf{H}_{\nu \beta \rho }  \label{aux51c}
\end{equation}%
and the $H$--field equations (\ref{aux51b}) In this case, the sources for
the equations (\ref{ep1})--(\ref{ep4}) are defined by the corresponding h--
and v--projections of
\begin{equation*}
\mathbf{\Upsilon }_{\alpha \beta }^{H}=-\frac{1}{4}\mathbf{H}_{\alpha }^{\
\nu \rho }\mathbf{H}_{\nu \beta \rho }.
\end{equation*}%
It was possible to generate exact solutions for (\ref{aux51c}) for such
ansats for $\mathbf{H}_{\nu \lambda \rho }$ when $\mathbf{\Upsilon }_{\alpha
\beta }^{H}$ is diagonal with respect to N--adapted frames \cite{vnces,v2}.

\section{Off--Diagonal Ansatz}

\label{appa}

In a series of papers \cite{vjhep2,vsbd,vs,vp,vt,vnces,valg,v2} we
elaborated and developed the anholonomic frame method of constructing exact
solutions with generic off--diagonal metrics (depending on 2-4 variables) in
general relativity, gauge gravity and various generalizations to string,
brane and generalized Finsler--affine gravity models. In this section, we
outline the necessary results and present some details on proofs of such
results.

\subsection{A Theorem for the 5D Ricci d--tensors}

Let us consider a 5D manifold of necessary smooth class provided with
N--connection structure $\mathbf{N}=[N_{i}^{4}=w_{i}(u^{\alpha }),$ $%
N_{i}^{5}=n_{i}(u^{\alpha })]$ and a d--metric of type (\ref{block2}). We
compute the components of the Ricci and Einstein tensors for a particular
ansatz for d--metric
\begin{eqnarray}
\delta s^{2} &=&g_{1}{(dx^{1})}^{2}+g_{2}(x^{2},x^{3}){(dx^{2})}%
^{2}+g_{3}\left( x^{k}\right) {(dx^{3})}^{2}  \notag \\
&&+h_{4}\left( x^{k},v\right) {(\delta v)}^{2}+h_{5}\left( x^{k},v\right) {%
(\delta y)}^{2},  \notag \\
\delta v &=&dv+w_{i}\left( x^{k},v\right) dx^{i},\ \delta y=dy+n_{i}\left(
x^{k},v\right) dx^{i}  \label{ans4d}
\end{eqnarray}%
with $g_{1}=const,$ $N_{i}^{4}=w_{i}\left( x^{k},v\right) $ and $%
N_{i}^{5}=n_{i}\left( x^{k},v\right) .$ The local coordinates are labelled
in the form $u^{\alpha }=(x^{i},y^{4}=v,y^{5}),$ for $i=1,2,3.$ Every
coordinate from a set $u^{\alpha }$ can may be time like, 3D space like, or
extra dimensional. We note that the metric (\ref{ans4d}) does not depend on
variable $y^{5}=y,$ but emphasize the dependence on the so--caled
''anisotropic'' variable $y^{4}=v.$ For simplicity, the partial derivatives
will be written in the form $a^{\times }=\partial a/\partial
x^{1},a^{\bullet }=\partial a/\partial x^{2},a^{\prime }=\partial a/\partial
x^{3},a^{\ast }=\partial a/\partial v.$

In equivalent form, we can re--write (\ref{ans4d}) \ in off--diagonal form (%
\ref{ansatz}) when
\begin{equation}
\delta s^{2}=\underline{g}_{\alpha \beta }\left( x^{i},v\right) du^{\alpha
}du^{\beta }  \label{metric5}
\end{equation}%
has the metric coefficients $\underline{g}_{\alpha \beta }$ are
parametrized by the matrix {\small
\begin{equation*}
\left[
\begin{array}{ccccc}
g_{1}+w_{11}h_{4}+n_{11}h_{5} & w_{12}h_{4}+n_{12}h_{5} &
w_{13}h_{4}+n_{13}h_{5} & w_{1}h_{4} & n_{1}h_{5} \\
w_{21}h_{4}+n_{21}h_{5} & g_{2}+w_{22}h_{4}+n_{22}h_{5} &
w_{23}h_{4}+n_{23}h_{5} & w_{2}h_{4} & n_{2}h_{5} \\
w_{31}h_{4}+n_{31}h_{5} & w_{32}h_{4}+n_{32}h_{5} &
g_{3}+w_{33}h_{4}+n_{33}h_{5} & w_{3}h_{4} & n_{3}h_{5} \\
w_{1}h_{4} & w_{2}h_{4} & w_{3}h_{4} & h_{4} & 0 \\
n_{1}h_{5} & n_{2}h_{5} & n_{3}h_{5} & 0 & h_{5}%
\end{array}%
\right]
\end{equation*}%
} where $w_{ij}=w_{i}w_{j}$ $\ $and $n_{ij}=n_{i}n_{j},$ with the
coefficients defined by some necessary smoothly class functions of type
\begin{eqnarray}
g_{1} &=&\pm 1,g_{2,3}=g_{2,3}(x^{2},x^{3}),h_{4,5}=h_{4,5}(x^{i},v),  \notag
\\
w_{i} &=&w_{i}(x^{i},v),n_{i}=n_{i}(x^{i},v).  \notag
\end{eqnarray}

\begin{theorem}
\label{t5dr}The nontrivial components of the 5D Ricci d--tensors (\ref%
{dricci}), $\widehat{\mathbf{R}}_{\alpha \beta }=(\widehat{R}_{ij},\widehat{R%
}_{ia},$ $\widehat{R}_{ai},\widehat{S}_{ab}),$ for the d--metric (\ref{ans4d}%
) and the canonical d--connection $\widehat{\mathbf{\Gamma }}_{\ \alpha
\beta }^{\gamma }$(\ref{candcon}), all components being  computed with
respect to the corresponding N--anholonomic frames (\ref{dder}) and (\ref%
{ddif}) are stated by formulas
\begin{eqnarray}
R_{2}^{2}=R_{3}^{3}=-\frac{1}{2g_{2}g_{3}}[g_{3}^{\bullet \bullet }-\frac{%
g_{2}^{\bullet }g_{3}^{\bullet }}{2g_{2}}-\frac{(g_{3}^{\bullet })^{2}}{%
2g_{3}}+g_{2}^{^{\prime \prime }}-\frac{g_{2}^{^{\prime }}g_{3}^{^{\prime }}%
}{2g_{3}}-\frac{(g_{2}^{^{\prime }})^{2}}{2g_{2}}], &&  \label{ricci1a} \\
S_{4}^{4}=S_{5}^{5}=-\frac{1}{2h_{4}h_{5}}\left[ h_{5}^{\ast \ast
}-h_{5}^{\ast }\left( \ln \sqrt{|h_{4}h_{5}|}\right) ^{\ast }\right] , &&
\label{ricci2a} \\
R_{4i}=-w_{i}\frac{\beta }{2h_{5}}-\frac{\alpha _{i}}{2h_{5}}, &&
\label{ricci3a} \\
R_{5i}=-\frac{h_{5}}{2h_{4}}\left[ n_{i}^{\ast \ast }+\gamma n_{i}^{\ast }%
\right] , &&  \label{ricci4a}
\end{eqnarray}%
where
\begin{eqnarray}
\alpha _{i} &=&\partial _{i}{h_{5}^{\ast }}-h_{5}^{\ast }\partial _{i}\ln
\sqrt{|h_{4}h_{5}|},\beta =h_{5}^{\ast \ast }-h_{5}^{\ast }[\ln \sqrt{%
|h_{4}h_{5}|}]^{\ast },  \label{abc} \\
\gamma  &=&3h_{5}^{\ast }/2h_{5}-h_{4}^{\ast }/h_{4},h_{4}^{\ast }\neq 0,%
\text{ }h_{5}^{\ast }\neq 0.  \notag
\end{eqnarray}
\end{theorem}

We note that d--metrics with vanishing $h_{4}^{\ast }$ or $h_{5}^{\ast }$
should be analyzed as special cases. In general form, a such theorem is
proved for Finsler--affine speces in Ref. \cite{v2}, see also the Appendix
to \cite{vnces} where the results are summarized for Riemann--Cartan and
Einstein spaces, in general, with noncommutative symmetries.

\subsection{Proof of Theorem \ref{t5dr}}

In this appendix, we give the details of the proof for the
(pseudo) Riemann--Cartan nonholonomic manifolds. It contains a
number of examples of N--adapted differential calculus not
presented in physical literature on gravity and strings. We note
that such results can be not obtained with standard analytic
programs (Mathematica 5 and/or Maple) because in our case we work
with the canonical d--connection and N--adapted frames.

It is a cumbersome task to perform tensor calculations (for instance, of the
curvature and Ricci tensors) for the off--diagonal ansatz (\ref{metric5})
but the formulas simplify substantially for the effectively diagonalized
metric (\ref{ans4d}). In this case, the N--adapted frames of type (\ref{ddif}%
) and (\ref{dder}) and are defined respectively by the pentads (frames,
funfbeins)
\begin{equation}
e^{i}=dx^{i},e^{4}=\delta v=dv+w_{i}\left( x^{k},v\right)
dx^{i},e^{5}=\delta y=dy+n_{i}\left( x^{k},v\right) dx^{i}  \label{ddif4a}
\end{equation}%
and the N--elongated partial derivative operators,
\begin{eqnarray}
e_{i} &=&\delta _{i}=\frac{\partial }{\partial x^{i}}-N_{i}^{a}\frac{%
\partial }{\partial y^{a}}=\frac{\partial }{\partial x^{i}}-w_{i}\frac{%
\partial }{\partial v}-n_{i}\frac{\partial }{\partial y},  \label{dder4a} \\
e_{4} &=&\frac{\partial }{\partial y^{4}}=\frac{\partial }{\partial v},\
e_{5}=\frac{\partial }{\partial y^{5}}=\frac{\partial }{\partial y}.  \notag
\end{eqnarray}

The N--elongated partial derivatives of a function $f\left( u^{\alpha
}\right) =f\left( x^{i},y^{a}\right) $ are computed in the form
\begin{eqnarray*}
\delta _{2}f &=&\frac{\delta f}{\partial u^{2}}=\frac{\delta f}{\partial
x^{2}}=\frac{\delta f}{\partial x}=\frac{\partial f}{\partial x}-N_{2}^{a}%
\frac{\partial f}{\partial y^{a}} \\
&=&\frac{\partial f}{\partial x^{2}}-w_{2}\frac{\partial f}{\partial v}-n_{2}%
\frac{\partial f}{\partial y}=f^{\bullet }-w_{2}\ f^{\prime }-n_{2}\ f^{\ast
}
\end{eqnarray*}%
where
\begin{equation*}
f^{\bullet }=\frac{\partial f}{\partial x^{2}}=\frac{\partial f}{\partial x}%
,\ f^{\prime }=\frac{\partial f}{\partial x^{3}}=\frac{\partial f}{\partial
x^{3}},\ f^{\ast }=\frac{\partial f}{\partial y^{4}}=\frac{\partial f}{%
\partial v}.
\end{equation*}%
The N--elongated differential is expressed
\begin{equation*}
\delta f=\frac{\delta f}{\partial u^{\alpha }}\mathbf{e}^{\alpha }.
\end{equation*}%
One perform a N--adapted differential calculus if we work with respect to
N--elongated frames and partial derivatives and differentials.

\subsubsection{Calculation of the N--connection curvature}

We compute the coefficients (\ref{ncurv}), for the d--metric (\ref{ans4d})
(equivalently, the ansatz (\ref{metric5})), defining the curvature of
N--connection $N_{i}^{a},$ by substituting $N_{i}^{4}=w_{i}\left(
x^{k},v\right) $ and $N_{i}^{5}=n_{i}\left( x^{k},v\right) ,$ where $i=2,3$
and $a=4,5$ (for our ansatz, we do not have dependence on $x^{1}$ and $%
y^{5}).$ The results for nontrivial values are
\begin{eqnarray}
\Omega _{23}^{4} &=&-\Omega _{23}^{4}=w_{2}^{\prime }-w_{3}^{\bullet
}-w_{3}w_{2}^{\ast }+w_{2}w_{3}^{\ast },  \label{omega} \\
\Omega _{23}^{5} &=&-\Omega _{23}^{5}=n_{2}^{\prime }-n_{3}^{\bullet
}-w_{3}n_{2}^{\ast }+w_{2}n_{3}^{\ast }.  \notag
\end{eqnarray}%
Such values must be zero if we wont to generate a solution for the
Levi--Civita connection. This is a necessary but not enough condition, see (%
\ref{lcsyma}). Even in this case, the nonholonomy coefficients (\ref{anhc})
can be nonzero if there are certain values $\partial _{a}N_{i}^{b}\neq 0.$

\subsubsection{Calculation of the canonical d--connection}

We compute the coefficients $\widehat{\mathbf{\Gamma }}_{\ \alpha \beta
}^{\gamma }=\left( \widehat{L}_{jk}^{i},\widehat{L}_{bk}^{a},\widehat{C}%
_{jc}^{i},\widehat{C}_{bc}^{a}\right) $ (\ref{candcon}) for the d--metric (%
\ref{ans4d}) (equivalently, the ansatz (\ref{metric5})) when $%
g_{jk}=\{g_{j}\}$ and $h_{bc}=\{h_{b}\}$ are diagonal and $g_{ik}$ depend
only on $x^{2}$ and $x^{3}$ but not on $y^{a}.$

We have
\begin{eqnarray}
\delta _{k}g_{ij} &=&\partial _{k}g_{ij}-w_{k}g_{ij}^{\ast }=\partial
_{k}g_{ij},\ \delta _{k}h_{b}=\partial _{k}h_{b}-w_{k}h_{b}^{\ast }
\label{aux01} \\
\delta _{k}w_{i} &=&\partial _{k}w_{i}-w_{k}w_{i}^{\ast },\ \delta
_{k}n_{i}=\partial _{k}n_{i}-w_{k}n_{i}^{\ast }  \notag
\end{eqnarray}%
resulting in formulas
\begin{equation*}
\widehat{L}_{jk}^{i}=\frac{1}{2}g^{ir}\left( \frac{\delta g_{jk}}{\delta
x^{k}}+\frac{\delta g_{kr}}{\delta x^{j}}-\frac{\delta g_{jk}}{\delta x^{r}}%
\right) =\frac{1}{2}g^{ir}\left( \frac{\partial g_{jk}}{\delta x^{k}}+\frac{%
\partial g_{kr}}{\delta x^{j}}-\frac{\partial g_{jk}}{\delta x^{r}}\right)
\end{equation*}%
The nontrivial values of $\widehat{L}_{jk}^{i}$ are%
\begin{eqnarray}
\widehat{L}_{22}^{2} &=&\frac{g_{2}^{\bullet }}{2g_{2}}=\alpha _{2}^{\bullet
},\ \widehat{L}_{23}^{2}=\frac{g_{2}^{\prime }}{2g_{2}}=\alpha
_{2}^{^{\prime }},\ \widehat{L}_{33}^{2}=-\frac{g_{3}^{\bullet }}{2g_{2}}
\label{aux02} \\
\widehat{L}_{22}^{3} &=&-\frac{g_{2}^{\prime }}{2g_{3}},\ \widehat{L}%
_{23}^{3}=\frac{g_{3}^{\bullet }}{2g_{3}}=\alpha _{3}^{\bullet },\ \
\widehat{L}_{33}^{3}=\frac{g_{3}^{\prime }}{2g_{3}}=\alpha _{3}^{^{\prime }}.
\notag
\end{eqnarray}%
In a similar form we compute the components%
\begin{equation*}
\widehat{L}_{bk}^{a}=\partial _{b}N_{k}^{a}+\frac{1}{2}h^{ac}\left( \partial
_{k}h_{bc}-N_{k}^{d}\frac{\partial h_{bc}}{\partial y^{d}}-h_{dc}\partial
_{b}N_{k}^{d}-h_{db}\partial _{c}N_{k}^{d}\right)
\end{equation*}%
having nontrivial values
\begin{eqnarray}
\widehat{L}_{42}^{4} &=&\frac{1}{2h_{4}}\left( h_{4}^{\bullet
}-w_{2}h_{4}^{\ast }\right) =\delta _{2}\ln \sqrt{|h_{4}|}\doteqdot \delta
_{2}\beta _{4},  \label{aux02a} \\
\widehat{L}_{43}^{4} &=&\frac{1}{2h_{4}}\left( h_{4}^{^{\prime
}}-w_{3}h_{4}^{\ast }\right) =\delta _{3}\ln \sqrt{|h_{4}|}\doteqdot \delta
_{3}\beta _{4}  \notag
\end{eqnarray}%
\begin{equation}
\widehat{L}_{5k}^{4}=-\frac{h_{5}}{2h_{4}}n_{k}^{\ast },\ \widehat{L}%
_{bk}^{5}=\partial _{b}n_{k}+\frac{1}{2h_{5}}\left( \partial
_{k}h_{b5}-w_{k}h_{b5}^{\ast }-h_{5}\partial _{b}n_{k}\right) ,
\label{aux02b}
\end{equation}

\begin{eqnarray}
\widehat{L}_{4k}^{5} &=&n_{k}^{\ast }+\frac{1}{2h_{5}}\left(
-h_{5}n_{k}^{\ast }\right) =\frac{1}{2}n_{k}^{\ast },  \label{aux02c} \\
\widehat{L}_{5k}^{5} &=&\frac{1}{2h_{5}}\left( \partial
_{k}h_{5}-w_{k}h_{5}^{\ast }\right) =\delta _{k}\ln \sqrt{|h_{4}|}=\delta
_{k}\beta _{4}.  \notag
\end{eqnarray}

We note that
\begin{equation}
\widehat{C}_{jc}^{i}=\frac{1}{2}g^{ik}\frac{\partial g_{jk}}{\partial y^{c}}%
\doteqdot 0  \label{aux02cc}
\end{equation}%
because $g_{jk}=g_{jk}\left( x^{i}\right) $ for the considered ansatz.

The values
\begin{equation*}
\widehat{C}_{bc}^{a}=\frac{1}{2}h^{ad}\left( \frac{\partial h_{bd}}{\partial
y^{c}}+\frac{\partial h_{cd}}{\partial y^{b}}-\frac{\partial h_{bc}}{%
\partial y^{d}}\right)
\end{equation*}%
for $h_{bd}=diag[h_{4},h_{5}]$  have nontrivial components
\begin{equation}
\widehat{C}_{44}^{4}=\frac{h_{4}^{\ast }}{2h_{4}}\doteqdot \beta _{4}^{\ast
},\widehat{C}_{55}^{4}=-\frac{h_{5}^{\ast }}{2h_{4}},\widehat{C}_{45}^{5}=%
\frac{h_{5}^{\ast }}{2h_{5}}\doteqdot \beta _{5}^{\ast }.  \label{aux02d}
\end{equation}

The set of formulas (\ref{aux02})--(\ref{aux02d}) define the nontrivial
coefficients of the canonical d--connection $\widehat{\mathbf{\Gamma }}_{\
\alpha \beta }^{\gamma }=\left( \widehat{L}_{jk}^{i},\widehat{L}_{bk}^{a},%
\widehat{C}_{jc}^{i},\widehat{C}_{bc}^{a}\right) $ (\ref{candcon}) for the
5D ansatz (\ref{ans4d}).

\subsubsection{Calculation of d--torsions}

We should put the nontrivial values (\ref{aux02})-- (\ref{aux02d}) into the
formulas for d--torsion \ (\ref{dtorsb}). One holds $\widehat{T}_{\ jk}^{i}=0
$ and $\widehat{T}_{\ bc}^{a}=0,$ because the coefficients $\widehat{L}%
_{jk}^{i}$ and $\widehat{C}_{bc}^{a}$ are symmetric for the chosen ansatz.

We have computed the nontrivial values of $\Omega _{.ji}^{a},$ see \ (\ref%
{omega}) resulting in%
\begin{eqnarray}
\widehat{T}_{23}^{4} &=&\Omega _{23}^{4}=-\Omega _{23}^{4}=w_{2}^{\prime
}-w_{3}^{\bullet }-w_{3}w_{2}^{\ast }+w_{2}w_{3}^{\ast },  \label{aux11} \\
\widehat{T}_{23}^{5} &=&\Omega _{23}^{5}=-\Omega _{23}^{5}=n_{2}^{\prime
}-n_{3}^{\bullet }-w_{3}n_{2}^{\ast }+w_{2}n_{3}^{\ast }.  \notag
\end{eqnarray}%
One follows
\begin{equation*}
\widehat{T}_{jc}^{i}=-\widehat{T}_{cj}^{i}=\widehat{C}_{jc}^{i}=\frac{1}{2}%
g^{ik}\frac{\partial g_{jk}}{\partial y^{c}}\doteqdot 0,
\end{equation*}%
see (\ref{aux02cc}). For the components
\begin{equation*}
\widehat{T}_{\ bi}^{a}=-\widehat{T}_{\ ib}^{a}=\widehat{P}_{.bi}^{a}=\frac{%
\partial N_{i}^{a}}{\partial y^{b}}-\widehat{L}_{.bj}^{a},
\end{equation*}%
i. e. for
\begin{equation*}
\widehat{P}_{.bi}^{4}=\frac{\partial N_{i}^{4}}{\partial y^{b}}-\widehat{L}%
_{.bj}^{4}=\partial _{b}w_{i}-\widehat{L}_{.bj}^{4}\mbox{ and }\widehat{P}%
_{.bi}^{5}=\frac{\partial N_{i}^{5}}{\partial y^{b}}-\widehat{L}%
_{.bj}^{5}=\partial _{b}n_{i}-\widehat{L}_{.bj}^{5},
\end{equation*}%
we have the nontrivial values%
\begin{eqnarray}
\widehat{P}_{.4i}^{4} &=&w_{i}^{\ast }-\frac{1}{2h_{4}}\left( \partial
_{i}h_{4}-w_{i}h_{4}^{\ast }\right) =w_{i}^{\ast }-\delta _{i}\beta _{4},\
\widehat{P}_{.5i}^{4}=\frac{h_{5}}{2h_{4}}n_{i}^{\ast },  \notag \\
\ \widehat{P}_{.4i}^{5} &=&\frac{1}{2}n_{i}^{\ast },\ \widehat{P}_{.5i}^{5}=-%
\frac{1}{2h_{5}}\left( \partial _{i}h_{5}-w_{i}h_{5}^{\ast }\right) =-\delta
_{i}\beta _{5}.  \label{aux12}
\end{eqnarray}

The formulas\ (\ref{aux11}) and (\ref{aux12}) state the \ nontrivial
coefficients of the canonical d--connection for the chosen ansatz (\ref%
{ans4d}).

\subsubsection{Calculation of the Ricci d-tensors}

Let us compute the value $\widehat{R}_{ij}=\widehat{R}_{\ ijk}^{k}$  (\ref%
{dricci}) for
\begin{equation*}
\widehat{R}_{\ hjk}^{i}=\frac{\delta \widehat{L}_{\ hj}^{i}}{\delta x^{k}}-%
\frac{\delta \widehat{L}_{\ hk}^{i}}{\delta x^{j}}+\widehat{L}_{\ hj}^{m}%
\widehat{L}_{\ mk}^{i}-\widehat{L}_{\ hk}^{m}\widehat{L}_{\ mj}^{i}-\widehat{%
C}_{\ ha}^{i}\Omega _{jk}^{a},
\end{equation*}%
from (\ref{dcurv}). It should be noted that $\widehat{C}_{ha}^{i}=0$ for the
ansatz under consideration, see (\ref{aux02cc}). We compute

\begin{equation*}
\frac{\delta \widehat{L}_{\ hj}^{i}}{\delta x^{k}}=\partial _{k}\widehat{L}%
_{\ hj}^{i}+N_{k}^{a}\partial _{a}\widehat{L}_{\ hj}^{i}=\partial _{k}%
\widehat{L}_{\ hj}^{i}+w_{k}\left( \widehat{L}_{\ hj}^{i}\right) ^{\ast
}=\partial _{k}\widehat{L}_{\ hj}^{i}
\end{equation*}%
because $\widehat{L}_{\ hj}^{i}$ do not depend on variable $y^{4}=v.$

Taking the derivative of (\ref{aux02}), we obtain
\begin{eqnarray*}
\partial _{2}\widehat{L}_{\ 22}^{2} &=&\frac{g_{2}^{\bullet \bullet }}{2g_{2}%
}-\frac{\left( g_{2}^{\bullet }\right) ^{2}}{2\left( g_{2}\right) ^{2}}%
,\partial _{2}\widehat{L}_{\ 23}^{2}=\frac{g_{2}^{\bullet ^{\prime }}}{2g_{2}%
}-\frac{g_{2}^{\bullet }g_{2}^{^{\prime }}}{2\left( g_{2}\right) ^{2}}%
,\partial _{2}\widehat{L}_{\ 33}^{2}=-\frac{g_{3}^{\bullet \bullet }}{2g_{2}}%
+\frac{g_{2}^{\bullet }g_{3}^{\bullet }}{2\left( g_{2}\right) ^{2}}, \\
\ \partial _{2}\widehat{L}_{\ 22}^{3} &=&-\frac{g_{2}^{\bullet ^{\prime }}}{%
2g_{3}}+\frac{g_{2}^{\bullet }g_{3}^{^{\prime }}}{2\left( g_{3}\right) ^{2}}%
,\partial _{2}\widehat{L}_{\ 23}^{3}=\frac{g_{3}^{\bullet \bullet }}{2g_{3}}-%
\frac{\left( g_{3}^{\bullet }\right) ^{2}}{2\left( g_{3}\right) ^{2}}%
,\partial _{2}\widehat{L}_{\ 33}^{3}=\frac{g_{3}^{\bullet ^{\prime }}}{2g_{3}%
}-\frac{g_{3}^{\bullet }g_{3}^{^{\prime }}}{2\left( g_{3}\right) ^{2}}, \\
\partial _{3}\widehat{L}_{\ 22}^{2} &=&\frac{g_{2}^{\bullet ^{\prime }}}{%
2g_{2}}-\frac{g_{2}^{\bullet }g_{2}^{^{\prime }}}{2\left( g_{2}\right) ^{2}}%
,\partial _{3}\widehat{L}_{\ 23}^{2}=\frac{g_{2}^{\prime \prime }}{2g_{2}}-%
\frac{\left( g_{2}^{\prime }\right) ^{2}}{2\left( g_{2}\right) ^{2}}%
,\partial _{3}\widehat{L}_{\ 33}^{2}=-\frac{g_{3}^{\bullet ^{\prime }}}{%
2g_{2}}+\frac{g_{3}^{\bullet }g_{2}^{^{\prime }}}{2\left( g_{2}\right) ^{2}},
\\
\ \partial _{3}\widehat{L}_{\ 22}^{3} &=&-\frac{g_{2}^{^{\prime \prime }}}{%
2g_{3}}+\frac{g_{2}^{\bullet }g_{2}^{^{\prime }}}{2\left( g_{3}\right) ^{2}}%
,\partial _{3}\widehat{L}_{\ 23}^{3}=\frac{g_{3}^{\bullet ^{\prime }}}{2g_{3}%
}-\frac{g_{3}^{\bullet }g_{3}^{^{\prime }}}{2\left( g_{3}\right) ^{2}}%
,\partial _{3}\widehat{L}_{\ 33}^{3}=\frac{g_{3}^{^{\prime \prime }}}{2g_{3}}%
-\frac{\left( g_{3}^{^{\prime }}\right) ^{2}}{2\left( g_{3}\right) ^{2}}.
\end{eqnarray*}

For these values and (\ref{aux02}), there are only 2 nontrivial components,

\begin{eqnarray*}
\widehat{R}_{\ 323}^{2} &=&\frac{g_{3}^{\bullet \bullet }}{2g_{2}}-\frac{%
g_{2}^{\bullet }g_{3}^{\bullet }}{4\left( g_{2}\right) ^{2}}-\frac{\left(
g_{3}^{\bullet }\right) ^{2}}{4g_{2}g_{3}}+\frac{g_{2}^{^{\prime \prime }}}{%
2g_{2}}-\frac{g_{2}^{^{\prime }}g_{3}^{^{\prime }}}{4g_{2}g_{3}}-\frac{%
\left( g_{2}^{^{\prime }}\right) ^{2}}{4\left( g_{2}\right) ^{2}}, \\
\widehat{R}_{\ 223}^{3} &=&-\frac{g_{3}^{\bullet \bullet }}{2g_{3}}+\frac{%
g_{2}^{\bullet }g_{3}^{\bullet }}{4g_{2}g_{3}}+\frac{\left( g_{3}^{\bullet
}\right) ^{2}}{4(g_{3})^{2}}-\frac{g_{2}^{^{\prime \prime }}}{2g_{3}}+\frac{%
g_{2}^{^{\prime }}g_{3}^{^{\prime }}}{4(g_{3})^{2}}+\frac{\left(
g_{2}^{^{\prime }}\right) ^{2}}{4g_{2}g_{3}},
\end{eqnarray*}%
with%
\begin{equation*}
\widehat{R}_{22}=\widehat{R}_{\ 223}^{3}\mbox{ and }\widehat{R}_{33}=-%
\widehat{R}_{\ 323}^{2},
\end{equation*}%
or%
\begin{equation*}
\widehat{R}_{2}^{2}=\widehat{R}_{3}^{3}=-\frac{1}{2g_{2}g_{3}}\left[
g_{3}^{\bullet \bullet }-\frac{g_{2}^{\bullet }g_{3}^{\bullet }}{2g_{2}}-%
\frac{\left( g_{3}^{\bullet }\right) ^{2}}{2g_{3}}+g_{2}^{\prime \prime }-%
\frac{g_{2}^{l}g_{3}^{l}}{2g_{3}}-\frac{\left( g_{2}^{l}\right) ^{2}}{2g_{2}}%
\right] ,
\end{equation*}%
which is just (\ref{ricci1a}).

Now, we consider
\begin{eqnarray*}
\widehat{P}_{\ bka}^{c} &=&\frac{\partial \widehat{L}_{\ bk}^{c}}{\partial
y^{a}}-\left( \frac{\partial \widehat{C}_{\ ba}^{c}}{\partial x^{k}}+%
\widehat{L}_{\ dk}^{c\,}\widehat{C}_{\ ba}^{d}-\widehat{L}_{\ bk}^{d}%
\widehat{C}_{\ da}^{c}-\widehat{L}_{\ ak}^{d}\widehat{C}_{\ bd}^{c}\right) +%
\widehat{C}_{\ bd}^{c}\widehat{P}_{\ ka}^{d} \\
&=&\frac{\partial \widehat{L}_{\ bk}^{c}}{\partial y^{a}}-\widehat{C}_{\
ba|k}^{c}+\widehat{C}_{\ bd}^{c}\widehat{P}_{\ ka}^{d}
\end{eqnarray*}%
from (\ref{dcurv}), where $\widehat{C}_{\ ba|k}^{c}$ is the covariant
h--derivative. Contracting indices, we have%
\begin{equation*}
\widehat{R}_{bk}=\widehat{P}_{\ bka}^{a}=\frac{\partial \widehat{L}_{\
bk}^{a}}{\partial y^{a}}-\widehat{C}_{\ ba|k}^{a}+\widehat{C}_{\ bd}^{a}%
\widehat{P}_{\ ka}^{d}
\end{equation*}%
Let us denote $\widehat{C}_{b}=\widehat{C}_{\ ba}^{c}$ and write%
\begin{equation*}
\widehat{C}_{.b|k}=\delta _{k}\widehat{C}_{b}-\widehat{L}_{\ bk}^{d\,}%
\widehat{C}_{d}=\partial _{k}\widehat{C}_{b}-N_{k}^{e}\partial _{e}\widehat{C%
}_{b}-\widehat{L}_{\ bk}^{d\,}\widehat{C}_{d}=\partial _{k}\widehat{C}%
_{b}-w_{k}\widehat{C}_{b}^{\ast }-L_{\ bk}^{d\,}\widehat{C}_{d}.
\end{equation*}%
We express%
\begin{equation*}
\widehat{R}_{bk}=\ _{[1]}R_{bk}+\ _{[2]}R_{bk}+\ _{[3]}R_{bk}
\end{equation*}%
where%
\begin{eqnarray*}
\ _{[1]}R_{bk} &=&\left( \widehat{L}_{bk}^{4}\right) ^{\ast },\
_{[2]}R_{bk}=-\partial _{k}\widehat{C}_{b}+w_{k}\widehat{C}_{b}^{\ast }+%
\widehat{L}_{\ bk}^{d\,}\widehat{C}_{d}, \\
\ _{[3]}R_{bk} &=&\widehat{C}_{\ bd}^{a}\widehat{P}_{\ ka}^{d}=\widehat{C}%
_{\ b4}^{4}\widehat{P}_{\ k4}^{4}+\widehat{C}_{\ b5}^{4}\widehat{P}_{\
k4}^{5}+\widehat{C}_{\ b4}^{5}\widehat{P}_{\ k5}^{4}+\widehat{C}_{\ b5}^{5}%
\widehat{P}_{\ k5}^{5},
\end{eqnarray*}%
and
\begin{eqnarray}
\widehat{C}_{4} &=&\widehat{C}_{44}^{4}+\widehat{C}_{45}^{5}=\frac{%
h_{4}^{\ast }}{2h_{4}}+\frac{h_{5}^{\ast }}{2h_{5}}=\beta _{4}^{\ast }+\beta
_{5}^{\ast },  \label{aux6a} \\
\widehat{C}_{5} &=&\widehat{C}_{54}^{4}+\widehat{C}_{55}^{5}=0,  \notag
\end{eqnarray}%
see (\ref{aux02d}) .

We compute%
\begin{equation*}
R_{4k}=\ _{[1]}R_{4k}+\ _{[2]}R_{4k}+\ _{[3]}R_{4k}
\end{equation*}%
with
\begin{eqnarray*}
\ _{[1]}R_{4k} &=&\left( \widehat{L}_{4k}^{4}\right) ^{\ast }=\left( \delta
_{k}\beta _{4}\right) ^{\ast }, \\
\ _{[2]}R_{4k} &=&-\partial _{k}\widehat{C}_{4}+w_{k}\widehat{C}_{4}^{\ast }+%
\widehat{L}_{\ 4k}^{4\,}\widehat{C}_{4},\widehat{L}_{\ 4k}^{4\,}=\delta
_{k}\beta _{4} \\
&=&-\partial _{k}\left( \beta _{4}^{\ast }+\beta _{5}^{\ast }\right)
+w_{k}\left( \beta _{4}^{\ast }+\beta _{5}^{\ast }\right) ^{\ast }+L_{\
4k}^{4\,}\left( \beta _{4}^{\ast }+\beta _{5}^{\ast }\right) , \\
\ _{[3]}R_{4k} &=&\widehat{C}_{.44}^{4}\widehat{P}_{.k4}^{4}+\widehat{C}%
_{.45}^{4}\widehat{P}_{.k4}^{5}+\widehat{C}_{.44}^{5}\widehat{P}_{.k5}^{4}+%
\widehat{C}_{.45}^{5}\widehat{P}_{.k5}^{5} \\
&=&\beta _{4}^{\ast }\left( w_{k}^{\ast }-\delta _{k}\beta _{4}\right)
-\beta _{5}^{\ast }\delta _{k}\beta _{5},
\end{eqnarray*}%
Summarizing, we get%
\begin{equation*}
\widehat{R}_{4k}=w_{k}\left[ \beta _{5}^{\ast \ast }+\left( \beta _{5}^{\ast
}\right) ^{2}-\beta _{4}^{\ast }\beta _{5}^{\ast }\right] +\beta _{5}^{\ast
}\partial _{k}\left( \beta _{4}+\beta _{5}\right) -\partial _{k}\beta
_{5}^{\ast },
\end{equation*}%
or, for
\begin{equation*}
\beta _{4}^{\ast }=\frac{h_{4}^{\ast }}{2h_{4}},\partial _{k}\beta _{4}=%
\frac{\partial _{k}h_{4}}{2h_{4}},\beta _{5}^{\ast }=\frac{h_{5}^{\ast }}{%
2h_{5}},\beta _{5}^{\ast \ast }=\frac{h_{5}^{\ast \ast }h_{5}-\left(
h_{5}^{\ast }\right) ^{2}}{2\left( h_{5}\right) ^{5}},
\end{equation*}%
we can write%
\begin{equation*}
2h_{5}\widehat{R}_{4k}=w_{k}\left[ h_{5}^{\ast \ast }-\frac{\left(
h_{5}^{\ast }\right) ^{2}}{2h_{5}}-\frac{h_{4}^{\ast }h_{5}^{\ast }}{2h_{4}}%
\right] +\frac{h_{5}^{\ast }}{2}\left( \frac{\partial _{k}h_{4}}{h_{4}}+%
\frac{\partial _{k}h_{5}}{h_{5}}\right) -\partial _{k}h_{5}^{\ast }
\end{equation*}%
which is equivalent to (\ref{ricci3a}).

In a similar way, we compute%
\begin{equation*}
\widehat{R}_{5k}=\ _{[1]}R_{5k}+\ _{[2]}R_{5k}+\ _{[3]}R_{5k}
\end{equation*}%
with
\begin{eqnarray*}
\ _{[1]}R_{5k} &=&\left( \widehat{L}_{5k}^{4}\right) ^{\ast },\
_{[2]}R_{5k}=-\partial _{k}\widehat{C}_{5}+w_{k}\widehat{C}_{5}^{\ast }+%
\widehat{L}_{\ 5k}^{4\,}\widehat{C}_{4}, \\
\ _{[3]}R_{5k} &=&\widehat{C}_{.54}^{4}\widehat{P}_{.k4}^{4}+\widehat{C}%
_{.55}^{4}\widehat{P}_{.k4}^{5}+\widehat{C}_{.54}^{5}\widehat{P}_{.k5}^{4}+%
\widehat{C}_{.55}^{5}\widehat{P}_{.k5}^{5}.
\end{eqnarray*}%
We have
\begin{eqnarray*}
\widehat{R}_{5k} &=&\left( \widehat{L}_{5k}^{4}\right) ^{\ast }+\widehat{L}%
_{\ 5k}^{4\,}\widehat{C}_{4}+C_{.55}^{4}\widehat{P}_{.k4}^{5}+C_{.54}^{5}%
\widehat{P}_{.k5}^{4} \\
&=&\left( -\frac{h_{5}}{h_{4}}n_{k}^{\ast }\right) ^{\ast }-\frac{h_{5}}{%
h_{4}}n_{k}^{\ast }\left( \frac{h_{4}^{\ast }}{2h_{4}}+\frac{h_{5}^{\ast }}{%
2h_{5}}\right) +\frac{h_{5}^{\ast }}{2h_{5}}\frac{h_{5}}{2h_{4}}n_{k}^{\ast
}-\frac{h_{5}^{\ast }}{2h_{4}}\frac{1}{2}n_{k}^{\ast },
\end{eqnarray*}%
or equivalently,%
\begin{equation*}
2h_{4}\widehat{R}_{5k}=h_{5}n_{k}^{\ast \ast }+\left( \frac{h_{5}}{h_{4}}%
h_{4}^{\ast }-\frac{3}{2}h_{5}^{\ast }\right) n_{k}^{\ast }
\end{equation*}%
which is just the formula (\ref{ricci4a}).

For the values
\begin{equation*}
\widehat{P}_{\ jka}^{i}=\frac{\partial \widehat{L}_{.jk}^{i}}{\partial y^{k}}%
-\left( \frac{\partial \widehat{C}_{.ja}^{i}}{\partial x^{k}}+\widehat{L}%
_{.lk}^{i}\widehat{C}_{.ja}^{l}-\widehat{L}_{.jk}^{l}\widehat{C}_{.la}^{i}-%
\widehat{L}_{.ak}^{c}\widehat{C}_{.jc}^{i}\right) +\widehat{C}_{.jb}^{i}%
\widehat{P}_{.ka}^{b}
\end{equation*}%
from (\ref{dcurv}), we obtain zeros because $\widehat{C}_{.jb}^{i}=0$ and $%
\widehat{L}_{.jk}^{i}$ do not depend on $y^{k}.$ So,
\begin{equation*}
\widehat{R}_{ja}=\widehat{P}_{\ jia}^{i}=0.
\end{equation*}

Taking
\begin{equation*}
\widehat{S}_{\ bcd}^{a}=\frac{\partial \widehat{C}_{\ bc}^{a}}{\partial y^{d}%
}-\frac{\partial \widehat{C}_{\ bd}^{a}}{\partial y^{c}}+\widehat{C}_{\
bc}^{e}\widehat{C}_{\ ed}^{a}-\widehat{C}_{\ bd}^{e}\widehat{C}_{\ ec}^{a}.
\end{equation*}%
from (\ref{dcurv}) and contracting the indices in order to define the Ricci
coefficients,%
\begin{equation*}
\widehat{R}_{bc}=\frac{\partial \widehat{C}_{\ bc}^{d}}{\partial y^{d}}-%
\frac{\partial \widehat{C}_{\ bd}^{d}}{\partial y^{c}}+\widehat{C}_{\ bc}^{e}%
\widehat{C}_{\ ed}^{d}-\widehat{C}_{\ bd}^{e}\widehat{C}_{\ ec}^{d}
\end{equation*}%
with $\widehat{C}_{\ bd}^{d}=\widehat{C}_{b}$ already computed, see (\ref%
{aux6a}), we obtain
\begin{equation*}
\widehat{R}_{bc}=\left( \widehat{C}_{.bc}^{4}\right) ^{\ast }-\partial _{c}%
\widehat{C}_{b}+\widehat{C}_{.bc}^{4}\widehat{C}_{4}-\widehat{C}_{.b4}^{4}%
\widehat{C}_{.4c}^{4}-\widehat{C}_{.b5}^{4}\widehat{C}_{.4c}^{5}-\widehat{C}%
_{.b4}^{5}\widehat{C}_{.5c}^{4}-\widehat{C}_{.b5}^{5}\widehat{C}_{.5c}^{5}.
\end{equation*}%
There are nontrivial values,
\begin{eqnarray*}
\widehat{R}_{44} &=&\left( \widehat{C}_{44}^{4}\right) ^{\ast }-\widehat{C}%
_{4}^{\ast }+\widehat{C}_{44}^{4}(\widehat{C}_{4}-\widehat{C}%
_{44}^{4})-\left( \widehat{C}_{45}^{5}\right) ^{2} \\
&=&\beta _{4}^{\ast \ast }-\left( \beta _{4}^{\ast }+\beta _{5}^{\ast
}\right) ^{\ast }+\beta _{4}^{\ast }\left( \beta _{4}^{\ast }+\beta
_{5}^{\ast }-\beta _{4}^{\ast }\right) -\left( \beta _{5}^{\ast }\right)
^{\ast } \\
\widehat{R}_{55} &=&\left( \widehat{C}_{55}^{4}\right) ^{\ast }-\widehat{C}%
_{55}^{4}\left( -\widehat{C}_{4}+2\widehat{C}_{45}^{5}\right)  \\
&=&-\left( \frac{h_{5}^{\ast }}{2h_{4}}\right) ^{\ast }+\frac{h_{5}^{\ast }}{%
2h_{4}}\left( 2\beta _{5}^{\ast }+\beta _{4}^{\ast }-\beta _{5}^{\ast
}\right)
\end{eqnarray*}%
Introducing
\begin{equation*}
\beta _{4}^{\ast }=\frac{h_{4}^{\ast }}{2h_{4}},\beta _{5}^{\ast }=\frac{%
h_{5}^{\ast }}{2h_{5}}
\end{equation*}%
we get%
\begin{equation*}
\widehat{R}_{4}^{4}=\widehat{R}_{5}^{5}=\frac{1}{2h_{4}h_{5}}\left[
-h_{5}^{\ast \ast }+\frac{\left( h_{5}^{\ast }\right) ^{2}}{2h_{5}}+\frac{%
h_{4}^{\ast }h_{5}^{\ast }}{2h_{4}}\right]
\end{equation*}%
which is just (\ref{ricci2a}).

Theorem \ref{t5dr} is proven.

\subsection{Some consequences of the Theorem \ref{t5dr}}

There are certain geometrical properties of the ansatz (\ref{metric5}) and
its canonical d--connection (\ref{candcon}).

\begin{corollary}
\label{ceint}The non--trivial components of the Einstein tensor $\widehat{%
\mathbf{G}}_{\ \beta }^{\alpha }=\widehat{\mathbf{R}}_{\ \beta }^{\alpha }-%
\frac{1}{2}\overleftarrow{\mathbf{\hat{R}\ }}\delta _{\beta }^{\alpha }$ for
the d--metric (\ref{ans4d}) given with respect to the N--adapted (co) frames
(\ref{ddif}) and (\ref{dder}) are
\begin{equation}
\widehat{G}_{1}^{1}=-\left( \widehat{R}_{2}^{2}+\widehat{S}_{4}^{4}\right) ,%
\widehat{G}_{2}^{2}=\widehat{G}_{3}^{3}=-\widehat{S}_{4}^{4},\widehat{G}%
_{4}^{4}=\widehat{G}_{5}^{5}=-\widehat{R}_{2}^{2}.  \label{einstdiag}
\end{equation}
\end{corollary}

The relations (\ref{einstdiag}) can be derived following the formulas for
the Ricci tensor (\ref{ricci1a})--(\ref{ricci4a}). They impose the condition
that the nonholonomically constrained dynamics of such gravitational fields
is defined by two independent components $\widehat{R}_{2}^{2}$ and $\widehat{%
S}_{4}^{4}$ and result in:

\begin{corollary}
\label{cors}The system of 5D Einstein equations is compatible for the
generic off--diagonal ansatz (\ref{metric5}) if the energy--momentum tensor $%
\mathbf{\Upsilon }_{\alpha \beta },$ given with respect to N--adapted frames
(\ref{ddif}) and (\ref{dder}), is diagonal and satisfies the conditions
\begin{equation}
\Upsilon _{2}^{2}=\Upsilon _{3}^{3}=\Upsilon _{2}(x^{2},x^{3},v),\ \Upsilon
_{4}^{4}=\Upsilon _{5}^{5}=\Upsilon _{4}(x^{2},x^{3}),\mbox{ and }\Upsilon
_{1}=\Upsilon _{2}+\Upsilon _{4}.  \label{emcond}
\end{equation}
\end{corollary}

Following the Corollaries \ref{ceint} and \ref{cors} and formulas (\ref%
{ricci1a})--(\ref{ricci4a}), we can write the Einstein equations for the
ansatz (\ref{ans4d}), equivalently for (\ref{metric5}), in the form
\begin{eqnarray}
\widehat{R}_{2}^{2} &=&\widehat{R}_{3}^{3}  \label{ep1a} \\
&=&\frac{1}{2g_{2}g_{3}}[\frac{g_{2}^{\bullet }g_{3}^{\bullet }}{2g_{2}}+%
\frac{(g_{3}^{\bullet })^{2}}{2g_{3}}-g_{3}^{\bullet \bullet }+\frac{%
g_{2}^{^{\prime }}g_{3}^{^{\prime }}}{2g_{3}}+\frac{(g_{2}^{^{\prime }})^{2}%
}{2g_{2}}-g_{2}^{^{\prime \prime }}]=-\Upsilon _{4}(x^{2},x^{3}),  \notag \\
\widehat{S}_{4}^{4} &=&\widehat{S}_{5}^{5}=\frac{1}{2h_{4}h_{5}}\left[
h_{5}^{\ast }\left( \ln \sqrt{|h_{4}h_{5}|}\right) ^{\ast }-h_{5}^{\ast \ast
}\right] =-\Upsilon _{2}(x^{2},x^{3},v),  \label{ep2a} \\
\widehat{R}_{4i} &=&-w_{i}\frac{\beta }{2h_{5}}-\frac{\alpha _{i}}{2h_{5}}=0,
\label{ep3a} \\
\widehat{R}_{5i} &=&-\frac{h_{5}}{2h_{4}}\left[ n_{i}^{\ast \ast }+\gamma
n_{i}^{\ast }\right] =0.  \label{ep4a}
\end{eqnarray}

This system of equations can be solved in general forms for different types
of vacuum and non--vacuum configurations.

\section{General Off--Diagonal Solutions}

One holds:

\begin{theorem}
\label{texs}The system of second order nonlinear partial differential
equations (\ref{ep1a})--(\ref{ep4a}) can be solved in general form if there
are given certain values of functions $g_{2}(x^{2},x^{3})$ (or, inversely, $%
g_{3}(x^{2},x^{3})),\ h_{4}\left( x^{i},v\right) $ (or, inversely, $%
h_{5}\left( x^{i},v\right) ),$ $\omega \left( x^{i},v\right) $ and of
sources $\Upsilon _{2}(x^{2},x^{3},v)$ and $\Upsilon _{4}(x^{2},x^{3}).$
\end{theorem}

We outline the main steps of constructing exact solutions which consists the
proof of this Theorem.

\begin{itemize}
\item The general solution of equation (\ref{ep1a}) can be written in the
form
\begin{equation}
\varpi =g_{[0]}\exp [a_{2}\widetilde{x}^{2}\left( x^{2},x^{3}\right) +a_{3}%
\widetilde{x}^{3}\left( x^{2},x^{3}\right) ],  \label{solricci1a}
\end{equation}%
were $g_{[0]},a_{2}$ and $a_{3}$ are some constants and the functions $%
\widetilde{x}^{2,3}\left( x^{2},x^{3}\right) $ define any coordinate
transforms $x^{2,3}\rightarrow \widetilde{x}^{2,3}$ for which the 2D line
element becomes conformally flat, i. e.
\begin{equation}
g_{2}(x^{2},x^{3})(dx^{2})^{2}+g_{3}(x^{2},x^{3})(dx^{3})^{2}\rightarrow
\varpi (x^{2},x^{3})\left[ (d\widetilde{x}^{2})^{2}+\epsilon (d\widetilde{x}%
^{3})^{2}\right] ,  \label{con10}
\end{equation}%
where $\epsilon =\pm 1$ for a corresponding signature. In terms of
coordinates $\widetilde{x}^{2,3},$ the equation (\ref{ep1a}) transform into%
\begin{equation*}
\varpi \left( \ddot{\varpi}+\varpi ^{\prime \prime }\right) -\dot{\varpi}%
-\varpi ^{\prime }=2\varpi ^{2}\Upsilon _{4}(\tilde{x}^{2},\tilde{x}^{3})
\end{equation*}%
or%
\begin{equation}
\ddot{\psi}+\psi ^{\prime \prime }=2\Upsilon _{4}(\tilde{x}^{2},\tilde{x}%
^{3}),  \label{auxeq01}
\end{equation}%
for $\psi =\ln |\varpi |.$ The integrals of (\ref{auxeq01}) depends on the
source $\Upsilon _{4}.$ As a particular case we can consider that $\Upsilon
_{4}=0.$ There are three alternative possibilities to generate solutions of (%
\ref{ep1a}). For instance, we can prescribe that $g_{2}=g_{3}$ and get the
equation (\ref{auxeq01}) for $\psi =\ln |g_{2}|=\ln |g_{3}|.$ If we suppose
that $g_{2}^{^{\prime }}=0,$ for a given $g_{2}(x^{2}),$ we obtain from (\ref%
{ep1a})%
\begin{equation*}
g_{3}^{\bullet \bullet }-\frac{g_{2}^{\bullet }g_{3}^{\bullet }}{2g_{2}}-%
\frac{(g_{3}^{\bullet })^{2}}{2g_{3}}=2g_{2}g_{3}\Upsilon _{4}(x^{2},x^{3})
\end{equation*}%
which can be integrated explicitly for given values of $\Upsilon _{4}.$
Similarly, we can generate solutions for a prescribed $g_{3}(x^{3})$ in the
equation
\begin{equation*}
g_{2}^{^{\prime \prime }}-\frac{g_{2}^{^{\prime }}g_{3}^{^{\prime }}}{2g_{3}}%
-\frac{(g_{2}^{^{\prime }})^{2}}{2g_{2}}=2g_{2}g_{3}\Upsilon
_{4}(x^{2},x^{3}).
\end{equation*}%
We note that a transform (\ref{con10}) is always possible for 2D metrics and
the explicit form of solutions depends on chosen system of 2D coordinates
and on the signature $\epsilon =\pm 1.$ In the simplest case with $\Upsilon
_{4}=0$ the equation (\ref{ep1a}) is solved by arbitrary two functions $%
g_{2}(x^{3})$ and $g_{3}(x^{2}).$

\item For $\Upsilon _{2}=0,$ the equation (\ref{ep2a}) relates two functions
$h_{4}\left( x^{i},v\right) $ and $h_{5}\left( x^{i},v\right) $ following
two possibilities:

a) to compute
\begin{eqnarray}
\sqrt{|h_{5}|} &=&h_{5[1]}\left( x^{i}\right) +h_{5[2]}\left( x^{i}\right)
\int \sqrt{|h_{4}\left( x^{i},v\right) |}dv  \notag \\
&=&h_{5[1]}\left( x^{i}\right) +h_{5[2]}\left( x^{i}\right) v,\ h_{4}^{\ast
}\left( x^{i},v\right) =0,  \label{p2}
\end{eqnarray}%
for $h_{4}^{\ast }\left( x^{i},v\right) \neq 0$ and some functions $%
h_{5[1,2]}\left( x^{i}\right) $ stated by boundary conditions;

b) or, inversely, to compute $h_{4}$ for a given $h_{5}\left( x^{i},v\right)
,h_{5}^{\ast }\neq 0,$%
\begin{equation}
\sqrt{|h_{4}|}=h_{[0]}\left( x^{i}\right) (\sqrt{|h_{5}\left( x^{i},v\right)
|})^{\ast },  \label{p1}
\end{equation}%
with $h_{[0]}\left( x^{i}\right) $ given by boundary conditions. We note
that the sourceless equation (\ref{ep2a}) is satisfied by arbitrary pairs of
coefficients $h_{4}\left( x^{i},v\right) $ and $h_{5[0]}\left( x^{i}\right)
. $

Solutions with $\Upsilon _{2}\neq 0$ can be found for an ansatz of type
\begin{equation}
h_{5}[\Upsilon _{2}]=h_{5},h_{4}[\Upsilon _{2}]=\varsigma _{4}\left(
x^{i},v\right) h_{4},  \label{auxf02}
\end{equation}%
where $h_{4}$ and $h_{5}$ are related by formula (\ref{p2}), or (\ref{p1}).
Substituting (\ref{auxf02}), we obtain%
\begin{equation}
\varsigma _{4}\left( x^{i},v\right) =\varsigma _{4[0]}\left( x^{i}\right)
-\int \Upsilon _{2}(x^{2},x^{3},v)\frac{h_{4}h_{5}}{4h_{5}^{\ast }}dv,
\label{auxf02a}
\end{equation}%
where $\varsigma _{4[0]}\left( x^{i}\right) $ are arbitrary integration
unctions.

\item The exact solutions of (\ref{ep3a}) for $\beta \neq 0$ are defined
from an algebraic equation, $w_{i}\beta +\alpha _{i}=0,$ where the
coefficients $\beta $ and $\alpha _{i}$ are computed as in formulas (\ref%
{abc}) by using the solutions for (\ref{ep1a}) and (\ref{ep2a}). The general
solution is
\begin{equation}
w_{k}=\partial _{k}\ln [\sqrt{|h_{4}h_{5}|}/|h_{5}^{\ast }|]/\partial
_{v}\ln [\sqrt{|h_{4}h_{5}|}/|h_{5}^{\ast }|],  \label{w}
\end{equation}%
with $\partial _{v}=\partial /\partial v$ and $h_{5}^{\ast }\neq 0.$ If $%
h_{5}^{\ast }=0,$ or even $h_{5}^{\ast }\neq 0$ but $\beta =0,$ the
coefficients $w_{k}$ could be any functions on $\left( x^{i},v\right) .$ \
For the vacuum Einstein equations this is a degenerated case imposing the
the compatibility conditions $\beta =\alpha _{i}=0,$ which are satisfied,
for instance, if the $h_{4}$ and $h_{5}$ are related as in the formula (\ref%
{p1}) but with $h_{[0]}\left( x^{i}\right) =const.$

\item Having defined $h_{4}$ and $h_{5}$ and computed $\gamma $ from (\ref%
{abc}), we can solve the equation (\ref{ep4a}) by integrating on variable ''$%
v$'' the equation $n_{i}^{\ast \ast }+\gamma n_{i}^{\ast }=0.$ The exact
solution is
\begin{eqnarray}
n_{k} &=&n_{k[1]}\left( x^{i}\right) +n_{k[2]}\left( x^{i}\right) \int h_{4}(%
\sqrt{|h_{5}|})^{-3}dv,~h_{5}^{\ast }\neq 0;  \notag \\
&=&n_{k[1]}\left( x^{i}\right) +n_{k[2]}\left( x^{i}\right) \int
h_{4}dv,\qquad ~h_{5}^{\ast }=0;  \label{n} \\
&=&n_{k[1]}\left( x^{i}\right) +n_{k[2]}\left( x^{i}\right) \int (\sqrt{%
|h_{5}|})^{-3}dv,~h_{4}^{\ast }=0,  \notag
\end{eqnarray}%
for some functions $n_{k[1,2]}\left( x^{i}\right) $ stated by boundary
conditions.

The Theorem \ref{texs} is proven.
\end{itemize}

\ Summarizing the results for the nondegenerated cases when $h_{4}^{\ast
}\neq 0$ and $h_{5}^{\ast }\neq 0,$ we derive an important result for 5D
exact solutions parametrized by ansatz of type (\ref{ans4d}), with local
coordinates $u^{\alpha }=\left( x^{i},y^{a}\right) $ when $x^{i}=\left(
x^{1},x^{\widehat{i}}\right) ,x^{\widehat{i}}=\left( x^{2},x^{3}\right)
,y^{a}=\left( y^{4}=v,y^{a}\right) ,$ and for arbitrary signatures $\epsilon
_{\alpha }=\left( \epsilon _{1},\epsilon _{2},\epsilon _{3},\epsilon
_{4},\epsilon _{5}\right) $ (where $\epsilon _{\alpha }=\pm 1):$

\begin{corollary}
\label{corgsol1}Any off--diagonal metric
\begin{eqnarray}
\delta s^{2} &=&\epsilon _{1}(dx^{1})^{2}+\epsilon _{\widehat{k}}g_{\widehat{%
k}}\left( x^{\widehat{i}}\right) (dx^{\widehat{k}})^{2}+\epsilon
_{4}h_{0}^{2}(x^{i})\times  \notag \\
&&\left[ f^{\ast }\left( x^{i},v\right) \right] ^{2}|\varsigma _{4}\left(
x^{i},v\right) |\left( \delta v\right) ^{2}+\epsilon _{5}\left[ f\left(
x^{i},v\right) -f_{0}(x^{i})\right] ^{2}\left( \delta y^{5}\right) ^{2},
\notag \\
\delta v &=&dv+w_{k}\left( x^{i},v\right) dx^{k},\ \delta
y^{5}=dy^{5}+n_{k}\left( x^{i},v\right) dx^{k},  \label{gensol1}
\end{eqnarray}%
with coefficients of necessary smooth class, where\ \ $g_{\widehat{k}}\left(
x^{\widehat{i}}\right) $ is a solution of the 2D equation (\ref{ep1a}) for a
given source $\Upsilon _{4}\left( x^{\widehat{i}}\right) ,$%
\begin{equation*}
\varsigma _{4}\left( x^{i},v\right) =\varsigma _{4[0]}\left( x^{i}\right) -%
\frac{\epsilon _{4}}{8}h_{0}^{2}(x^{i})\int \Upsilon _{2}(x^{\widehat{k}%
},v)f^{\ast }\left( x^{i},v\right) \left[ f\left( x^{i},v\right)
-f_{0}(x^{i})\right] dv,
\end{equation*}%
and the N--connection coefficients $N_{i}^{4}=w_{i}(x^{k},v)$ and $%
N_{i}^{5}=n_{i}(x^{k},v)$ are
\begin{equation}
w_{i}=-\frac{\partial _{i}\varsigma _{4}\left( x^{k},v\right) }{\varsigma
_{4}^{\ast }\left( x^{k},v\right) }  \label{gensol1w}
\end{equation}%
and
\begin{equation}
n_{k}=n_{k[1]}\left( x^{i}\right) +n_{k[2]}\left( x^{i}\right) \int \frac{%
\left[ f^{\ast }\left( x^{i},v\right) \right] ^{2}}{\left[ f\left(
x^{i},v\right) -f_{0}(x^{i})\right] ^{3}}\varsigma _{4}\left( x^{i},v\right)
dv,  \label{gensol1n}
\end{equation}%
define an exact solution of the system of Einstein equations (\ref{ep1a})--(%
\ref{ep4a}) for arbitrary nontrivial functions $f\left( x^{i},v\right) $
(with $f^{\ast }\neq 0),$ $f_{0}(x^{i}),$ $h_{0}^{2}(x^{i})$, $\ \varsigma
_{4[0]}\left( x^{i}\right) ,$ $n_{k[1]}\left( x^{i}\right) $ and $\
n_{k[2]}\left( x^{i}\right) ,$ and sources $\Upsilon _{2}(x^{\widehat{k}%
},v),\Upsilon _{4}\left( x^{\widehat{i}}\right) $ and any integration
constants and signatures $\epsilon _{\alpha }=\pm 1$ to be defined by
certain boundary conditions and physical considerations.
\end{corollary}

Any metric (\ref{gensol1}) with $h_{4}^{\ast }\neq 0$ and $h_{5}^{\ast }\neq
0$ has the property to be generated by a function of four variables $f\left(
x^{i},v\right) $ with emphasized dependence on the anisotropic coordinate $v,
$ because $f^{\ast }\doteqdot \partial _{v}f\neq 0$ and by arbitrary sources
$\Upsilon _{2}(x^{\widehat{k}},v),\Upsilon _{4}\left( x^{\widehat{i}}\right)
.$ The rest of arbitrary functions which do not depend on $v$ have been
obtained in result of integration of partial differential equations. This
fix a specific class of metrics generated by using the relation (\ref{p1})
and the first formula in (\ref{n}). We can generate also a different class
of solutions with $h_{4}^{\ast }=0$ by considering the second formula in (%
\ref{p2}) and respective formulas in (\ref{n}). The ''degenerated'' cases
with $h_{4}^{\ast }=0$ but $h_{5}^{\ast }\neq 0$ and inversely, $h_{4}^{\ast
}\neq 0$ but $h_{5}^{\ast }=0$ are more special and request a proper
explicit construction of solutions. Nevertheless, such type of solutions are
also generic off--diagonal and they could be of substantial interest.

The sourceless case with vanishing $\Upsilon _{2}$ and $\Upsilon _{4}$ is
defined by the following:

\begin{remark}
Any off--diagonal metric (\ref{gensol1}) with $\varsigma _{\Upsilon }=1,$ $%
h_{0}^{2}(x^{i})=$ $h_{0}^{2}=const,$ $w_{i}=0$ and $n_{k}$ computed as in (%
\ref{gensol1n}) but for $\varsigma _{\Upsilon }=1,$ defines a vacuum
solution of 5D Einstein equations for the canonical d--connection (\ref%
{candcon}).
\end{remark}

By imposing additional constraints on arbitrary functions from $%
N_{i}^{5}=n_{i}$ and $N_{i}^{5}=w_{i},$ we can select off--diagonal
gravitational configurations with such distorsions of the Levi--Civita
connection to the canonical d--connections when both classes of linear
connections result in the same solutions of the vacuum Einstein equations,
see next Appendix \ref{ss4d}.

\section{4D Nonholonomic Manifolds and Einstein Gravity}

\label{ss4d}The method of constructing 5D solutions with nontrivial torsion
can be restricted to generate 4D nonholonomic configurations and generic
off--diagonal solutions in general relativity.

\subsection{Reductions from 5D to 4D}

To construct a $5D\rightarrow 4D$ reduction for the ansatz (\ref{ans4d}) and
(\ref{metric5}) is to eliminate from formulas the variable $x^{1}$ and to
consider a 4D space (parametrized by local coordinates $\left(
x^{2},x^{3},v,y^{5}\right) )$ being trivially embedded into 5D space
(parametrized by local coordinates $\left( x^{1},x^{2},x^{3},v,y^{5}\right) $
with $g_{11}=\pm 1,g_{1\widehat{\alpha }}=0,\widehat{\alpha }=2,3,4,5)$ with
possible \ 4D conformal and anholonomic transforms depending only on
variables $\left( x^{2},x^{3},v\right) .$ We suppose that the 4D metric $g_{%
\widehat{\alpha }\widehat{\beta }}$ could be of arbitrary signature. In
order to emphasize that some coordinates are stated just for a such 4D space
we put ''hats'' on the Greek indices, $\widehat{\alpha },\widehat{\beta },...
$ \ and on the Latin indices from the middle of the alphabet, $\widehat{i},%
\widehat{j},...=2,3,$ where $u^{\widehat{\alpha }}=\left( x^{\widehat{i}%
},y^{a}\right) =\left( x^{2},x^{3},y^{4}=v,y^{5}\right) .$

In result, the Theorem \ref{t5dr} and Corollaries \ref{ceint} and \ref{cors}
and Theorem \ref{texs} can be reformulated for 4D gravity with mixed
holonomic--anholonomic variables. We outline here the most important
properties of a such reduction.

\begin{itemize}
\item The metric (\ref{block2}) (equivalently, (\ref{ansatz})), or, in a
more restricted case, of metric (\ref{ans4d}) (equivalently, with (\ref%
{metric5})) can be transformed into a 4D one (trivially embedded into a 5D
spacetime),

\begin{equation}
\mathbf{g}=\mathbf{g}_{\widehat{\alpha }\widehat{\beta }}\left( x^{\widehat{i%
}},v\right) du^{\widehat{\alpha }}\otimes du^{\widehat{\beta }}
\label{metric4}
\end{equation}%
with the metric coefficients\textbf{\ }$g_{\widehat{\alpha }\widehat{\beta }%
} $ parametrized in the form%
\begin{equation*}
{\left[
\begin{array}{cccc}
g_{2}+w_{2}^{\ 2}h_{4}+n_{2}^{\ 2}h_{5} & w_{2}w_{3}h_{4}+n_{2}n_{3}h_{5} &
w_{2}h_{4} & n_{2}h_{5} \\
w_{2}w_{3}h_{4}+n_{2}n_{3}h_{5} & g_{3}+w_{3}^{\ 2}h_{4}+n_{3}^{\ 2}h_{5} &
w_{3}h_{4} & n_{3}h_{5} \\
w_{2}h_{4} & w_{3}h_{4} & h_{4} & 0 \\
n_{2}h_{5} & n_{3}h_{5} & 0 & h_{5}%
\end{array}%
\right] ,}
\end{equation*}%
where the coefficients are some necessary smoothly class functions of type:
\begin{eqnarray}
g_{2,3} &=&g_{2,3}(x^{2},x^{3}),h_{4,5}=h_{4,5}(x^{\widehat{k}},v),  \notag
\\
w_{\widehat{i}} &=&w_{\widehat{i}}(x^{\widehat{k}},v),n_{\widehat{i}}=n_{%
\widehat{i}}(x^{\widehat{k}},v);~\widehat{i},\widehat{k}=2,3.  \notag
\end{eqnarray}

\item We obtain a quadratic line element
\begin{equation}
\delta s^{2}=g_{2}(dx^{2})^{2}+g_{3}(dx^{3})^{2}+h_{4}(\delta
v)^{2}+h_{5}(\delta y^{5})^{2},  \label{dmetric4}
\end{equation}%
written with respect to the anholonomic co--frame $\left( dx^{\widehat{i}%
},\delta v,\delta y^{5}\right) ,$ where
\begin{equation}
\delta v=dv+w_{\widehat{i}}dx^{\widehat{i}}\mbox{ and }\delta
y^{5}=dy^{5}+n_{\widehat{i}}dx^{\widehat{i}}  \label{ddif4}
\end{equation}%
is the dual of $\left( \delta _{\widehat{i}},\partial _{4},\partial
_{5}\right) ,$ where
\begin{equation}
\delta _{\widehat{i}}=\partial _{\widehat{i}}+w_{\widehat{i}}\partial
_{4}+n_{\widehat{i}}\partial _{5}.  \label{dder4}
\end{equation}

\item If the conditions of the 4D variant of the Theorem \ref{t5dr} are
satisfied, we have the same equations (\ref{ep1a}) --(\ref{ep4a}) were we
substitute\\ $h_{4}=h_{4}\left( x^{\widehat{k}},v\right) $ and $%
h_{5}=h_{5}\left( x^{\widehat{k}},v\right) .$ As a consequence, we have $%
\alpha _{i}\left( x^{k},v\right) \rightarrow \alpha _{\widehat{i}}\left( x^{%
\widehat{k}},v\right) ,\beta =\beta \left( x^{\widehat{k}},v\right) $ and $%
\gamma =\gamma \left( x^{\widehat{k}},v\right) $ resulting in $w_{\widehat{i}%
}=w_{\widehat{i}}\left( x^{\widehat{k}},v\right) $ and $n_{\widehat{i}}=n_{%
\widehat{i}}\left( x^{\widehat{k}},v\right) .$

\item One holds the same formulas (\ref{p2})-(\ref{n}) from the Theorem \ref%
{texs} on the general form of exact solutions with that difference that
their 4D analogs are to be obtained by reductions of holonomic indices, $%
\widehat{i}\rightarrow i,$ and holonomic coordinates, $x^{i}\rightarrow x^{%
\widehat{i}},$ i. e. in the 4D solutions there is not contained the variable
$x^{1}.$

\item The formulae (\ref{einstdiag}) for the nontrivial coefficients of the
Einstein tensor in 4D, stated by the Corollary \ref{ceint}, are \ written
\begin{equation}
\widehat{G}_{2}^{2}=\widehat{G}_{3}^{3}=-\widehat{S}_{4}^{4},\widehat{G}%
_{4}^{4}=\widehat{G}_{5}^{5}=-\widehat{R}_{2}^{2}.  \label{einstdiag4}
\end{equation}

\item For symmetries of the Einstein tensor (\ref{einstdiag4}), \ we can
introduce a matter field source with a diagonal energy momentum tensor,
like\ it is stated in the Corollary \ref{cors} by the conditions (\ref%
{emcond}), which in 4D are transformed into
\begin{equation}
\Upsilon _{2}^{2}=\Upsilon _{3}^{3}=\Upsilon _{2}(x^{2},x^{3},v),\ \Upsilon
_{4}^{4}=\Upsilon _{5}^{5}=\Upsilon _{4}(x^{2},x^{3}).  \label{emcond4}
\end{equation}
\end{itemize}

\subsection{Off--diagonal metrics in general relativity}

For 4D configurations, we can generate a class of off--diagonal metrics in
general relativity if we impose the conditions that the nonholonomic
transforms and correspoindingly generated distorsions $\widehat{\mathbf{Z}}%
_{\alpha \beta }$ (\ref{aux53}) of the Levi--Civita connection to the
canonical d--connection induce such distorsions of the curvature tensor $%
^{\bigtriangledown }\mathcal{Z}_{\ \beta }^{\alpha }$ (\ref{dist1}) that
there are satisfied the conditions $^{\bigtriangledown }\mathcal{Z}_{\ \beta
\gamma \alpha }^{\alpha }=0$ (\ref{cdist}). In this case, the Ricci tensor
has the same nontrivial values for the Levi--Civita connection and the
canonical d--connection, both computed with respect to N--adapted bases (\ref%
{ddif4}) and (\ref{dder4}). In this subsection, we demonstrate how the
conditions (\ref{cdist}) can be solved for a 4D d--metric (\ref{dmetric4}),
equivalently, for (\ref{metric4}).

The h-- and v--coefficients of distorsion $\widehat{\mathbf{Z}}_{\ \beta
\gamma }^{\tau }$ are defined from (\ref{lcsyma}),

\begin{equation}
\widehat{\mathbf{Z}}_{\ \beta \gamma }^{\tau }=\left( \widehat{Z}_{\
jk}^{i}=0,\widehat{Z}_{\ bk}^{a}=-\frac{\partial N_{k}^{a}}{\partial y^{b}},%
\widehat{Z}_{\ jc}^{i}=\frac{1}{2}g^{ik}\Omega _{jk}^{a}h_{ca},\widehat{Z}%
_{\ bc}^{a}=0\right) .  \label{aux21}
\end{equation}%
Introducing these values in (\ref{dist1}), for
\begin{equation*}
\widehat{P}_{\ bk}^{a}=\frac{\partial N_{k}^{a}}{\partial y^{b}}-\widehat{L}%
_{\ bk}^{a}=\ ^{\bigtriangledown }\widehat{P}_{\ bk}^{a}-\widehat{Z}_{\
bk}^{a},
\end{equation*}%
we compute the distorsions of the d--curvatures (\ref{dcurv}) which are used
for definition of the Ricci d--tensors:
\begin{equation}
\widehat{R}_{\ hjk}^{i}=\ ^{\bigtriangledown }\widehat{R}_{\ hjk}^{i}-%
\widehat{Z}_{.ha}^{i}\Omega _{.jk}^{a},  \notag
\end{equation}%
\begin{eqnarray*}
\widehat{P}_{\ jka}^{i} &=&\ ^{\bigtriangledown }P_{\ jka}^{i}-(\frac{%
\partial \widehat{Z}_{.ja}^{i}}{\partial x^{k}}+\widehat{L}_{.lk}^{i}%
\widehat{Z}_{.ja}^{l}-\widehat{L}_{.jk}^{l}\widehat{Z}_{.la}^{i}-\widehat{Z}%
_{.ak}^{c}\widehat{C}_{.jc}^{i}-\widehat{Z}_{.ak}^{c}\widehat{Z}_{.jc}^{i} \\
&&-\widehat{L}_{.ak}^{c}\widehat{Z}_{.jc}^{i})-\widehat{C}_{.jb}^{i}\widehat{%
Z}_{.ka}^{b}-\widehat{C}_{.jb}^{i}\widehat{Z}_{.ka}^{b},
\end{eqnarray*}%
\begin{equation*}
\widehat{P}_{\ bka}^{c}=\ ^{\bigtriangledown }P_{\ bka}^{c}-\left( \frac{%
\partial \widehat{Z}_{.bk}^{c}}{\partial y^{a}}+\widehat{Z}_{.dk}^{c\,}%
\widehat{C}_{.ba}^{d}-\widehat{Z}_{.bk}^{d}\widehat{C}_{.da}^{c}-\widehat{Z}%
_{.ak}^{d}\widehat{C}_{.bd}^{c}\right) -\widehat{C}_{.bd}^{c}\widehat{Z}%
_{.ak}^{d},
\end{equation*}%
\begin{equation*}
\widehat{S}_{\ bcd}^{a}=\ ^{\bigtriangledown }S_{\ bcd}^{a}.
\end{equation*}%
Contracting the indices and re--grouping the terms, we get the distorsions
for the Ricci d--tensors (\ref{dricci})%
\begin{eqnarray*}
\widehat{R}_{ji} &=&\ ^{\bigtriangledown }\widehat{R}_{ji}-\widehat{Z}%
_{.ha}^{i}\Omega _{.ji}^{a}, \\
\widehat{R}_{ja} &=&\ ^{\bigtriangledown }\widehat{R}_{ja}+\frac{\partial
\widehat{Z}_{.ja}^{i}}{\partial x^{i}}+\widehat{L}_{.li}^{i}\widehat{Z}%
_{.ja}^{l}-\widehat{L}_{.ji}^{l}\widehat{Z}_{.la}^{i}-\widehat{L}_{.ak}^{c}%
\widehat{Z}_{.jc}^{i}, \\
\widehat{R}_{bk} &=&\ ^{\bigtriangledown }\widehat{R}_{bk}+\frac{\partial
\widehat{Z}_{.bk}^{c}}{\partial y^{c}}+\widehat{Z}_{.bk}^{d\,}\widehat{C}%
_{.dc}^{c}-\widehat{C}_{.bd}^{a}\widehat{Z}_{.ak}^{d}, \\
\widehat{R}_{ba} &=&\ ^{\bigtriangledown }\widehat{R}_{ba}.
\end{eqnarray*}

If we consider a subclass of solutions with
\begin{equation}
\Omega _{.ji}^{a}=0,  \label{oeq}
\end{equation}%
when $\widehat{Z}_{.ja}^{l}=0,$ see (\ref{aux21}), we obtain that
\begin{equation*}
\widehat{R}_{ji}=\ ^{\bigtriangledown }\widehat{R}_{ji},\widehat{R}_{ja}=\
^{\bigtriangledown }\widehat{R}_{ja},\widehat{R}_{ba}=\ ^{\bigtriangledown }%
\widehat{R}_{ba}.
\end{equation*}%
One follows that
\begin{equation*}
\widehat{R}_{\alpha \beta }=\ ^{\bigtriangledown }\widehat{R}_{\alpha \beta }
\end{equation*}%
if $\widehat{R}_{bk}=\ ^{\bigtriangledown }\widehat{R}_{bk}$ which hold when
\begin{equation}
\frac{\partial \widehat{Z}_{.bk}^{c}}{\partial y^{c}}+\widehat{Z}_{.bk}^{d\,}%
\widehat{C}_{.dc}^{c}-\widehat{C}_{.bd}^{a}\widehat{Z}_{.ak}^{d}=0.
\label{lccr}
\end{equation}

Let us analyze the equations (\ref{lccr}) for the ansatz (\ref{dmetric4})
(they hold also for the 5D d--metric (\ref{ans4d})). Such metrics do not
depend on $y^{5}.$ \ We have
\begin{equation*}
\frac{\partial \widehat{Z}_{.bk}^{4}}{\partial v}+\widehat{Z}_{.bk}^{d\,}%
\widehat{C}_{.dc}^{c}-\widehat{C}_{.bd}^{a}\widehat{Z}_{.ak}^{d}=0,
\end{equation*}%
or, for $\left( \widehat{Z}_{.bk}^{4}\right) ^{\ast }=\partial \widehat{Z}%
_{.bk}^{4}/\partial v,$
\begin{equation*}
\left( \widehat{Z}_{4k}^{4}\right) ^{\ast }+\widehat{Z}_{4k}^{d\,}\widehat{C}%
_{bc}^{c}-\widehat{C}_{4d}^{a}\widehat{Z}_{ak}^{d}=0\mbox{ and }\left(
\widehat{Z}_{5k}^{4}\right) ^{\ast }+\widehat{Z}_{5k}^{d\,}\widehat{C}%
_{dc}^{c}-\widehat{C}_{5d}^{a}\widehat{Z}_{.ak}^{d}=0,
\end{equation*}%
where $\widehat{Z}_{5k}^{d}=\partial N_{k}^{d}/\partial y^{5}=0,$ see (\ref%
{aux21}). We get
\begin{equation*}
\left( \widehat{Z}_{4k}^{4}\right) ^{\ast }+\widehat{Z}_{4k}^{4\,}\widehat{C}%
_{4c}^{c}-\widehat{C}_{44}^{a}\widehat{Z}_{ak}^{4}=0\mbox{ and }\widehat{Z}%
_{4k}^{c\,}\widehat{C}_{5c}^{4}=0.
\end{equation*}%
These formulas,  for $N_{k}^{4}=w_{k}$ and $N_{k}^{5}=n_{k},$ see also
formulas (\ref{aux02d}) and (\ref{aux6a}), result in%
\begin{eqnarray}
w_{k}^{\ast \ast }+\frac{w_{k}^{\ast }}{2}\frac{h_{5}^{\ast }}{h_{5}} &=&0,
\label{aux32} \\
n_{k}^{\ast }\frac{h_{5}^{\ast }}{h_{5}} &=&0.  \notag
\end{eqnarray}%
There are two possibilities to satisfy this system: 1) to take
\begin{equation*}
h_{5}^{\ast }=0\text{\mbox{ which impose }}w_{k}^{\ast \ast }=0
\end{equation*}%
or 2) to consider that $h_{5}^{\ast }\neq 0$ imposing that
\begin{equation*}
n_{k}^{\ast }=0\mbox{ for any }w_{k}^{\ast \ast }+\frac{w_{k}^{\ast }}{2}%
\frac{h_{5}^{\ast }}{h_{5}}=0.
\end{equation*}

The next step is to see how the constraints (\ref{oeq}) may be solved. As a
matter of principle, they are satisfied for any $w_{k}$ and $n_{k}$ for
which
\begin{eqnarray}
w_{2}^{\prime }-w_{3}^{\bullet }+w_{3}w_{2}^{\ast }-w_{2}w_{3}^{\ast } &=&0,
\label{aux31} \\
n_{2}^{\prime }-n_{3}^{\bullet }+w_{3}n_{2}^{\ast }-w_{2}n_{3}^{\ast } &=&0
\notag
\end{eqnarray}%
where we put $n_{1}=0$ and $w_{1}=0$ in order tho have a limit to the 4D
configurations, see (\ref{omega}) and (\ref{aux11}). For instance, if $%
w_{i}=0,$ one reduces (\ref{aux31}) \ to
\begin{equation*}
n_{2}^{\prime }-n_{3}^{\bullet }=0
\end{equation*}%
which can be solved for any $n_{2}=0$ and $n_{3}=n_{3}(x^{3},v),$ or,
inversely, for any $n_{2}=n_{2}(x^{2},v)$ and $n_{3}=0.$

In \ general, we conclude that any set of functions $h_{4,5}(x^{2},x^{3},v)$
and $w_{2,3}(x^{2},x^{3},v)$ and $n_{2,3}(x^{2},x^{3},v)$ for which the
system of equations (\ref{aux32}) and (\ref{aux31}) is compatible and has
nontrivial solutions, defined also as solutions of (\ref{ep1a})--(\ref{ep4a}%
), can be used for generation of exact solutions of the vacuum 4D (or 5D)
solutions of the vacuum Einstein equations. In this case, the Ricci tensors
for the Levi--Civita and the canonical d--connections have the same
coefficients for the N--adapted bases (\ref{ddif4}) and (\ref{dder4}) (with
extensions to 5D (\ref{ddif}) and (\ref{dder})). Such solutions can be
extended to certain matter and string like corrections if the effective
sources do not depend on covariant derivatives.

Finally, one should be emphasized that even we can nonholonomically
constrain the solutions in order to have zero distorsions of the Ricci
tensor (indused by deforming  the Levi Civita to the canonical d--connection
one) the distorsions of the d--curvature tensors do not vanish. For
instance, the components $\widehat{R}_{\ bjk}^{a}$ distinguished in (\ref%
{dcurv}) may be not zero, but they are not used for the definition of the
Ricci and Einstein tensor. We can have alow nontrivial values of the
nonholonomically induced canonical torsion (\ref{dtorsb}) because $\partial
N_{k}^{a}/\partial y^{b}$ may be not zero even $\Omega _{.jk}^{a}=0$. This
is an anholonomic frame effect which states that a generic off--diagonal
vacuum metric may be characterized by certain effective torsion coefficient
induced by the off--diagonal terms and related distorsions of curvature
tensor but with zero distorsions of the Ricci tensor.

\end{document}